# Spin Glasses: Disorder, Frustration, and Nonequilibrium Complexity

Naeimeh Tahriri,[1] Vahid Mahdikhah,[2] Jahanfar Abouie, [1,*] Daryoosh Vashaee[2,3,*]

[1]Department of Physics, Institute for Advanced Studies in Basic Sciences (IASBS), Zanjan 45137-66731, Iran

[2]Department of Electrical and Computer Engineering, North Carolina State University, Raleigh, NC 27606, USA

[3]Department of Materials Science and Engineering, North Carolina State University, Raleigh, NC 27606, USA

## Abstract

Spin glasses occupy a unique place in condensed matter: they freeze collectively while remaining structurally disordered, and they exhibit slow, history-dependent dynamics that reflect an exceptionally rugged free-energy landscape. This review provides an integrated account of spin-glass physics, emphasizing how microscopic ingredients (quenched randomness, frustration, competing exchange interactions, and random fields) conspire to produce macroscopic glassiness. We begin with the canonical Edwards-Anderson and Sherrington-Kirkpatrick formulations to introduce the central theoretical ideas that recur across the literature: extensive degeneracy, metastability, and the emergence of long relaxation times that manifest as aging, memory, and rejuvenation under standard experimental protocols. We then summarize the principal routes used to characterize spin-glass freezing, combining thermodynamic signatures with dynamical probes that reveal the separation of timescales and the sensitivity to thermal and magnetic histories. Building on these foundations, we draw connections across experimental material classes (metallic alloys, insulating oxides, and geometrically frustrated systems) by emphasizing how intrinsic versus induced disorder and competing interaction networks shape the observed phenomenology. Recent advances in reentrant and room-temperature spin-glass materials are highlighted as a rapidly developing direction that tests the limits of established paradigms and motivates new materials-driven questions. The review concludes by connecting modern computational developments, including machine-learning phase identification and neural-network analogies, to longstanding challenges in classification, universality, and out-of-equilibrium behavior, and by outlining emerging opportunities at the boundary between classical and quantum spin glasses.

## Contents

* jahan@iasbs.ac.ir, dvashae@ncsu.edu

# 1 Introduction

## 1.1 Spin Glasses: From Materials to Complex Systems

Spin glass materials have earned significant attention in materials science and condensed matter physics, with their unique properties leading to diverse applications and fundamental insights. The importance of this field was underscored by the 2021 Nobel Prize in Physics, which recognized groundbreaking contributions to the understanding of complex physical systems. Half of the prize was awarded to Giorgio Parisi of Sapienza University of Rome, Italy, for his revolutionary work on disorder and random phenomena in complex materials, particularly his studies on spin glasses.[1]

Spin glasses serve as prototype complex systems, characterized by randomness and disorder that make them challenging to comprehend. Parisi's work on these disordered magnetic systems has provided crucial insights into the principles underlying their behavior.[2] In materials science, the unique magnetic properties of spin glasses have been exploited for the development of magnetic recording media with enhanced stability. A recent breakthrough involves manganese-doped germanium telluride (MnGeTe), which exhibits a spin glass state that can be manipulated with extremely low energy input. This property could lead to computers that consume only one-millionth of the energy currently required to switch a bit, potentially revolutionizing energy-efficient computing and data storage.[3] Notably, Mn-doped GeTe compounds are also promising thermoelectric materials, offering multifunctionality for energy conversion and memory applications.[4]

In condensed matter physics, spin glasses serve as model systems for studying complex phenomena in disordered materials. They provide a platform for investigating concepts such as frustration, broken ergodicity, and replica symmetry breaking, which have implications for understanding other glassy systems and complex materials.[5] The study of spin glasses has contributed to our understanding of other complex systems, including structural glasses, neural networks, and even financial markets.[6] The concepts developed in spin glass theory, such as the replica method and mean-field theory[2], have found applications in these diverse fields, demonstrating the far-reaching impact of spin glass research.

Spin glasses have shown promising potential in thermoelectric applications, with both metallic and oxide spin glasses exhibiting interesting thermoelectric properties. Metallic spin glasses, such as CuMn alloys, exhibit interesting thermoelectric behavior related to their complex magnetic interactions.[7] However, oxide and insulating spin glasses have gained particular attention for their high thermopower. In oxide spin glasses, such as $La_{0.5}Ca_{0.5}Mn_{O3}$, a colossal thermoelectric power of -80 mV/K has been observed at 58 K, attributed to the coexistence of charge-ordered and spin glass states.[8] This observation has been explained by incorporating Kondo properties of the spin glass along with magnon scattering. Manganites have also demonstrated remarkable thermoelectric behavior. For instance, Gd-Sr manganites, specifically $Gd_{0.5}Sr_{0.5}MnO_3$, exhibit a colossal thermoelectric power of 35 mV/K at 40 K.[9] This colossal thermopower in oxide spin glasses has been attributed to the highly correlated electron spectra characteristic of manganites.[9] Furthermore, chemically complex alloys like Fe-Co-Ni-Mn have shown exceptional thermal stability in their spin glass state, with freezing temperatures above room temperature, making them promising candidates for practical thermoelectric applications.[10]

In the field of quantum computing, spin glass research has shown promise, particularly in the context of developing more robust and efficient qubits. Spin glasses, which are disordered magnetic systems, have attracted attention due to their unique properties that can be leveraged for quantum computation.[11] The ability to control and manipulate spin states in spin glass materials is crucial for their application in quantum computing, as it allows researchers to create and manipulate qubits with potentially longer coherence times, a critical factor in the performance of quantum computers.

Recent breakthroughs in this area include D-Wave's demonstration of a 5,000-qubit programmable spin glass system, which has shown quantum advantage on optimization problems.[12,13] This system employs quantum annealing, where a transverse-field term introduces quantum tunneling between metastable states, allowing the system to explore the energy landscape more efficiently than classical algorithms that rely solely on thermal activation.[14] Researchers have also made progress in understanding the critical properties of quantum spin glasses, including the critical transverse field at which the spin glass transition occurs, the critical exponents associated with the correlation length and spin glass susceptibility, and the spin-overlap distribution within the spin glass phase.[15] These findings have significant implications for developing quantum annealers, devices that use quantum mechanics to solve complex optimization problems.

While spin glasses themselves are not typically used as qubits, the study of their complex magnetic behavior and disorder can provide insights into managing and controlling quantum states in disordered systems. For example, a recent publication introduced the concept of topological quantum spin glass order, which combines aspects of spin glass order and topological order.[16] This new concept opens up avenues for both statistical mechanics and quantum computer science, further emphasizing the ongoing exploration and potential of spin glasses in quantum computing.

In the field of optics, researchers have implemented an optical simulation of spin glasses, which allows for the study of spin glass dynamics with tunable complexity. This approach offers advantages in scalability and the ability to vary constraints, potentially leading to new insights and applications in optical computing and information processing.[17]

The discovery of room-temperature spin glasses could potentially lead to the development of novel sensors, memory devices, and spintronics applications that operate under ambient conditions.[18] The ongoing exploration of spin glass materials and their properties promises to yield further innovative applications in technology and materials science.[19]

In computer science, spin glass models have been instrumental in developing algorithms for complex optimization problems. The inherent complexity and frustration in spin glass systems make them ideal for modeling and solving challenging computational tasks, particularly in artificial neural networks where they have been used to understand associative memory and storage capacity.[20]

Spin glass concepts have also found applications in biology, particularly in studying protein folding and RNA sequence evolution. The energy landscape of spin glasses shares similarities with the folding landscape of proteins, providing insights into protein folding mechanisms and potential implications for understanding and treating diseases related to protein misfolding.[21]

As we continue to resolve the complexities of spin glasses, their study remains at the forefront of physics, offering a window into the behavior of disordered systems and paving the way for technological advancements across multiple disciplines.

## 1.2 The Birth of Spin Glasses

In the early 1960s, scientists were studying magnetic properties of metal alloys. They focused on isolated magnetic impurities within non-magnetic metals. Dilute magnetic alloys such as $AuFe_x$ or $CuMn_x$ exhibit three-dimensional (3D) random-site spin glass behavior, where magnetic impurities are randomly distributed within a non-magnetic metallic matrix. Unlike the Ising model's restricted up-down spin orientations, the magnetic spins here possess full rotational freedom within the three-dimensional space characteristic of the Heisenberg universality class.[22,23] Later, they explored how these impurities interact with the metal's electrons. This led to the discovery of spin glasses in the late 1960s.[24,25,26,27] Spin glasses, first introduced by Bryan Coles, are a type of magnetic alloy where magnetic atoms are scattered throughout a non-magnetic metal.[28,29] These atoms freeze in random orientations, similar to how atoms freeze in a glass. This results in a unique set of properties, including a linear relationship between temperature and specific heat. In other word spin glasses are a magnetic phase characterized by random spin orientations without long-range order, similar to paramagnets. However, unlike paramagnets, spin glasses exhibit a unique property: individual spins are frozen into specific orientations, resulting in a non-zero local magnetization but a net zero magnetization for the entire system,[30] as depicted in Figure 1.

Spin glasses exhibit unique properties distinct from traditional glasses. To create a spin glass state, two key elements are required: 1) Competing Interactions: Magnetic moments within the material must experience conflicting interactions, preventing any single configuration from being favored.[31] 2) Random Interactions: These interactions should be somewhat random, leading to a disordered state. Due to these conditions, spin glasses exhibit a fundamentally different structure compared to conventional ordered materials.

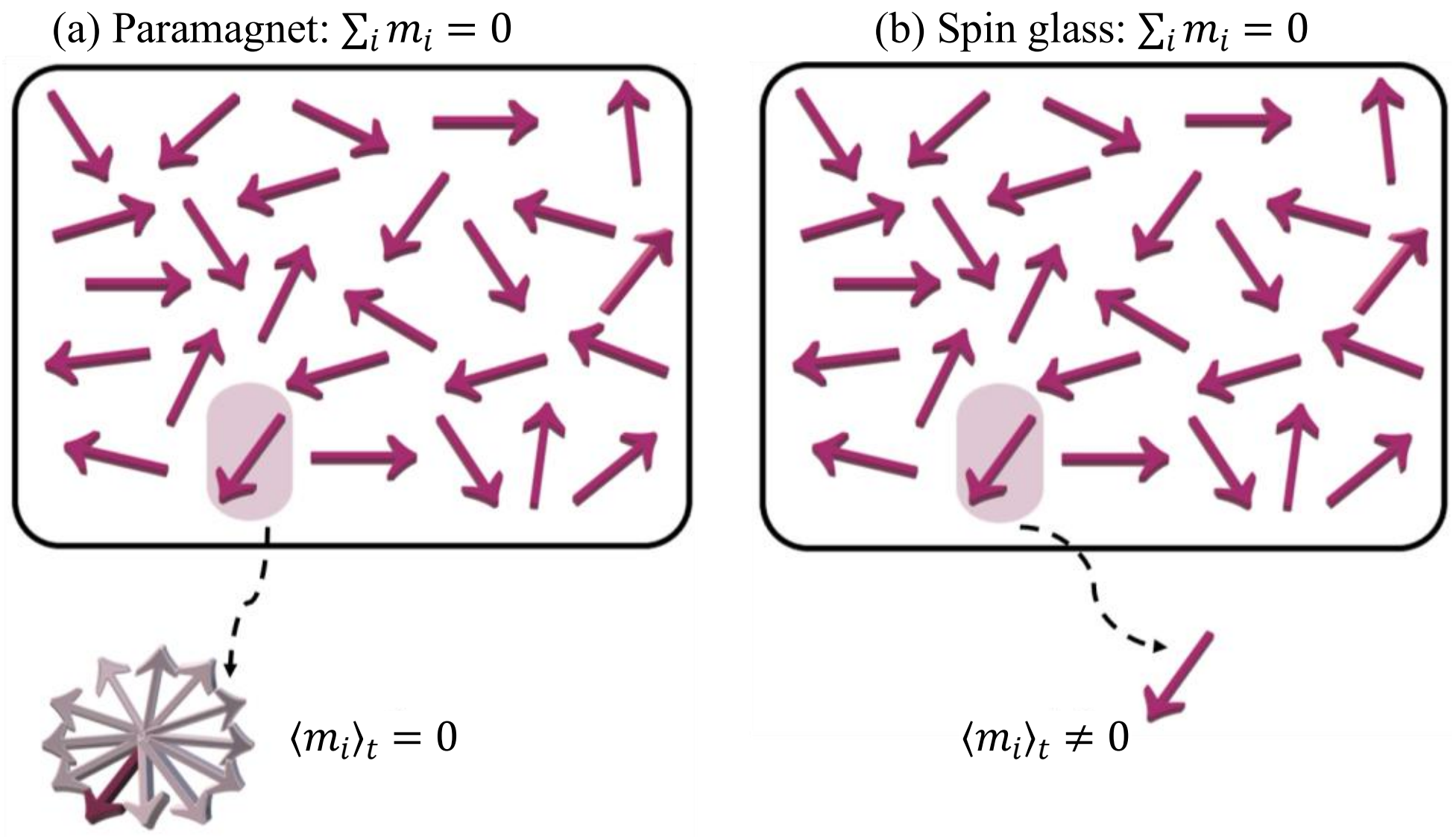

Figure 1. Magnetic moments of (a) paramagnetic and (b) spin glass systems. Both paramagnets and spin glasses exhibit a zero net magnetization. However, their magnetic behaviors differ significantly. In paramagnets, the magnetic moments of individual atoms are randomly oriented and can be easily aligned by an external magnetic field. In spin glasses, the magnetic moments are frozen in random directions, creating a disordered state. Despite the overall zero magnetization, local regions within the material can exhibit localized magnetic moments. $\langle ... \rangle_t$ is, the average over a sufficiently long observation time.

## 1.3 Phase Transitions in Spin Glasses

Spin glasses exhibit a complex phase diagram with multiple phases and transitions. Understanding the order parameter and phase transitions in spin glasses is crucial for understanding their unique behavior. The order parameter is a quantity that characterizes the state of a system and its phase transitions. In conventional magnetic systems, the magnetization is a suitable order parameter.[32] However, in spin glasses, the situation is more complex due to the disordered nature of the system. Several order parameters have been proposed for spin glasses, each with its advantages and limitations (Theoretical perspectives will be discussed in Section 4). Phase transition occurs as the temperature is lowered below the spin glass freezing temperature ($T_f$). At this temperature, the system enters the spin glass phase, characterized by a frozen, disordered state.[33] In some systems, a transition can occur from the paramagnetic phase to the spin glass phase, resembling the ferromagnetic-paramagnetic transition found in conventional magnets. In certain materials, a *reentrant spin glass transition* may also take place: the system first evolves from a magnetically ordered state (ferromagnetic or antiferromagnetic) into a spin glass state upon cooling, followed by a return to magnetic order at still lower temperatures.[34]

In spin glass systems, the term freezing temperature $T_f$ refers to the temperature below which spins become dynamically arrested in disordered orientations, marking the onset of spin glass behavior. This temperature is also commonly denoted as the spin glass transition temperature $T_{SG}$, and the two terms are used interchangeably in the spin glass literature. Below $T_f$, the system exhibits a non-ergodic, glassy magnetic state without long-range order but with frozen local moments. In contrast, the term glass transition temperature $T_g$ originates from the field of structural glasses and denotes the temperature at which a supercooled liquid becomes a rigid amorphous solid. While $T_g$ and $T_f$ may sometimes appear interchangeably in discussions of disordered systems, their physical contexts differ: $T_g$ describes kinetic arrest in molecular or structural glasses, whereas $T_f = T_{SG}$ represents the magnetic freezing point in spin glasses governed by frustrated spin interactions and randomness.

## 1.4 Magnetic Interactions in Spin Glasses

These interactions, including direct exchange,[35,36,†] Ruderman-Kittel-Kasuya-Yosida (RKKY),[37] dipolar,[38-41,‡] and super-exchange,[42,43,§] influence the behavior of the magnetic moments within the material. The RKKY interaction in spin glasses is a long-range interaction mediated by conduction electrons. It arises from the coupling between the localized magnetic moments of the spin glass atoms and the itinerant electrons in the metal. This interaction can be both ferromagnetic or antiferromagnetic, depending on the distance between the magnetic atoms and the properties of the conduction electrons. In spin glasses, the RKKY interaction plays a crucial role in determining the overall magnetic behavior of the system. The competition between ferromagnetic and antiferromagnetic RKKY interactions can lead to the formation of a disordered spin glass state. The strength and sign of the RKKY interaction can vary depending on the specific material and the arrangement of magnetic atoms within the lattice.

Figure 2 shows the exchange interaction parameter $J$ as a function of the interatomic distance $r$ (in units of the lattice constant) for four dilute magnetic alloys: CuMn, AuMn, AuFe, and PtMn. Positive $J$ values correspond to ferromagnetic (FM) coupling, favoring parallel alignment of magnetic moments (yellow region), whereas negative $J$ values correspond to antiferromagnetic (AFM) coupling, favoring antiparallel alignment (blue region). The curves exhibit an oscillatory dependence of $J(r)$, characteristic of the RKKY interaction, where the sign of the exchange alternates with distance. For CuMn, AuMn, and AuFe, the nearest-neighbor interaction ($r \approx 1$) is ferromagnetic ($J > 0$), while for PtMn it is antiferromagnetic ($J < 0$). At larger separations, all systems show a change in sign of $J$, indicating the coexistence of competing FM and AFM couplings that lead to magnetic frustration and the emergence of spin glass behavior.

As we mentioned the sign and magnitude of $J$ oscillate with distance for all alloys and these oscillations are attributed to the RKKY interaction, which arises from the polarization of conduction electrons surrounding the impurities. However, the observed periodicity of these oscillations deviates from the predictions of a simple RKKY model for isolated impurities (dotted line in Figure 2). This discrepancy suggests that the polarization of d-electrons, which are delocalized across several lattice spacings, plays a significant role in mediating impurity-impurity interactions. To fully understand the distance dependence of these interactions, cluster band structure calculations incorporating anisotropy may be necessary.[44]

---

† Examples of spin glasses dominated by direct exchange include $Eu_{1-x}Sr_xS$, $CdCr_{2x}In_{2(1-x)}S_4$ and $LiHo_xY_{1-x}F_4$[36].

‡ Dipolar spin glasses include insulating rare earth systems like $CuFe_{0.5}V_{0.5}O_2$[38], $LiHoF_4$[39], Gd-based system[40] and molecular nanomagnets[41] with large spin.

§ The nature of spin glass behavior can be seen in Cobalt-Doped Iron Disulfide[42]. Also, in layered oxyselenides ($La_2O_3(Mn_{1-x}Co_x)_2Se_2$)[43], the interactions between Mn/Co ions are mediated via super-exchange interaction through oxygen atoms.

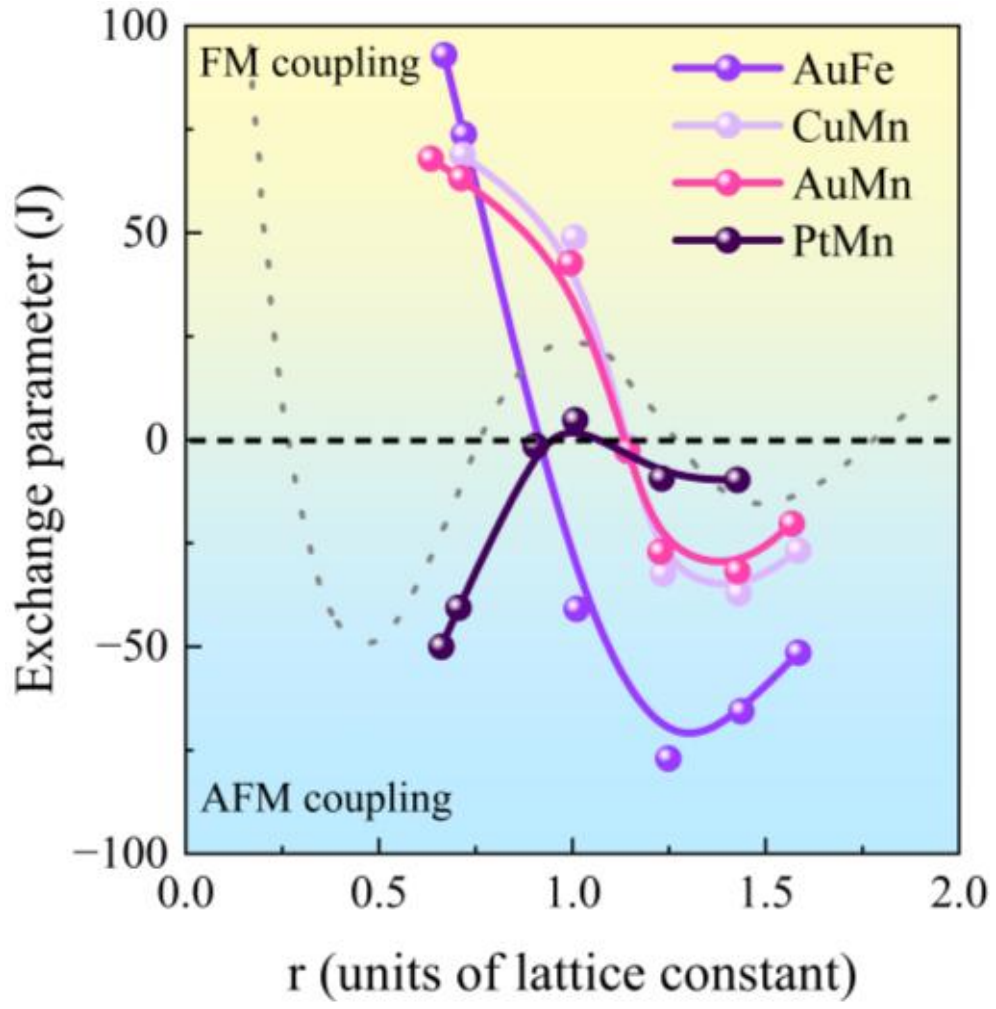

Figure 2. Estimated exchange parameters $J$ as a function of distance for four spin glass systems, derived from experimental susceptibility data through theoretical analysis. Positive $J$ values correspond to FM interactions (yellow region), while negative values correspond to AFM interactions (blue region). The dotted line represents the RKKY-type conduction electron polarization induced by a Mn impurity in Cu. Figure reconstructed from data reported in Ref. [44].

## 2 Origins of Spin Glass Behavior

Figure 3 outlines the primary factors contributing to the formation of a spin glass phase. These factors include frustration, random fields, disorder, and charge ordering, which will be discussed in detail in the following sections.

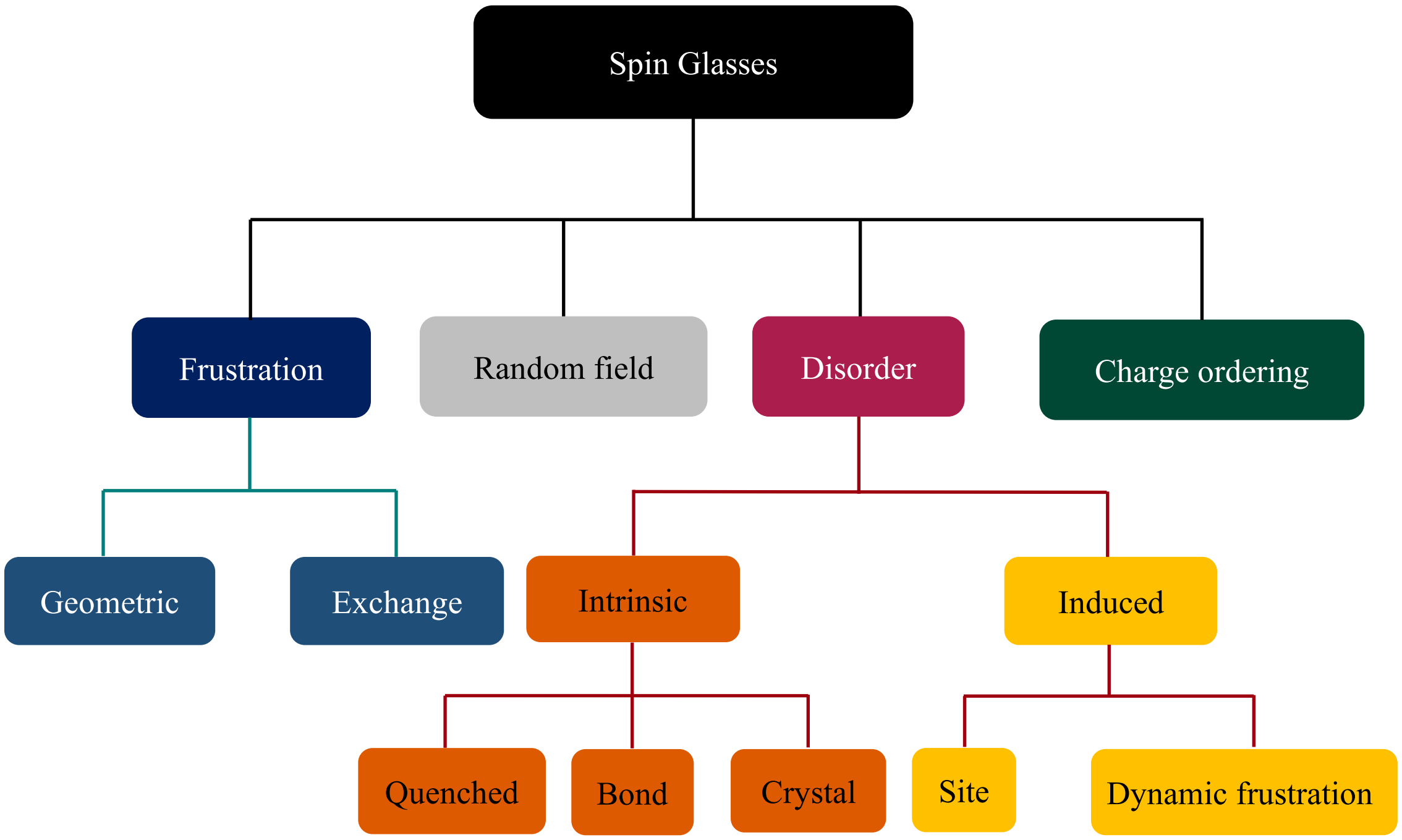

Figure 3. This diagram illustrates the key factors that can lead to the formation of a spin glass phase.

## 2.1 Disorder

Disorder is a crucial ingredient in the formation of a spin glass phase. Disorder can arise from random impurities in the material, variations in atomic positions, or even competing magnetic interactions. As a result, the spins find themselves in a state of frustration, unable to find a low-energy configuration that satisfies all interactions simultaneously. This leads to a complex, frozen state where spins are oriented randomly, but with a strong correlation between their orientations. This unique state, characterized by slow dynamics and a lack of long-range order, defines the spin glass phase.

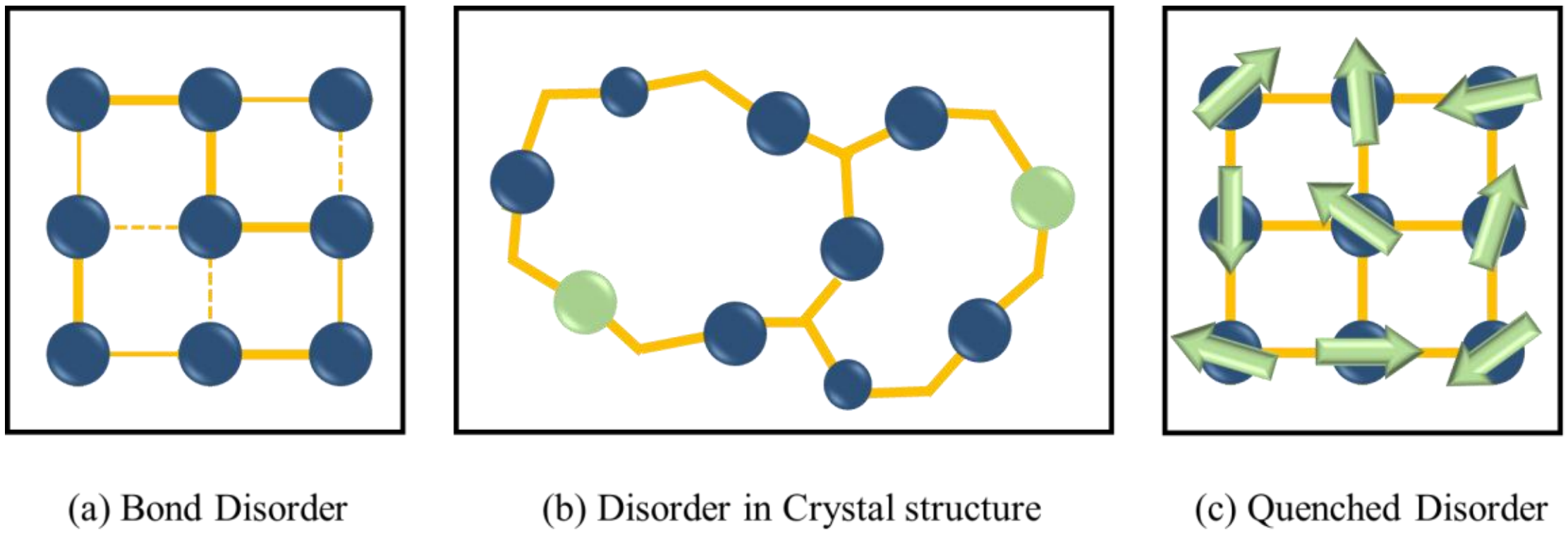

Figure 4. Schematic illustration of three types of disorder relevant to spin glass systems: (a) bond disorder, arising from random variations in interaction strengths between neighboring spins; (b) structural disorder, reflecting an irregular atomic arrangement due to defects, impurities, or random site occupancy; and (c) quenched disorder, originating from a frozen-in random spatial distribution of magnetic moments.

### 2.1.1 Intrinsic Disorder

#### 2.1.1.1 Disorder in Crystal Structure

Unlike the meticulously ordered arrangement of atoms in perfect crystals, amorphous materials exhibit a disordered, random structure. This irregularity stems from defects, impurities, or the random distribution of different atomic species. In the realm of spin glasses, this atomic-level disarray mirrors the structural randomness observed in conventional glasses. $Y_2Mo_2O_7$,[45] $Sr_2FeMoO_6$,[46] and some amorphous metallic glasses can be considered in this group. Notably, the concept of "self-induced spin glassiness" underscores the critical role of intrinsic disorder and competing interactions in the formation of the spin glass state. This phenomenon is exemplified by the anomalous spin glass behavior observed in single-crystalline elemental neodymium (Nd). The key materials in this category are diluted magnetic semiconductors (semi magnetic semiconductors), with the chemical formula of the form $A^{II}_{1-x}Mn_xB^{VI}$ (A= Hg, Cd, Zn, … and B=S, Se, Te, …) where Mn atoms are substituted at A sites, randomly. The random distribution of magnetic ions over the cation sublattice causes significant magnetic effects, such as the formation of a spin-glass-like phase at low temperatures.[47]

#### 2.1.1.2 Quenched Disorder

Quenched disorder arises from the random spatial distribution of magnetic moments or the random strengths of interactions between them, both of which remain fixed once the material is formed. A prime example is the CuMn alloy, where manganese atoms are randomly dispersed within the copper lattice. This random distribution of magnetic manganese atoms, established during the alloy's formation, persists indefinitely. The resulting spatial randomness in magnetic interactions contributes significantly to the spin glass behavior observed in both CuMn alloys and AuFe systems.[48,49]

#### 2.1.1.3 Bond Disorder

Bond disorder arises from variations in the strength of interactions (bonds) between magnetic spins within a material. Unlike ordered systems where bond strengths are uniform, disordered systems exhibit a random distribution of these strengths. This can occur due to a random distribution of ferromagnetic and antiferromagnetic interactions, leading to glassy behavior.[50] A prominent example of bond disorder is found in spin ice, where frustrated interactions between magnetic rare-earth ions on a pyrochlore lattice of corner-sharing tetrahedra give rise to a highly degenerate ground state. At low temperatures, spin ice systems exhibit a glass-like state due to the interplay of random interactions and geometric frustration.[51] This phenomenon, initially proposed for dilute pyrochlore spin ice, involves a glass transition driven by frustrated interactions among emergent degrees of freedom within the degenerate manifold. Examples of spin ice materials include pyrochlore compounds containing magnetic rare-earth ions such as $Ho_2Ti_2O_7$ and $Dy_2Ti_2O_7$.[52,53]

The main forms of disorder that give rise to spin glass behavior are summarized schematically in Figure 4.

### 2.1.2 Induced Disorder

#### 2.1.2.1 Dynamic Frustration

Dynamic frustration in spin glasses is caused by temporal fluctuations in the system, as opposed to static disorder which is time-invariant. It arises due to competing interactions that favor incompatible ground states. In the case of $PrAu_2Si_2$, it has a well-ordered crystal structure, which is not common for materials exhibiting spin glass behavior, as spin glasses are often associated with disorder. Here, the spin-glass-like behavior is not induced by disorder but by fluctuations in the crystal-field levels, which destabilize the induced magnetic moments. These fluctuations prevent the formation of long-range magnetic order and generate slow, glassy spin dynamics, effectively producing a spin-glass-like state through dynamic frustration rather than traditional disorder-induced freezing.[54]

#### 2.1.2.2 Site Disorder

Site disorder arises from irregularities in the occupancy of atomic positions within a crystal lattice, often disrupting the anticipated magnetic order. A prime example is the compound $URh_2Ge_2$, where observed spin glass behavior was attributed to site disorder within the rhodium and germanium sublattices. Crucially, extended annealing, which eliminated this disorder, transformed the material into an ordered antiferromagnet.[55] A specific manifestation of site disorder is the disorder-induced magnetic phase. In this scenario, the introduction of disorder triggers the emergence of a distinct magnetic phase. Such disorder can modify magnetic interactions within the material, leading to novel magnetic behaviors. For instance, in $Sr_6FeMoO_6$, significant magnetic disorder contributes to the formation of a spin glass phase.[46]

## 2.2 Frustration

Frustration, a cornerstone concept in the study of spin glasses, originates from the inherent inability to simultaneously satisfy all exchange interactions within a given spin configuration.[56] Frustration appears in two primary forms: geometrical and exchange-based. To illustrate the physics of frustration, we consider a two-dimensional Edward-Anderson Ising model on a square lattice with nearest-neighbor exchange couplings $J_{ij}$. In this model the exchange couplings are random which can take on either positive or negative values. A fundamental unit of the lattice is a plaquette, as depicted in Figure 5a, consisting of four spins and their four mutual couplings. If the number of negative bonds around a plaquette is even, it is always possible to find a pair of spin configurations that satisfy all the bonds. However, if the number of negative bonds is odd, there will be a conflict, and it becomes impossible to satisfy all the bonds simultaneously (Figure 5b). Another example is the kagome lattice, in which geometric frustration arises when antiferromagnetic spins on corner-sharing triangles cannot simultaneously minimize all nearest-neighbor interactions, leading to a highly degenerate ground state as depicted in Figure 5c and d.

This inability to satisfy all the bonds is known as frustration. Frustration leads to a degeneracy of ground states in the spin glass system, as there are many different ways to arrange the spins to minimize the energy.[30]

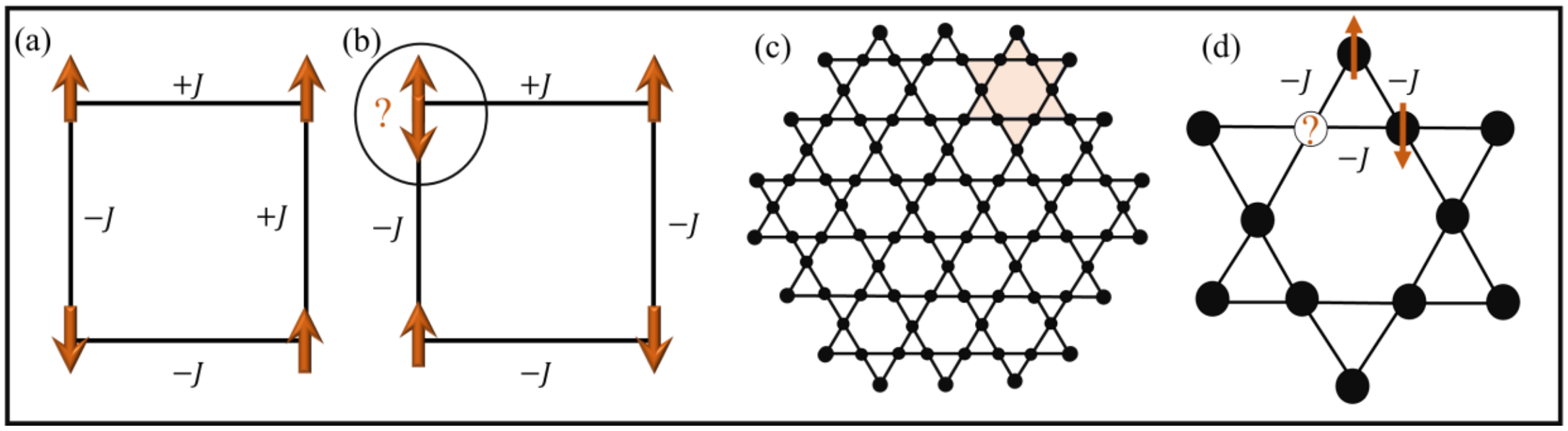

Figure 5. (a) A plaquette of the two-dimensional Edwards-Anderson Ising model with (a) an even number of negative couplings and (b) an odd number of negative couplings which lead to a frustrated state; (c) The kagome lattice and (d) a hexagram unit of the kagome lattice highlighting one of its constituent upward-pointing triangles; the antiferromagnetic interactions along the three edges of this triangle cannot all be satisfied simultaneously, illustrating the origin of geometric frustration in the lattice.

### 2.2.1 Geometrical Frustration

Originally introduced to describe real spin glass materials, this phenomenon arises from the specific geometrical arrangement of spins within the underlying lattice structure. It often manifests in magnetic systems and spin glasses, where the lattice imposes conflicting constraints on spin orientations. As a result, multiple energetically equivalent ground states emerge. Below is a categorization of the different types of geometrical frustration.

#### 2.2.1.1 Ising Antiferromagnet on Stacked Triangle Lattice

The anisotropic kagome antiferromagnet emerges as a potential candidate for glassiness without the need for disorder, owing to the inherent frustration arising from its geometric structure.[57] Unlike traditional spin glasses, where disorder plays a crucial role, the kagome lattice's unique arrangement of spins leads to competing interactions and a multitude of degenerate ground states. This concept is exemplified by the Ising model on a stacked triangular lattice, where antiferromagnetic interactions on a triangular motif prevent simultaneous energy minimization. The stacking of these frustrated layers amplifies the system's tendency towards glassiness.[58]

#### 2.2.1.2 Cluster Magnetism with Inner Cluster Geometry

Cluster magnetism with inner cluster geometry presents a fascinating interplay between geometric frustration and disorder, leading to complex magnetic behaviors. A sparse random graph model[**,59] has been developed to study this interplay, incorporating both triangular and tetrahedral structures where geometric restrictions make it challenging for spins to align optimally due to conflicting interactions.[60] As illustrated schematically in Figure 6, the triangular-cluster version of the model consists of strongly antiferromagnetic triangles whose internal geometric frustration persists even when the clusters are sparsely and randomly coupled through weak inter-cluster bonds. This model allows for the introduction of cluster network connectivity as a controllable parameter, providing insights into how the degree of interconnectedness affects magnetic properties.

[**] A random graph model is a theoretical framework in which spins (or clusters of spins) are placed on the nodes of a graph, and the magnetic interactions are represented by edges between nodes. The connectivity between nodes is assigned randomly, allowing one to systematically study how disorder in the interaction network influences magnetic phases. Such models are widely used in spin glass theory, especially in generalizations of the Bethe lattice, diluted lattices, and networks with controlled coordination number.[59]

In the case of triangular cluster geometry, the model reveals the emergence of a classical spin liquid state at a temperature $T_{SL}$, followed by a cluster spin glass (CSG) phase at a lower temperature $T_{CSG}$.[60] Notably, the CSG ground state remains robust even in conditions of very weak disorder or large negative intracluster interactions. While these phase transitions appear independent of network connectivity, variations in connectivity significantly influence the level of frustration for large positive intracluster interactions.[60] In contrast, the tetrahedral cluster geometry, which lacks intrinsic frustration, exhibits different behavior. The CSG ground state in this case can be suppressed under conditions of weak disorder or large negative intracluster interactions, and the CSG phase boundary displays a reentrance that depends on network connectivity. [60]

The study of cluster magnetism with inner cluster geometry extends beyond theoretical models to real materials. For instance, in amorphous $La_{0.7}Sr_{0.3}MnO_3$ films with a fractal structure, two distinct states of magnetic frustration have been observed, characteristic of CSG and SG states.[61] These states arise from competing FM and AFM interactions, as well as the geometry of fractal formations. The high density of clusters in these materials facilitates effective magnetic interactions between magnetic moments without the involvement of free charge carriers.

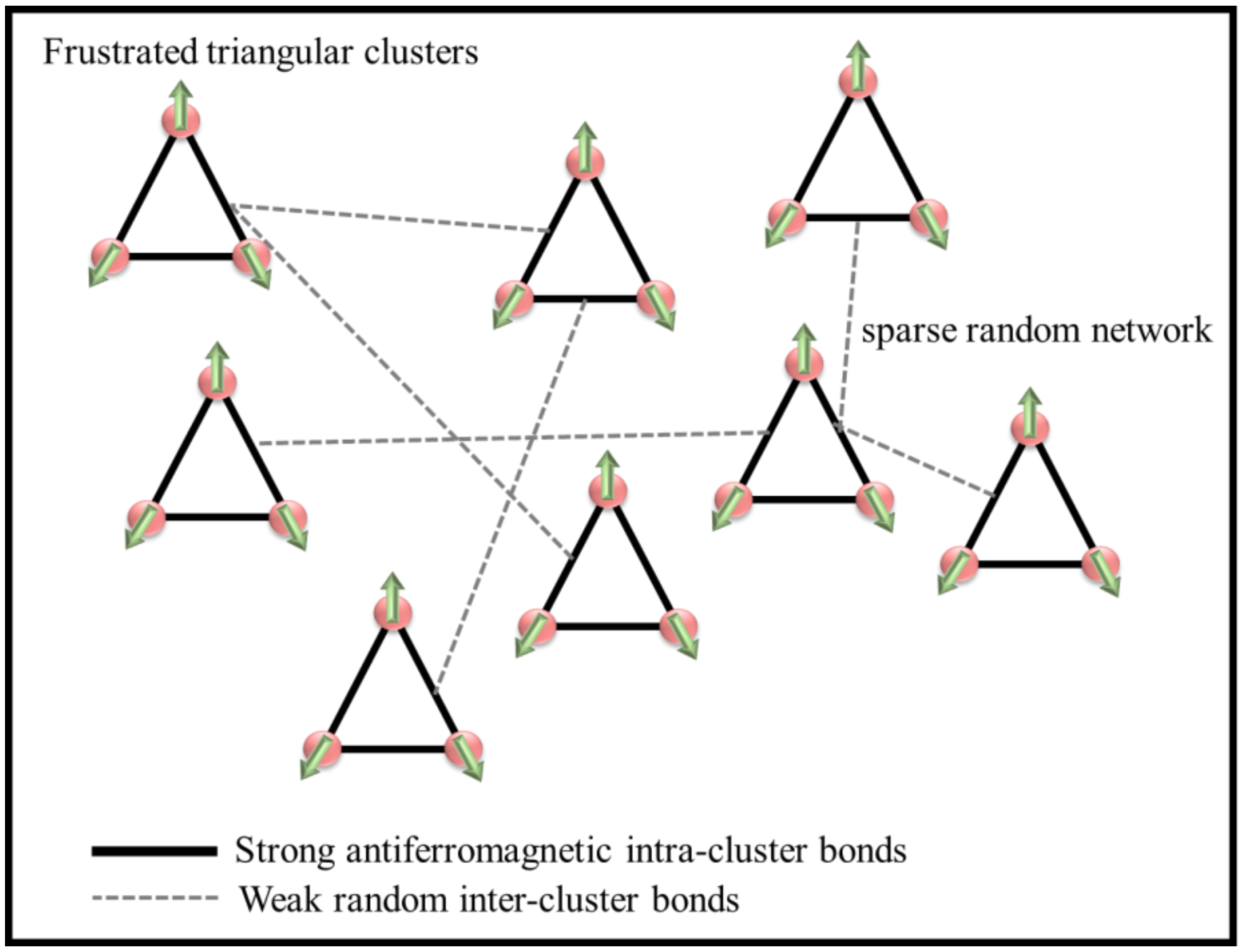

Figure 6. Schematic representation of the frustrated triangular-cluster model on a sparse random graph. Each triangle represents a rigidly coupled three-spin cluster with strong antiferromagnetic intra-cluster interactions, leading to intrinsic geometric frustration (shown by the incompatible 120° spin arrangement). The clusters are weakly and randomly interconnected via thin gray lines, forming a diluted network with tunable connectivity.

### 2.2.2 Exchange Frustration

#### 2.2.2.1 Square Lattice Ising Spin System

This theoretical model is based on a square lattice where each site hosts a spin that can take two states (up or down), and the interactions between neighboring spins are described by the Ising model. When the model is extended to include next-nearest-neighbor interactions, competing interactions can arise, leading to exchange frustration. This occurs not because of the geometry of the lattice, but because the different exchange couplings ($J_1, J_2$) favor incompatible spin alignments. In the context of spin glasses, such competition between interactions illustrates how frustration and disorder combine to produce complex magnetic behavior.[62]

#### 2.2.2.2 Exchange Bias Frustration

The interplay of magnetic moments can lead to exchange bias frustration within spin glass systems. This phenomenon arises from competing influences governing the orientation of neighboring spins. Studies on systems like NiMn/CoFeB bilayers offer insights into how spin glass behavior impacts exchange bias.[63] The observed decrease in exchange bias field with increasing cooling field highlights the role of spin glass frustration in influencing the magnetic properties of these bilayers. This type of frustration is well-documented in systems such as AgMn and CuMn dilute alloys, as well as FM/spin glass nanocomposites.[64,65]

Recent research has expanded our understanding of exchange bias frustration in spin glasses. For instance, a study on CuMn spin glasses demonstrated both conventional and spontaneous exchange bias effects.[66] The conventional exchange bias, typically observed in ferromagnet/antiferromagnet interfaces, was unexpectedly found in this spin glass system. More intriguingly, a spontaneous exchange bias was observed after zero-field cooling (field cooling and zero-field cooling are described in section 3.3). This effect is attributed to interactions between ferromagnetic clusters embedded within the spin glass matrix. This finding provides new insights into the relationship between exchange bias and spin glasses, offering a foundational understanding for exchange bias research.

Furthermore, the role of geometric frustration in producing exchange bias has been explored in materials like $Mn_3(C_6S_6)$.[67] Unlike traditional disorder-induced frustration, geometric frustration in these systems leads to exchange bias effects that are less sensitive to time and temperature perturbations compared to disordered spin glasses. This suggests that the observed exchange bias originates from geometric frustration and short-range magnetic correlations, rather than from structural disorder or long-range magnetic interactions. The ability to induce exchange bias isothermally in these systems, as opposed to only through field cooling in ferromagnet/antiferromagnet bilayers, opens up new possibilities for designing exchange bias materials with tailored properties for specific applications.

#### 2.2.2.3 Orbital Degrees of Freedom Induced Frustration

Orbital degrees of freedom significantly enrich the complexity of magnetic systems, often introducing frustration in spin glasses. The interplay between spin and orbital order parameters leads to a diverse range of magnetic phases, particularly in transition metal oxides with partially filled d-orbitals. For instance, in pyrochlore oxide crystals, the coupling of spin and orbital degrees of freedom can result in a glass transition without the need for extrinsic randomness or quenched disorder.[68] This phenomenon, known as a spin-orbital glass transition, occurs when distortions in the atomic lattice cause two types of rotational degrees of freedom of spins to become coupled, forming a glassy state at a specific critical temperature.

The complexity of orbital-induced frustration is further exemplified in perovskite manganates like $Pr_{1-x}Ca_xMnO_3$, where Jahn-Teller distortions and orbital ordering contribute to spin glass behavior.[69] In these systems, the charge-ordering transition temperature increases with hole concentration in certain regimes, and the long-range ordering of $Mn^{3+}$ and $Mn^{4+}$ ions is intricately linked to antiferromagnetic spin ordering and the ordering of $Mn^{3+}$ $e_g$ orbitals. This intricate interplay can lead to situations where charge ordering occurs at a higher temperature than spin ordering ($T_{\mathrm{CO}} > T_{\mathrm{N}}$), or where they occur concurrently ($T_{\mathrm{CO}} \geq T_{\mathrm{N}} \geq T_{\mathrm{N}}$).[69] The resulting frustrated state exhibits complex spin, charge, and orbital ordering patterns, including the CE-type antiferromagnetic phase. This configuration, first identified in half-doped manganites, consists of ferromagnetic zigzag chains that are coupled antiferromagnetically to neighboring chains and is typically accompanied by charge and orbital ordering.[70]

Recent studies have also revealed the dynamic nature of orbital-induced frustration in spin glasses. For example, in $PrAu_2Si_2$ (a stoichiometric compound with a well-ordered crystal structure) spin glass freezing has been observed to result from dynamic fluctuations of the crystal-field levels.[54] These fluctuations destabilize induced moments and frustrate the development of long-range magnetic correlations, providing a novel mechanism for producing a frustrated ground state. This dynamic aspect of orbital-induced frustration adds another layer of complexity to spin glass systems, challenging traditional views that both static frustration and disorder are essential ingredients in all spin glasses.[71]

## 2.3 Random Fields

Random fields, characterized by spatially varying magnetic fields in an Ising ferromagnet, constitute another pathway to spin glass behavior. These fields locally disrupt the system's rotational symmetry, leading to complex magnetic ordering. The random-field Ising model (RFIM) has been extensively studied as a prototypical system for understanding this phenomenon. In the RFIM, the Hamiltonian includes both exchange interactions and random local fields, resulting in a delicate balance between ferromagnetic order and disorder-induced frustration. The effect of quenched random fields is visualized in Figure 7: although the nearest-neighbor exchange still favors parallel spins, the local random fields $h_i$ (shown as small arrows acting on each site) overcome the ferromagnetic coupling at certain locations, nucleating reversed spins and breaking the system into a complex domain structure characteristic of the RFIM.

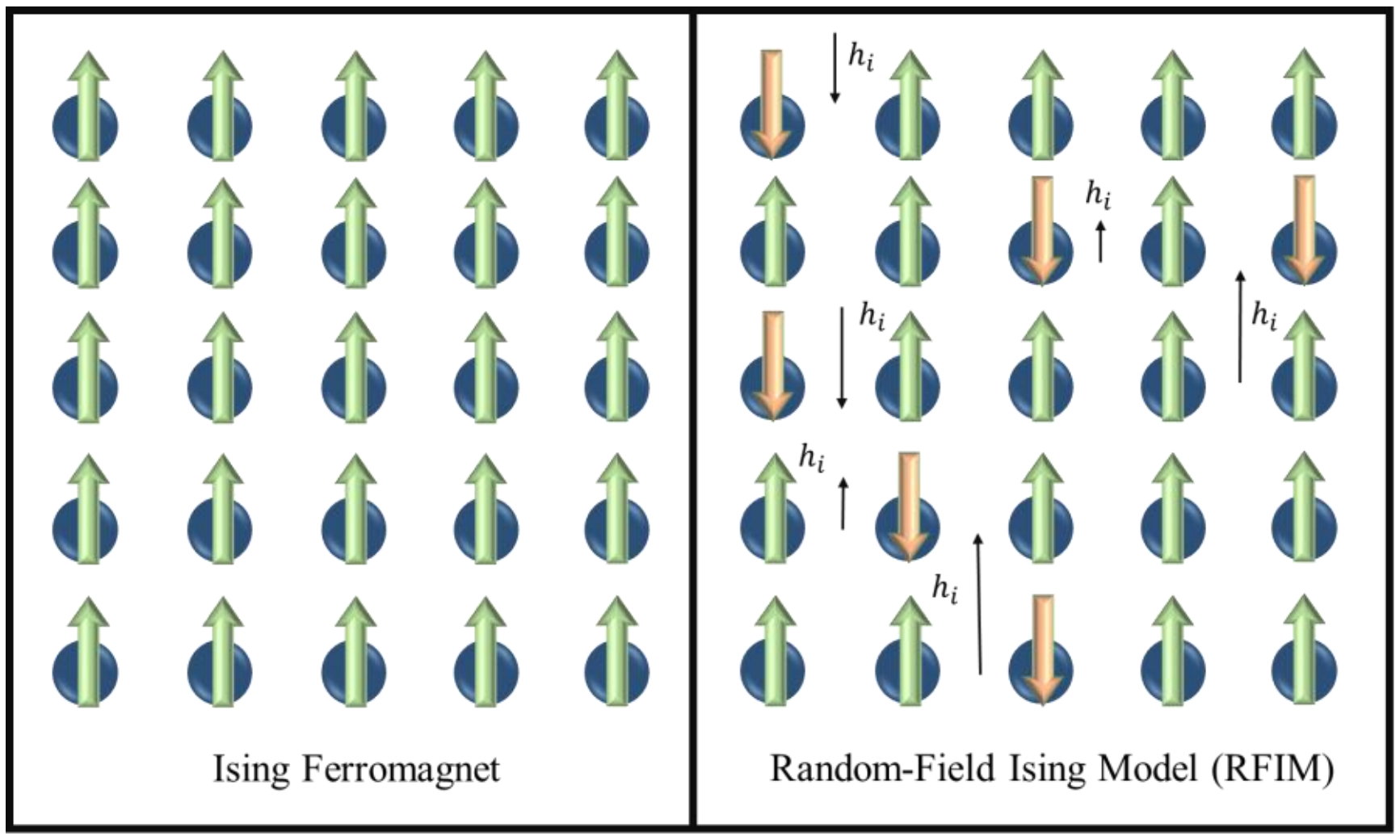

Figure 7. Schematic illustration of RFIM. Left: clean ferromagnet with all spins aligned in the absence of random fields. Right: the same ferromagnetic exchange interactions are now competing with site-dependent random fields of varying strength and direction, locally forcing some spins to point against the majority and destroying long-range order, leading to domain formation and spin-glass-like complexity.

Compounds like $Fe_xZn_{1-x}F_2$ exemplify the transition from a purely antiferromagnetic to a spin-glass-like state without altering the crystal structure. As the concentration of magnetic $Fe^{2+}$ ions decrease, the system exhibits a crossover from random field to spin glass behavior. For $x = 0.25$, the dc magnetization shows irreversibility, characteristic of spin glass systems.[71] Mössbauer studies have provided further evidence for this spin glass behavior in the $Fe_xZn_{1-x}F_2$ system.[72] The transition from long-range antiferromagnetic order to a spin glass state occurs as the iron concentration is reduced, with the spin glass phase emerging at lower concentrations due to the interplay between disorder and frustration.

Another example of random-field effect on antiferromagnets is, EuSe which exhibits spin-glass-like characteristics when subjected to an applied magnetic field. The competition between the antiferromagnetic exchange interactions and the Zeeman energy of the applied field leads to a complex magnetic phase diagram. This behavior can be understood in the context of the random-field model, where the applied field acts as a source of quenched disorder, disrupting the antiferromagnetic order and inducing glassy dynamics.

The study of random field systems has led to important theoretical developments in the understanding of disordered magnets. The droplet theory, proposed by Fisher and Huse, provides a framework for understanding the low-temperature properties of both spin glasses and random field systems.[73] According to this theory, the low-energy excitations in the spin glass or random-field phase are compact clusters of reversed spins, which dominate the system's thermodynamic and dynamic behavior. The interplay between random fields and geometric frustration can lead to novel magnetic states, as demonstrated in recent studies on cluster spin glasses with random graph structures.[60]

## 2.4 Charge Ordering

Charge ordering is a fascinating phenomenon that can lead to spin glass states in various materials, particularly in complex oxides and perovskites. This phenomenon involves the arrangement of metal ions with distinct oxidation states in specific lattice sites, often resulting in insulating or semiconducting behavior due to suppressed electron mobility. In rare-earth manganites such as $Ln_{1-x}A_xMnO_3$ and $Pr_{1-x}Sr_xMnO_3$, charge ordering is closely linked to complex magnetic and electronic properties, including the emergence of spin glass states.[74] As schematically shown in Figure 8, the regular checkerboard charge ordering of $Mn^{3+}$ and $Mn^{4+}$ ions in manganites creates a situation in which ferromagnetic and antiferromagnetic exchange paths compete on a rigid lattice, ultimately driving the system into a spin glass state with randomly frozen spin directions.

The interplay between charge ordering and spin glass behavior is particularly evident in systems with geometric frustration. For instance, in $La_{0.5}Ca_{0.5}MnO_3$, a colossal thermoelectric power of -80 mV/K has been observed at 58 K, attributed to the coexistence of charge-ordered and spin glass states.[8] This phenomenon is explained by incorporating Kondo properties of the spin glass along with magnon scattering. Similarly, in $Er_{0.1}Yb_{0.9}Fe_2O_4$, double two-dimensional charge ordering states have been observed at relatively high temperatures, with energy gaps of 170 meV and 193 meV between the 3D and the double 2D states.[75] These observations highlight the complex interplay between charge ordering and spin dynamics in frustrated systems.

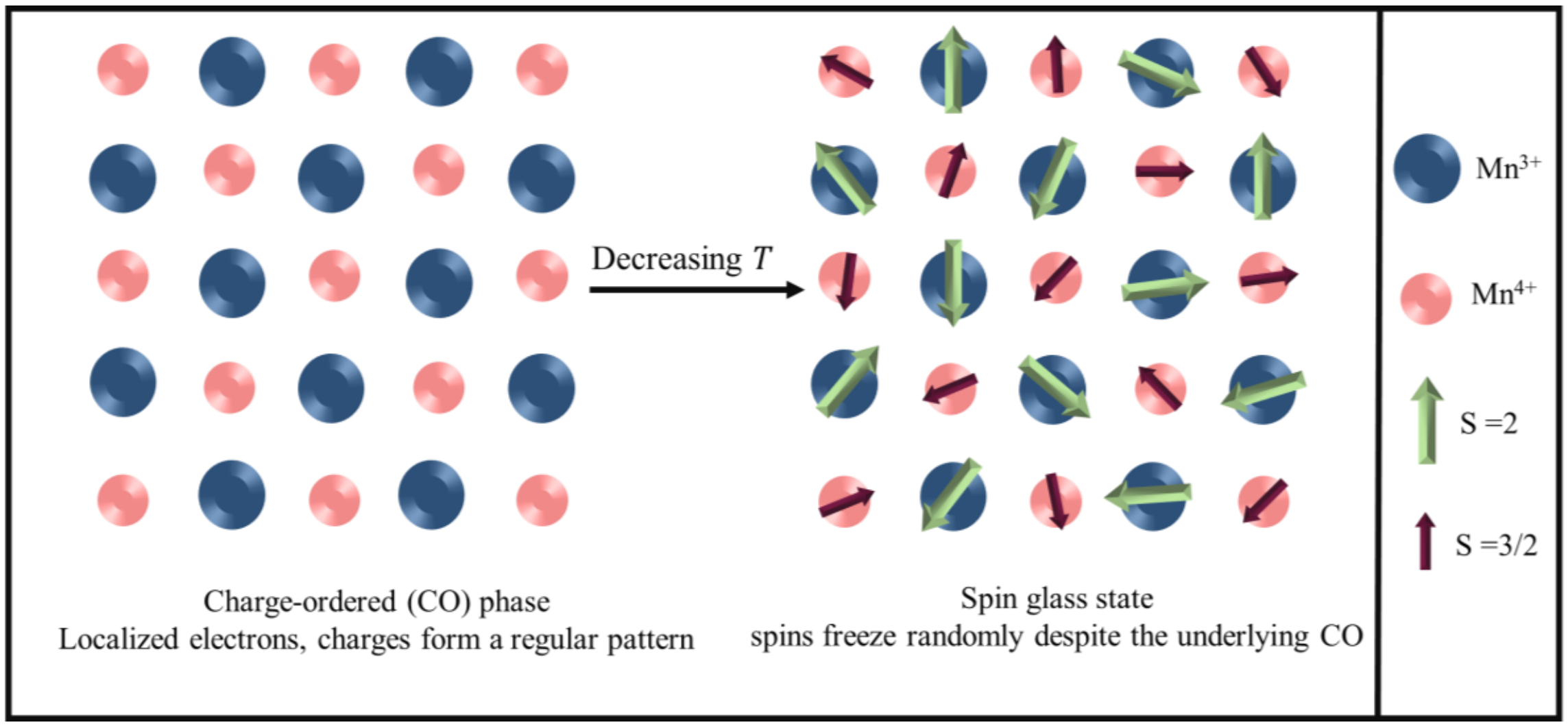

Figure 8. Schematic illustration of charge ordering inducing a spin glass state in manganites. Left: checkerboard charge-ordered arrangement of $Mn^{3+}$ (blue circles) and $Mn^{4+}$ (pink circles) at intermediate doping, localizing the $e_g$ electrons. Right: below the spin glass transition the Mn spins freeze in random orientations due to competing ferromagnetic double-exchange and antiferromagnetic superexchange interactions on the fixed charge-ordered background.

Double perovskites, such as $Sr_2FeMoO_6$ or $Sr_2CrReO_6$, also exhibit charge ordering and can host spin glass states. In these materials, the ordering of different cations on the B-site of the perovskite structure can lead to complex magnetic interactions and frustrated spin configurations.[76] For example, in $A_2BX_6$ vacancy-ordered halide double perovskites, the incorporation of trivalent B-site cations allows for a wide variety of magnetic cations to be included in the structure, leading to antiferromagnetic coupling of weak-to-moderate strength and signatures of frustrated long-range ordering.[77]

The relationship between charge ordering and spin glass behavior extends beyond oxide materials. Recent studies have shown that even in ionic liquids, charge ordering can lead to spin-glass-like states. In these systems, given the ionic positions from molecular simulations, the ionic charges minimize a "spin glass" Hamiltonian[††] for nearest-neighbor interactions.[78] This demonstrates that the principles of frustration and disorder that underlie spin glass formation in magnetic materials can also apply to charge degrees of freedom in non-magnetic systems, broadening our understanding of these complex phenomena.

# 3 Characterizations of Spin Glass Properties

Key experimental techniques for characterizing spin glass systems include specific heat capacity, entropy, nonequilibrium dynamics (aging, rejuvenation, and memory effects), DC magnetization (field-cooled and zero-field-cooled measurements), and AC susceptibility. These methods, as outlined by Mydosh and Binder,[79] provides essential insights into the unique properties of spin glasses. Spin glass behavior exhibits a strong time-dependent nature, highlighting the importance of the experimental timescale relative to the system's relaxation processes. The observed freezing temperature depends on the interplay between the material's relaxation time of magnetization ($\tau$), the characteristic time for spins to approach equilibrium, and the experimental observation time ($T$). This relationship is often quantified by the Deborah number ($\mathrm{De} = \tau/T$).[80] The complex energy landscape of spin glasses, characterized by numerous metastable states separated by significant energy barriers, results in a broad spectrum of relaxation times. Aging behavior arises in the system's response to perturbations, where the relaxation time depends on its prior history. This time-dependent response is further shaped by the depth and complexity of the energy landscape as well as the degree of disorder present.[81]

The experimental timescale plays a crucial role in determining the observed spin glass behavior. Techniques such as DC magnetization (seconds to hours) and AC susceptibility (microseconds) probe distinct aspects of the underlying dynamics. When measurements are performed rapidly, the system may become trapped in a metastable configuration, unable to fully relax, underscoring the need to account for timescale effects when interpreting experimental data.

If the experimental duration is much shorter than the characteristic spin relaxation time ($\mathrm{De} \gg 1$), the system remains effectively frozen in a nonequilibrium state, resembling a solid with limited responsiveness to external perturbations. This effect becomes particularly pronounced at low temperatures, where relaxation processes are slow. In contrast, experiments carried out over timescales much longer than the relaxation time ($\mathrm{De} << 1$) allow the system to explore its energy landscape more thoroughly, leading to fluid-like behavior that approaches equilibrium conditions.[80,81]

---

[††] This Hamiltonian, $H = \frac{1}{2}\sum_{i \neq j} J_{ij} z_i z_j$, is an effective, simplified representation where $z_i$ and $z_j$ represent the $\pm 1$ charge (or spin) of two interacting ions, and $J_{ij}$ is the interaction energy between them. In this context, $J_{ij}$ is proportional to the inverse of the distance between the ions, $J_{ij} = l_B/(r_i - r_j)$, making it the Coulomb energy of a pair of charges. The term $l_B$ in the paper refers to the Bjerrum length. It is defined by the formula: $l_B = \beta e^2/\epsilon$ where $e$ is the elementary charge and $\epsilon$ is the dielectric constant of the medium.[78]

Spin glasses exhibit a wide distribution of relaxation times, reflecting the diverse timescales over which individual spins or spin clusters reorient. As temperature decreases towards the freezing temperature, the energy landscape becomes rougher, leading to a broader distribution of energy barriers and, consequently, relaxation times.[82] As the system approaches the $T_f$ from above, a progressive slowdown occurs due to the increasing energy barriers that impede spin reorientation. At temperatures exceeding $T_f$, the relaxation time distribution peaks at shorter durations, indicating a larger fraction of spins capable of overcoming these barriers within the experimental timeframe. However, as the system is cooled through $T_f$, this distribution shifts towards longer timescales, reflecting the growing difficulty for spins to escape local energy minima. This transition signifies the onset of the spin glass phase, where the system becomes effectively frozen on experimental timescales due to the extremely long relaxation times associated with many spins. Incredibly slow relaxation processes represent a defining feature of spin glass behavior. When an external field is altered, the magnetization relaxes over exceptionally long timescales, leading to a pronounced delay in response to AC excitation.[83] This phenomenon (commonly referred to as *slow dynamics*) is typical of canonical spin glasses such as AuFe, AgMn, and CuMn.[84] The broad distribution of relaxation times prevents the system from reaching equilibrium within accessible experimental windows, especially below the freezing temperature ($T_f$).

As a result, the system exhibits *aging*, where its response depends on its prior history. Memory and rejuvenation effects, which reflect the system's capacity to retain information about past configurations, naturally arise from this nonequilibrium state. The strong frequency dependence observed in AC susceptibility measurements stems from this broad relaxation spectrum, as different frequencies probe distinct dynamical regimes.

As we will discuss the following sections, this transition from a paramagnetic state to a spin glass state is often characterized by several key features:

1. A cusp in the magnetic susceptibility
2. The onset of history-dependent magnetization behavior
3. A divergence of the nonlinear magnetic susceptibility
4. A residual entropy
5. The splitting of zero-field-cooled (ZFC) and field-cooled (FC) magnetization curves

It is important to note that in some spin glass systems, especially those with complex compositions or structures, there may be multiple characteristic temperatures or a range of temperatures over which the spin glass transition occurs. For instance, in chemically complex alloys, an ultrahigh freezing temperature above room temperature has been observed, which is unusual for conventional bulk spin glasses.[85]

## 3.1 Heat Capacity

Unlike conventional phase transitions, spin glasses exhibit either a broaden peak or no peak in specific heat capacity, without sharp discontinuities. For example, CuMn shows a broad maximum,[86] whereas spinel $Co_2SnO_4$ exhibits no peak.[87]

While the temperature derivative of specific heat shows a broad anomaly, its field derivative displays a sharper peak. For example, Figure 9a and b presents the specific heat and specific heat per temperature for CuMn ($T_f = 3.89$),[86] where the specific heat curve features a broad, rounded peak near $T_f$ rather than the sharp lambda-type anomaly typically observed in conventional phase transitions. Applying a magnetic field lowers the peak in $C$ and increases the specific heat at elevated temperatures. This occurs because the spins must now work against the field energy, making it harder for them to reorient freely.

At low temperatures, many spin glass materials show a specific heat that varies linearly with temperature. This behavior can be understood by considering low-energy spin excitations whose energies $\varepsilon$ are distributed with an approximately constant density of states as $\varepsilon \to 0$. If these modes are treated as bosonic excitations with occupation number $n(\varepsilon) = (\exp(\beta\varepsilon) - 1)^{-1}$ their contribution to the internal energy is $U = \int_0^\infty d\varepsilon \rho(\varepsilon) \varepsilon n(\varepsilon)$ where $\rho(\varepsilon)$ is the density of states. Assuming $\rho(\varepsilon) \approx \rho(0)$ for small energies and introducing the dimensionless variable $x = \beta\varepsilon$, the integral reduces to a term proportional to $(k_B T)^2 \rho(0) \int_0^\infty dx\, x n(x)$. The integral over $x$ is a constant, so the temperature dependence arises entirely from the prefactor $(k_B T)^2$. Differentiating with respect to temperature then gives a specific heat that grows linearly with $T$ as $T \to 0$.[30]

The temperature derivative of $C/T$ (Figure 10a right axis, $n = 1$) reveals a broad anomaly spanning several degrees around $T_f$, indicating a gradual change in the system's thermal properties. Most notably, the second temperature derivative of $C/T$ (Figure 10a right axis, $n = 2$) exhibits an anomaly at $T_f$[86] for $H = 0$, providing a clearer indication of the spin glass transition.

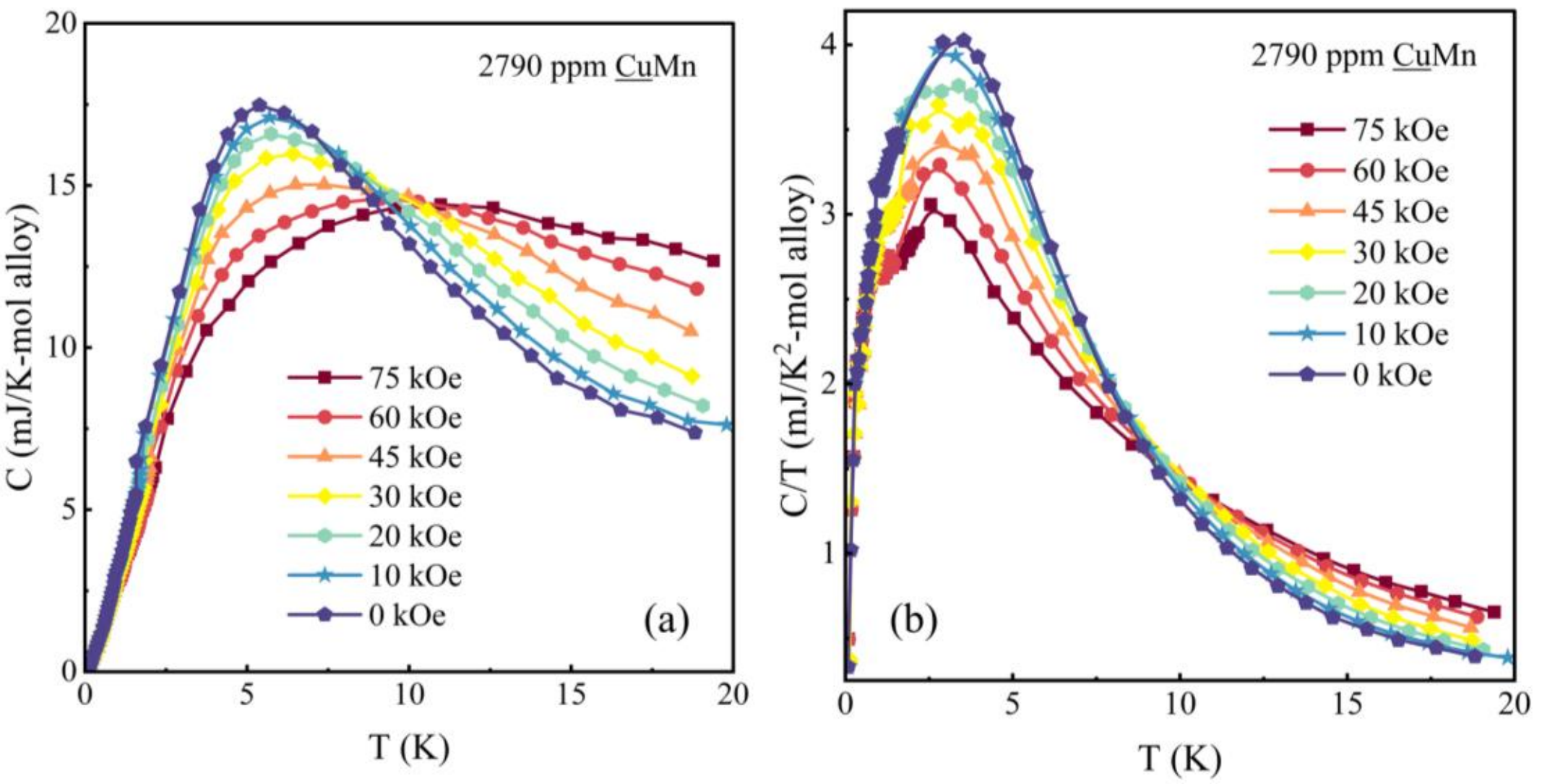

Figure 9. (a) Specific heat ($C$) and (b) $C/T$ of CuMn ($T_f$ = 3.89 K) showing a broad, rounded peak near $T_f$ rather than a sharp lambda-type anomaly. Applying a magnetic field suppresses the peak and increases $C$ at higher temperatures, as spins must work against the field. At low temperatures, the specific heat varies approximately linearly with temperature, typical of spin glass behavior. Figures reconstructed based on data extracted from Ref. [86].

These observations point out the complex nature of the spin glass state. The anomalous specific heat behavior was attributed to the non-ergodic nature inherent to disordered systems.[88] External magnetic fields disrupt the near-degenerate spin orientations within energy valleys, facilitating spin flips and consequently suppressing the spin glass state. These observations demonstrate the critical role of quenched disorder and a delicate interplay of interactions in stabilizing the spin glass phase. This anomalous behavior is indicative of a high degree of magnetic disorder, consistent with a spin glass state.[89]

As another example, a separate study characterized the specific heat of the spinel $Co_2SnO_4$, as shown in Figure 10b (left axis), at $H = 0$ Oe and $H = 1$ kOe.[87] No measurable difference is observed between the data for $H = 0$ Oe and $H = 1$ kOe, and there is no indication of a peak in $C_p(T)$ near $T_f = 39$ K, as would be expected for a conventional first- or second-order magnetic phase transition. Instead, only an inflection point is observed near 39 K, which results in a peak in $\partial C_p / \partial T$ at the same temperature, as shown in Figure 10b (right axis). If the contribution were purely magnetic, the specific heat would show a peak and then drop toward zero at higher temperatures. However, because the phonon contribution is also present, this peak is not observed, and the phonons drive the specific heat toward a saturated high-temperature value.

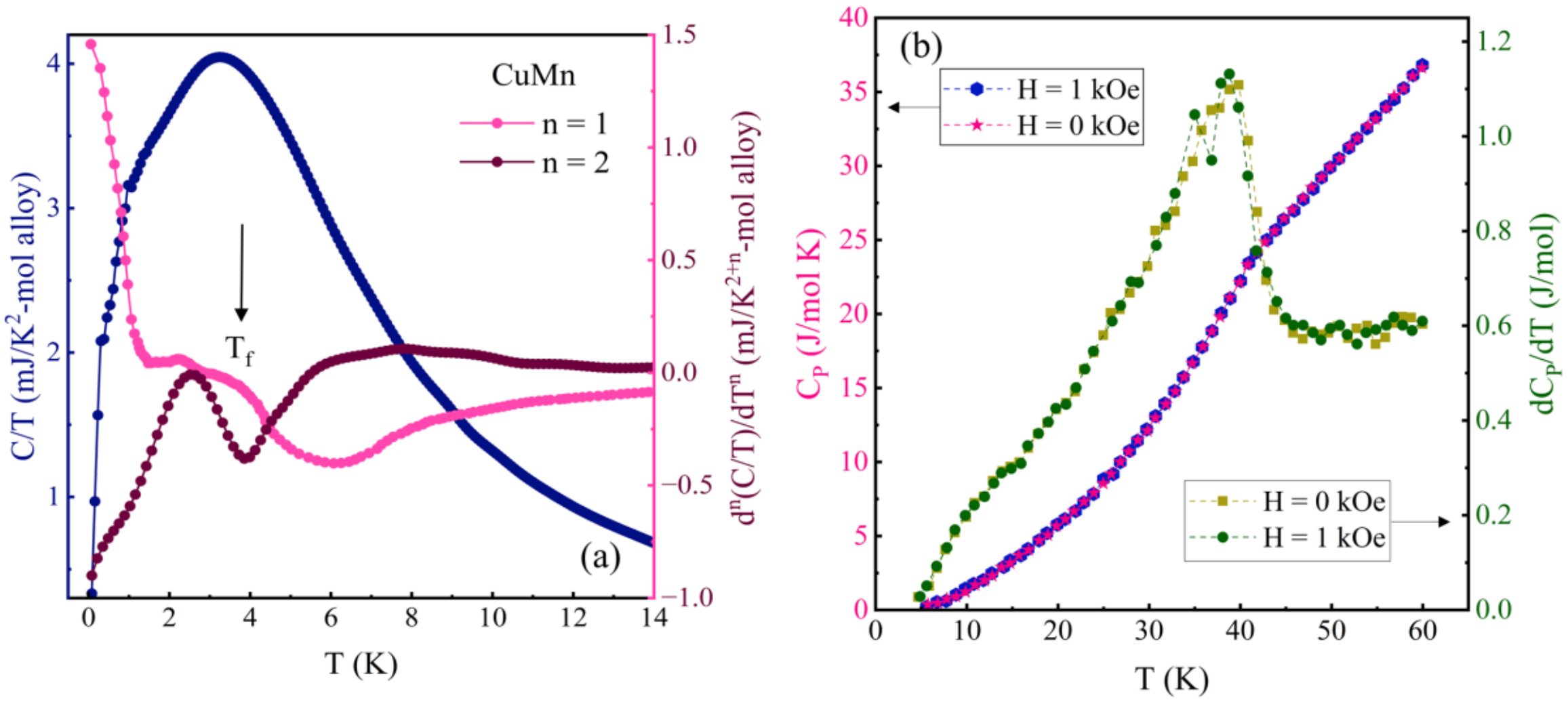

Figure 10. (a) (left axis) $C/T$ for $H = 0$, (right axis, $n = 1$) The temperature derivative of $C/T$, (right axis, n = 2) The second temperature derivative of $C/T$ as a function of temperature, exhibiting an anomaly at $T_f$ for $H = 0$. (b) Spin glass transition revealed by specific heat measurements in $Co_2SnO_4$. The left axis shows the temperature dependence of specific heat ($C_p$) measured during warming under zero magnetic field ($H = 0$ Oe) and an applied field of 1 kOe. The right y-axis and corresponding curve represent the temperature derivative of specific heat ($dC_p/dT$). Figure reconstructed based on data extracted from Ref. [87].

## 3.2 Entropy

A key consequence of the complex energy landscape of spin glasses is the existence of a non-zero entropy at absolute zero temperature, known as residual entropy or zero-point entropy (a finite amount of entropy that remains frozen in the system at a temperature of absolute zero).[90] The third law of thermodynamics states that the entropy of a system in thermal equilibrium should approach zero as the temperature approaches absolute zero. However, this does not apply to systems like glasses that are not in true thermodynamic equilibrium since the energy of the system may depend not only on the temperature, but also on the history of states of the sample.[90,91] Because they are out of equilibrium, the "thermodynamic" entropy, determined by heat flow, is not equal to the "statistical" entropy, which measures volumes in phase space.[92,93] As a spin glass is cooled, it becomes trapped in one of the many deep minima of its energy landscape, unable to explore the entire phase space and reach the true ground state. This trapping in a multitude of possible ground state configurations leads to a macroscopic degeneracy, resulting in a finite entropy at zero temperature.[94] Because the system is frozen into one of a vast number of possible configurations ($W > 1$), it possesses a non-zero entropy at $T = 0$, given by the Boltzmann-Planck equation: $S^0 = k_B \ln W$, where $W$ is the number of accessible ground states.[91]

This residual entropy is a key feature of frustrated systems, as it characterizes the complexity of the ground state. For spin glasses, the problem is not only of theoretical interest but has also been the subject of experimental investigation. Its experimental determination is challenging but provides a crucial test for theoretical models of spin glasses. The first step is to measure the total specific heat ($C_{\text{total}}$) of the spin glass sample over a wide temperature range, starting from a very low temperature (e.g., 0.5 K) up to well above the freezing temperature. This measurement includes contributions from lattice vibrations (phonons) and the magnetic spins. To isolate the magnetic contribution, the specific heat of a non-magnetic, structurally identical (isostructural) compound is also measured. This provides the lattice contribution ($C_{\text{lattice}}$). The magnetic specific heat is then found by subtraction: $C_{\text{mag}}(T) = C_{\text{total}}(T) - C_{\text{lattice}}(T)$.[95] The entropy released by the spins as they order upon cooling (or gained upon heating) is calculated by integrating the magnetic specific heat divided by temperature. This integration is performed from the lowest measurement temperature up to a high temperature where all magnetic ordering is gone and the spins are completely random, $\Delta S_{\mathrm{mag}} = \int [C_{\mathrm{mag}}(T)/T]\, dT$[92] (integrated from $T_{\mathrm{low}}$ to $T_{\mathrm{high}}$). For a mole of magnetic ions with a specific spin quantum number $S_q$, the total entropy associated with the complete disorder of these spins is a fixed theoretical value given by: $S_{\mathrm{total}} = R \ln(2S_q + 1)$ where $R$ is the ideal gas constant. The residual entropy is the portion of the total magnetic entropy that was not released as heat upon cooling, because the system became trapped in a disordered state. It is the difference between the theoretical maximum and the experimentally measured entropy change: $S_0 = S_{\mathrm{total}} - \Delta S_{\mathrm{mag}}$.[96]

For $CuGa_2O_4$ and $CuAl_2O_4$, researchers performed these measurements and found that the experimentally integrated entropy ($\Delta S_{\mathrm{mag}}$) was significantly less than the expected total entropy ($S_{\mathrm{total}}$). This "missing entropy" is the residual entropy, $S_0$, which remains locked in the glass at $T = 0$. The values of $S_0 = 4.96$ $JK^{-1}mol^{-1}$ for $CuGa_2O_4$ and $S_0 = 4.76$ $JK^{-1}mol^{-1}$ for $CuAl_2O_4$ indicate that these materials have a very high degeneracy in their ground states, even greater than predicted by some theoretical spin glass models.[91]

## 3.3 ZFC/FC Magnetization Splitting

Zero-field cooled (ZFC) and field-cooled (FC) magnetization measurements are fundamental techniques for characterizing magnetic materials. In a ZFC measurement, the sample is first cooled in the absence of an external magnetic field, and magnetization is subsequently measured upon applying a field. In contrast, FC measurements involve cooling the sample while a field is applied and then monitoring the magnetization as the temperature changes. Comparing ZFC and FC curves provides valuable insights into magnetic behavior, including phase transitions and irreversibility.

Different magnetic materials exhibit characteristic ZFC/FC behaviors. Paramagnets exhibit overlapping ZFC and FC curves due to the absence of long-range magnetic order, following Curie's law. [97] In a conventional ferromagnet the system develops a spontaneous magnetization below the Curie temperature $T_{\mathrm{C}}$ ; cooling in a finite field (FC) typically yields a larger magnetization than the ZFC protocol because the FC history aligns domains and reduces domain-wall pinning, producing an irreversible splitting of the ZFC and FC curves below $T_{\mathrm{C}}$ (Figure 11a).[98] In a simple antiferromagnet the susceptibility commonly exhibits a cusp at the Néel temperature $T_{\mathrm{N}}$ (a susceptibility maximum associated with the onset of antiparallel order). If the antiferromagnet is ideal and measured in small probing fields, ZFC and FC curves largely overlap (Figure 11b) because the ordered state is an equilibrium one and does not retain strong history dependence; however, real antiferromagnets with spin canting, domains, or disorder often show some weak ZFC-FC irreversibility or broadened peaks.[99]

A spin glass is distinct. It shows (1) a pronounced cusp in the low-field ZFC magnetization at the freezing temperature $T_f$, (2) a large ZFC–FC bifurcation below $T_f$ with the FC magnetization remaining relatively flat or only weakly temperature dependent on cooling (Figure 11c), and (3) characteristic nonequilibrium signatures such as frequency dependence of the cusp in AC susceptibility, aging, and memory effects. These dynamical fingerprints (frequency shift of $T_f$, slow relaxation/aging) distinguish spin glasses unambiguously from ordinary antiferromagnets and ferromagnets that can also display cusps or weak irreversibility due to domain or canting effects.[100]

The presence of a ZFC cusp and a bifurcation between ZFC and FC magnetization suggests that the system may be a spin glass, but further measurements are needed for confirmation. AC susceptibility should be measured at multiple frequencies to check for a frequency-dependent cusp (signature of glassy freezing). Additionally, applying a small DC field during susceptibility measurements can reveal field-dependent peak broadening and shifts: if the peak changes with the field, the system is likely a spin glass; if it remains stable, the system is not a spin glass. Complementary relaxation and aging experiments provide further confirmation.

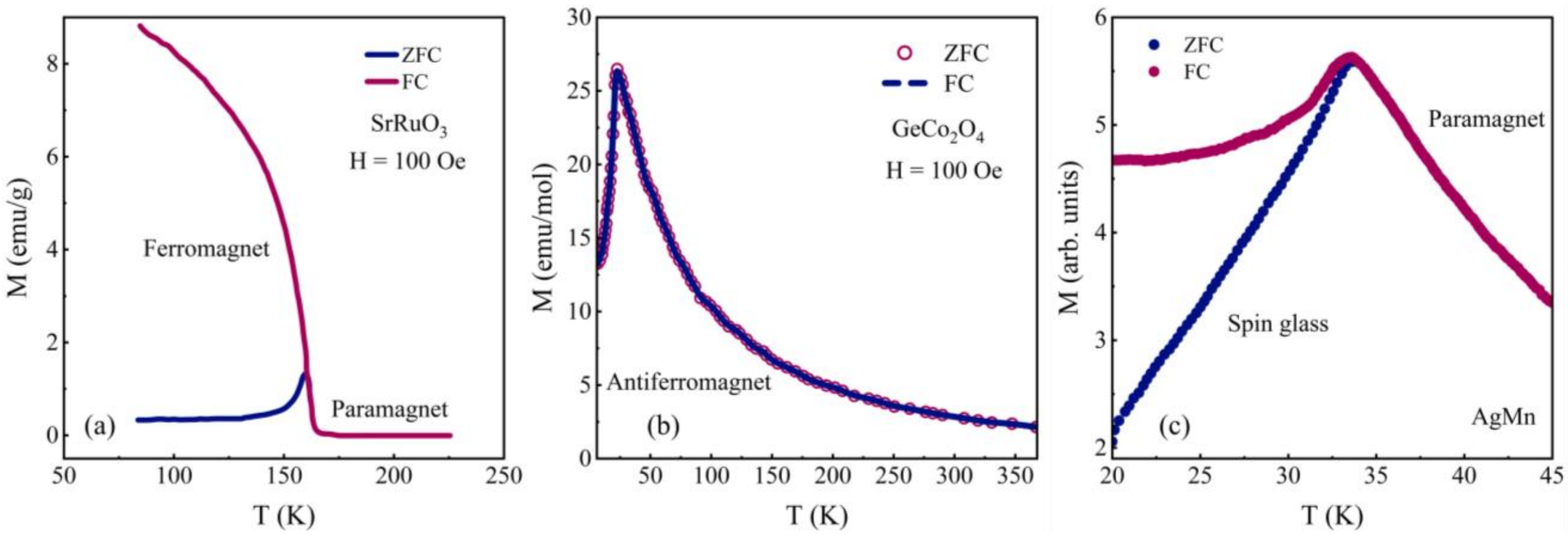

Figure 11. ZFC-FC magnetization measurements for (a) a ferromagnet ($SrRuO_3$). (b) An antiferromagnet ($GeCo_2O_4$) and (c) a spin glass (AgMn) under 0.1 Oe field. Figures reconstructed based on data extracted from refs. [98], [99] and [101] respectively.

## 3.4 Aging, Rejuvenation and Memory (Nonequilibrium Dynamics)

Temperature cycling experiments have been instrumental in uncovering the complex dynamics of glassy systems. By subjecting spin glasses to repeated temperature variations, researchers have observed intriguing phenomena such as aging, rejuvenation, and memory effects. Aging, the gradual slowing down of relaxation processes, is a well-established characteristic of glasses. Rejuvenation, a seemingly counter-intuitive reversal of aging, occurs upon reheating the system. Remarkably, the system retains a "memory" of its previous state, as evidenced by its return to the original relaxation behavior upon cooling back to the initial temperature. These observations evidence the hierarchical nature of the energy landscape in glassy systems and the importance of time-dependent processes in determining their behavior.[102] Aging effects have been extensively studied in canonical spin glasses such as CuMn[103] and AgMn[104]. Additionally, AuFe[105] alloys exhibit pronounced memory effects, highlighting the universality of these phenomena in this class of materials. A detailed description follows.

Spin glasses exhibit pronounced deviations from equilibrium behavior at low temperatures. Unlike conventional magnets, they fail to reach thermal equilibrium within experimentally accessible timescales. This nonequilibrium state manifests in phenomena such as aging, memory, and rejuvenation. Understanding the temporal evolution of spin glasses under varying temperature and magnetic field conditions necessitates a deep comprehension of these dynamic processes. The interplay between the system's history and its response to external stimuli is essential for unraveling the complexities of spin glass behavior.[106] Spin glasses exhibit characteristically slow dynamics, often becoming trapped in metastable states far from equilibrium. This sluggish relaxation gives rise to a rich array of nonequilibrium phenomena.[107,108,79] A hallmark of spin glasses is their pronounced dependence on experimental history. Unlike systems in equilibrium, where measurements are independent of prior conditions, spin glasses exhibit a strong correlation with waiting time ($t_w$), the duration of a system's exposure to specific conditions before measurement. This phenomenon, known as aging, manifests as a progressive slowing down of the system's response as $t_w$ increases. Essentially, the system's history influences its present state, demonstrating a clear deviation from equilibrium behavior.

Rejuvenation and memory effects are features of spin glass dynamics, reflecting the system's intricate response to changing external conditions. These phenomena arise from the system's history-dependent evolution within a complex energy landscape. Rejuvenation occurs when a previously aged system, upon experiencing a change in conditions, exhibits a reset of its relaxation processes. Conversely, memory effects manifest as the system's ability to "recall" its previous state upon returning to prior conditions. These behaviors underscore the profound influence of time on the spin glass state, emphasizing its deviation from equilibrium dynamics.[109]

An illustrative exemplification of this behavior is captured through the data of the $La_{2-x}Sm_xNiMnO_6$, as depicted in Figure 12a[110]. A comprehensive experimental protocol was employed to investigate the time-dependent magnetic response of the system. The sample was first cooled in zero field (ZFC) to 15 K, after which a 10 Oe magnetic field was applied and the magnetization was recorded for 3600 seconds ($t_1$). It was then rapidly cooled to 10 K and held for another 3600 seconds ($t_2$) before being reheated to 15 K for a final measurement period of 3600 seconds ($t_3$). The resulting magnetization curves for $t_1$ and $t_3$ exhibit a pronounced overlap, demonstrating the system's ability to recover its prior state following a temperature cycle. This experimental protocol provides a clear means to probe memory and rejuvenation effects in spin glasses.

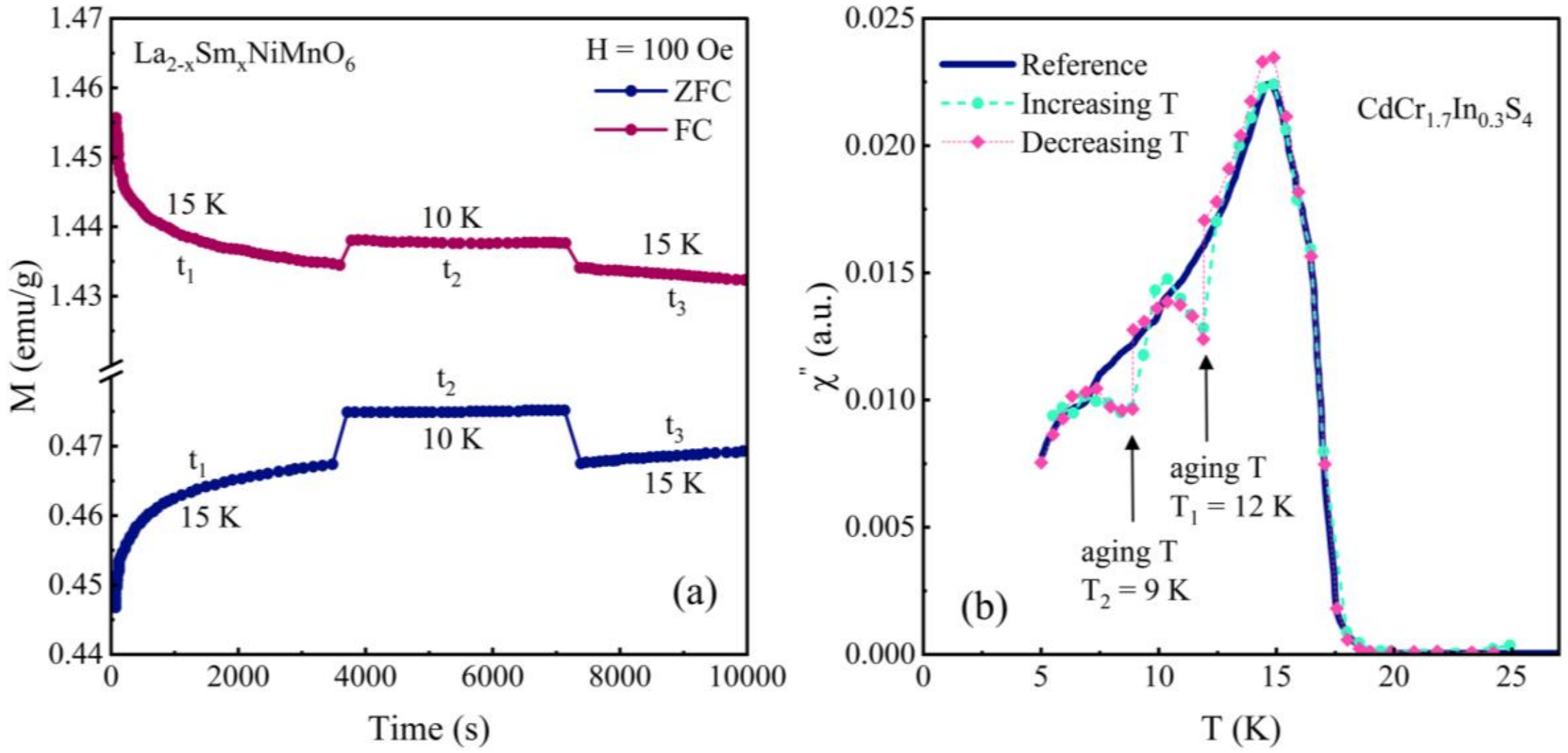

Figure 12. (a) Magnetic relaxation behavior of $La_{(2-x)}Sm_xNiMnO_6$ for $x = 0$, spin glass at 15 K under a 100 Oe magnetic field using ZFC and FC protocols. The relaxation behavior suggests that the characteristic curves for cycle $t_3$ represent a continuation of the trend observed in the curves for cycle $t_1$, thereby demonstrating a robust magnetic memory effect. (b) An example of multiple rejuvenation and memory steps for $CdCr_{1.7}In_{0.3}S_4$. The solid line is measured upon heating the sample at a constant rate of 0.1 K/min (reference curve). Pink symbols: the measurement is done during cooling at this same rate, except that the cooling procedure two stops during cooling, which allow the spin glass to age 7 h at 12 K and then 40 h at 9 K. Both aging memories are retrieved independently when heating back (cyan circles). Figures reconstructed based on data extracted from refs. [110] and [109] respectively.

AC susceptibility measurements provide insights into the memory and rejuvenation processes characteristic of spin glasses. Figure 12b illustrates the nonequilibrium dynamics of the insulating Heisenberg spin glass $CdCr_{1.7}In_{0.3}S_4$ below $T_f$.[109] The out-of-phase AC susceptibility ($\chi''$) was monitored during both cooling and heating cycles to investigate the system's dynamic response. A reference curve, obtained under continuous cooling, served as a baseline. Introducing deliberate temperature pauses ($T_1$, $T_2$) during cooling revealed aging effects, manifested as a decrease in $\chi''$. Subsequent warming led to rejuvenation, where $\chi''$ recovered towards the reference curve. Intriguingly, even without pauses, $\chi''$ exhibited aging-like behavior during continuous warming, indicating a system memory of its previous cooling history. These observations underscore the complex interplay of aging and rejuvenation in spin glasses, revealing their intricate nonequilibrium dynamics.

## 3.5 AC Susceptibility

As mentioned in section 3.3 and 3.4 the distinguishing mark of spin glasses, a sharp cusp in AC susceptibility, was first observed in AuFe and CuMn alloys during the early 1970s. This discovery marked a significant breakthrough in understanding these complex magnetic systems. Previous studies on these al-loys had been conducted in relatively high magnetic fields, which obscured this critical feature. By employing low-frequency (50-155 Hz) and low-field (5 Gauss) measurements, researchers identified a pronounced peak in susceptibility, signaling a new phase of matter.[111] The peak's magnitude increased with Fe concentration, deviating from expectations based on the traditional molecular-field model. While comparable in magnitude to antiferromagnetic transitions, the spin glass cusp lacked the characteristic step-like behavior and was more susceptible to experimental limitations at higher temperatures.

AC susceptibility measurements have since become a standard technique for characterizing spin glasses, as they are particularly sensitive to the freezing transition, which is not easily detectable through other methods such as specific heat measurements.[112] The frequency dependence of the AC susceptibility is a key feature that distinguishes spin glasses from other magnetic systems. For instance, in $Cu_{1-x}Mn_x$ alloys, a prototypical spin glass system, the location of the susceptibility cusp is dependent on the measurement frequency, a characteristic unique to spin glasses (Figure 13).[112] This frequency dependence is often quantified by the relative shift in freezing temperature per decade of frequency, which has been found to be around 0.005 for CuMn alloys with concentrations ranging from 1 to 6.3 at% Mn.[111]

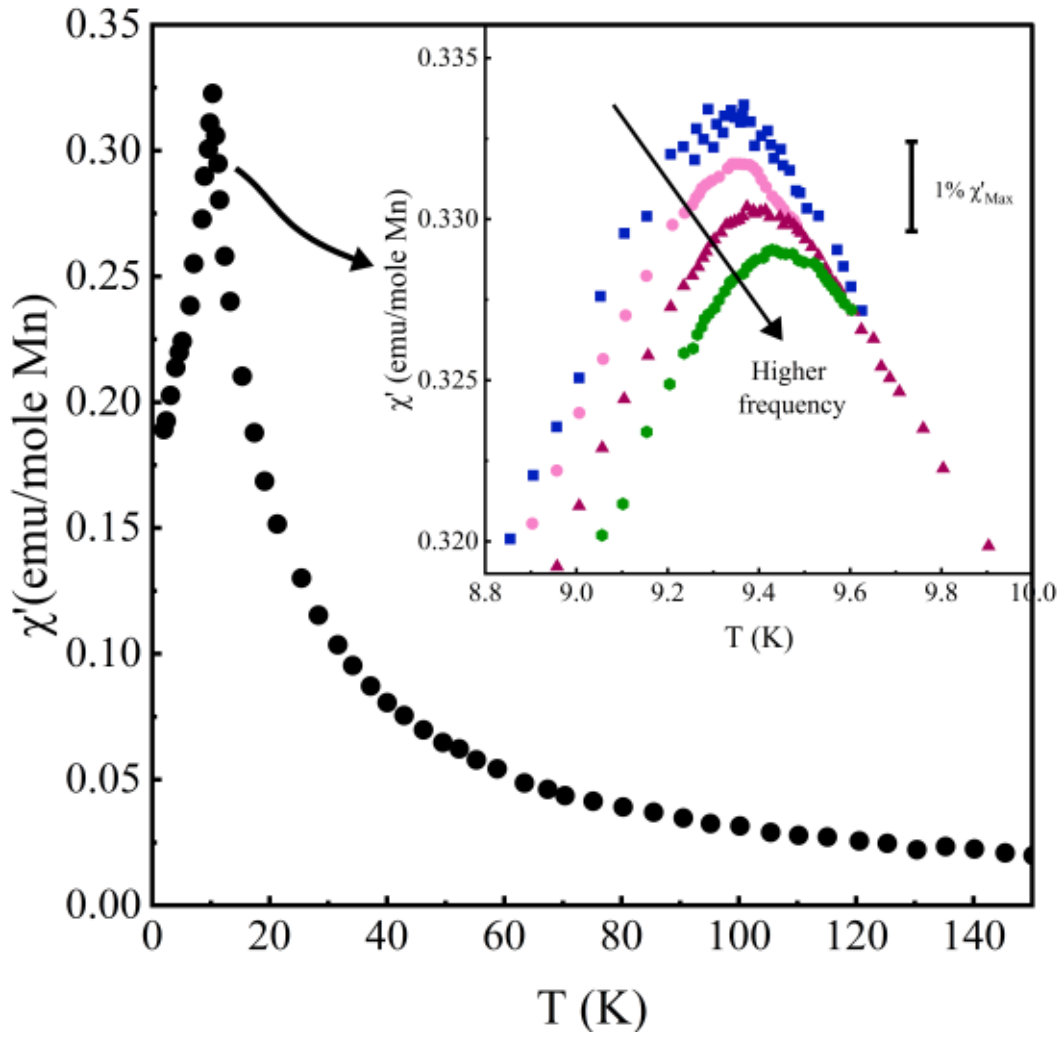

Figure 13: AC susceptibility (real part) of CuMn (1 at% Mn) showing the cusp at the freezing temperature. The inset shows the frequency dependence of the cusp from 2.6 Hz (triangles) to 1.33 kHz (squares). Figure reconstructed based on data extracted from Ref. [111].

The complex nature of spin glass dynamics is further revealed through detailed AC susceptibility studies. Both the real ($\chi'$) and imaginary ($\chi''$) components of the susceptibility provide valuable information about the spin glass state. The imaginary component, $\chi''$, which is nonzero below the freezing temperature, is indicative of the irreversibility and energy dissipation characteristic of spin glasses.[112] Moreover, AC susceptibility measurements have been instrumental in identifying different types of spin glass behavior, such as cluster spin glasses. For example, in $Zn_3V_3O_8$, a geometrically frustrated system, AC susceptibility measurements revealed a logarithmic variation of the freezing temperature with frequency, indicating the formation of a cluster spin glass state.[113] Such studies highlight the versatility of AC susceptibility in probing the intricate magnetic behavior of frustrated systems and distinguishing between various types of spin glass order.

# 4 Theories

## 4.1 Edwards-Anderson Model

The Edwards-Anderson (EA) model, introduced by S. F. Edwards and P. W. Anderson in 1975, is a spin glass model defined on a regular lattice (typically 2D or 3D). Each lattice site $i$ carries an Ising spin $S_i = \pm 1$, and the spins interact via a short-range spin-spin interaction described by the Hamiltonian:

$$H = -\sum_{\langle ij \rangle} J_{ij} S_i S_j,$$

where the sum runs over nearest neighbors, and the couplings $J_{ij}$ are quenched random variables (e.g., Gaussian or $\pm J$), thereby encoding disorder and frustration in local interactions.[24]

The EA model, unlike mean-field models such as the Sherrington-Kirkpatrick (SK) model (see section 4.2), operates in finite spatial dimensions with only short-range interactions, making it a more realistic representation of actual spin glass materials. It captures essential spin glass phenomena, including non-ergodicity[114], slow dynamics, aging[115], memory, history dependence[116], and a rugged energy landscape[117] with numerous metastable states, as confirmed by numerical simulations and experimental studies. The EA model serves as the canonical framework for Monte Carlo, parallel tempering, and finite-size scaling studies of glassy transitions in 2D and 3D systems.[118] However, unlike the SK model, the EA model lacks analytical tractability, with most insights derived from simulations or approximate methods. The nature of the spin glass phase in 3D remains contentious, with some simulations suggesting ultrametricity and multiple pure states consistent with mean-field predictions, while Boettcher (2005) posits a lower critical dimension of 2.5, implying that 2D systems are borderline and 3D systems may be near criticality.[119] The existence of a de Almeida-Thouless line, indicating a phase transition in a finite field, is well-supported in 4D but remains unresolved and controversial in 3D due to noisy data.[120] Notably, typical experimental signatures of spin glasses, such as temperature chaos, cooling rate dependence, and rejuvenation, are not well-reproduced in EA simulations, possibly due to limited timescales or missing physical ingredients.[121]

In the EA framework, one defines the spin glass order parameter from the dynamical time correlations of a single sample (fixed disorder). Consider a system evolving in time under fixed $J_{ij}$ . Then one may define[30]:

$$q_{\mathrm{EA}}^{\mathrm{time}} = \langle \tfrac{1}{N} \textstyle\sum_{i=1}^{N} \lim_{t\to\infty} \langle s_i(0) s_i(t) \rangle \rangle_{\mathrm{d.c.}}$$

where $\langle \cdots \rangle$ denotes thermal average over long-time equilibrium dynamics for a given disorder realization, while an additional average over disorder configurations $\langle \cdots \rangle_{\mathrm{d.c.}}$ is performed to obtain physically meaningful quantities. This quantity captures whether each spin tends to remain correlated with its initial state forever. In a paramagnetic phase, these correlations decay and $q_{\mathrm{EA}}^{\mathrm{time}} = 0$; in a spin glass phase certain spins remain "frozen" in time and $q_{\mathrm{EA}}^{\mathrm{time}} > 0$.[122] An alternative but related definition uses replicas, two independent equilibrium realizations of the same disorder sample. In the replica approach to spin glasses, "two independent equilibrium realizations" refer to two statistically independent configurations of the same disorder sample, denoted $a$ and $b$, obtained by equilibrating the system separately under identical couplings $J_{ij}$, temperature, and Hamiltonian.[79] In practice, one imagines preparing two copies of the system, starting from different random initial spin states and allowing each to reach thermal equilibrium on its own, for example through separate Monte Carlo chains or parallel-tempering trajectories. Because the free-energy landscape of a spin glass contains many metastable valleys, these two equilibrium states need not coincide even though they share the same quenched disorder. Their similarity is quantified by the replica overlap[122]

$$q_{ab} = \frac{1}{N} \sum_{i} \langle s_i^{(a)} s_i^{(b)} \rangle$$

which measures how correlated the two independently equilibrated configurations are. The Edwards–Anderson order parameter is then defined as

$$q_{\mathrm{EA}}^{\mathrm{config}} = \lim_{N\to\infty} E_J[q_{ab}]$$

where $E_J$ denotes an average over the quenched bond distribution. A nonzero value of $q_{\mathrm{EA}}^{\mathrm{config}}$ indicates that typical equilibrium configurations retain a finite mutual overlap, signaling broken ergodicity and the emergence of true spin glass order.

These two definitions are physically consistent: time averages in one heat-bath realization and configuration overlaps between replicas coincide in identifying frozen-in order.

## 4.2 Sherrington-Kirkpatrick Model

In 1975, Sherrington and Kirkpatrick introduced a fully connected, infinite-range Ising spin glass model. The SK Hamiltonian is given, for a system of $N$ spins $S_i = \pm 1$, by

$$H = -\frac{1}{\sqrt{N}}\sum_{i<j} J_{ij} S_i S_j,$$

where the couplings $J_{ij}$ are independent Gaussian random variables with zero mean and quenched variance $\langle J_{ij}^2 \rangle = J^2/N$. This choice of variance (equivalently the $1/N$ normalization in the Hamiltonian) ensures a well-defined thermodynamic limit in which the free energy remains extensive in $N$. The SK model therefore assumes random coupling strengths between all pairs of spins (quenched disorder) while deliberately neglecting any spatial structure, each spin interacts equally with every other spin (infinite range).[25]

The SK model represents a mean-field approximation to spin glasses in which each pair of spins interacts, either ferro- or antiferromagnetically, with identical statistics. Its key physical predictions include a transition at a finite temperature from a paramagnetically disordered high-$T$ phase to a complex glassy phase characterized by broken ergodicity, aging and memory, slow dynamics, and hierarchical metastable states.[123]

To solve the SK model, the replica trick was introduced. In this approach, one formally creates $n$ identical copies (replicas) of the system and averages over the disorder by computing the $n$-th moment of the partition function, $\overline{Z^n}$ with the overbar indicating the average over all possible configurations of the random couplings, before taking the analytical continuation $n \to 0$. In the original formulation by Sherrington and Kirkpatrick, it was assumed that all replicas behave identically, that is, replica symmetry holds. Under this assumption, all overlaps between replicas are considered identical. Later analyses revealed, however, that such simplification yields unphysical consequences (most notably negative entropy at low temperatures) because it overlooks the complex organization of the spin glass phase space. In reality, the phase is characterized by a multitude of metastable states and a highly intricate free-energy landscape composed of valleys separated by barriers. Distinct replicas may occupy different valleys, and as a result, their mutual overlaps are not necessarily identical.

This is where Giorgio Parisi introduced his revolutionary idea: replica symmetry breaking (RSB). In Parisi's solution, the symmetry between replicas is not merely broken; rather, a hierarchy of pure states emerges, characterized by a structured distribution of overlaps between these states. The replicas are organized into groups, subgroups, and so on, in a nested hierarchical fashion, where the degree of overlap changes at each level. This construction reflects the ultrametric organization of the spin glass state space: distances between pure states satisfy the ultrametric inequality, and valleys in the free-energy landscape branch into sub-valleys in a self-similar way. This hierarchical organization is not a mathematical artifact but a profound physical insight into the nature of disordered and frustrated systems. The Parisi RSB scheme, which can be implemented at one-step, two-step, or full (continuous) levels, is now regarded as one of the most significant developments in the theory of complex systems, with far-reaching applications in optimization theory, neural networks, error-correcting codes, and structural glasses.[26]

The SK model is exactly solvable in the mean-field limit, providing one of the rare examples of a disordered and frustrated system with an analytic low-temperature solution. Through Parisi's replica symmetry breaking, it reveals an emergent infinite hierarchy of glassy states, an ultrametric organization of free-energy minima, and a non-self-averaging overlap distribution $P(q)$, which describes the probability of finding two pure states with mutual overlap $q$. In contrast to simpler symmetric (replica-symmetric) approximations that yield a single overlap value, the RSB solution predicts a broad $P(q)$, reflecting the multiplicity of metastable states. Furthermore, the SK model exhibits aging phenomena in its out-of-equilibrium dynamics: the two-time spin autocorrelation function $C(t, t_w) = 1/N \sum_i \langle S_i(t) S_i(t_w) \rangle$ depends explicitly on the waiting time $t_w$, indicating slow relaxation and memory effects characteristic of glassy systems.[124] Mathematically, it has served as a rigorous testing ground: the existence of a phase transition, the thermodynamic limit, and exact solution via Parisi's free energy have all been proved using interpolation methods and concentration inequalities.[125]

However, the infinite-range connectivity inherent to the SK Hamiltonian is not physically realistic for most materials, where interactions tend to be short-range. As such, while the SK model provides fundamental insight, it does not directly describe finite-dimensional spin glasses as modeled by EA systems. Additionally, the formal replica trick and analytic continuation $n \to 0$ are mathematically subtle and initially heuristic in nature. Despite later proofs, the method remains conceptually subtle and technically demanding. The full RSB solution itself requires infinitely many hierarchical steps, and though it matches numerical experiments in the mean-field setting, its relevance to realistic systems (e.g. 3D) remains debated. Moreover, some qualitative aspects of real spin glasses, such as temperature chaos, cooling-rate dependence, and rejuvenation effects, are challenging to reproduce fully within the SK framework, largely because dynamics in clean mean-field models differ from finite-dimensional, kinetic behaviors observed experimentally.

## 4.3 Thouless-Anderson-Palmer (TAP) Approach

The TAP approach provides a framework for formulating the free-energy functional of infinite-range mean-field spin glasses in terms of local magnetizations, $m_i$. Instead of working directly with the microscopic spin variables $S_i = \pm 1$, this method introduces coarse-grained variables $m_i = \langle S_i \rangle$, expressing the free energy as a function of $\{m_i\}$ whose stationary points correspond to the system's physical (meta)stable states. Unlike a naive mean-field treatment, the TAP free energy includes an additional "Onsager reaction" term that compensates for the self-interaction effect arising in fully connected models, namely, the dependence of the local mean field on $m_i$ through the collective response of the rest of the system. For the SK model, this free energy is expressed as follows:

$$\mathcal{F}_{\mathrm{TAP}}[\{m_i\}] = -\sum_{i<j} J_{ij} m_i m_j - \sum_i \left[ \frac{1}{\beta} s(m_i) + \frac{\beta}{2}(1-q) m_i^2 \right] + \text{constant},$$

where $s(m) = -\frac{1+m}{2} \ln \frac{1+m}{2} - \frac{1-m}{2} \ln \frac{1-m}{2}$ is the Ising entropy and the term $\frac{\beta}{2}(1-q)m_i^2$ with $\beta = \frac{1}{k_B T}$ is the Onsager correction (here $q = \frac{1}{N} \sum_i m_i^2$). Stationarity $\partial \mathcal{F}_{\mathrm{TAP}} / \partial mi = 0$ yields the TAP equations: self-consistency relations for the local magnetizations that include the Onsager reaction term. The seminal derivation and presentation of this approach are in Thouless, Anderson and Palmer (1977), and numerous later works formalized and extended the TAP formalism and its mathematical status. The TAP formulation is valuable because its local minima can be interpreted as individual pure or metastable states; counting TAP solutions leads naturally to the concept of complexity (configurational entropy), i.e. the logarithmic number of metastable states at given free-energy level. The TAP equations can also be obtained as a systematic Plefka expansion (an expansion in inverse connectivity, small coupling $J_{ij}$ or a perturbative expansion of the free energy around a non-interacting limit, with local constraints (magnetizations) held fixed), which clarifies their range of formal validity and convergence conditions. The TAP framework therefore provides a concrete bridge between microscopic Hamiltonians and the picture of a rugged free-energy landscape populated by exponentially many states, and it is widely used to study complexity, marginal stability, and the distribution of minima in mean-field glassy systems.[126,127]

The TAP free energy provides an explicit functional whose stationary points correspond to possible mean-field configurations of the system. While true thermodynamic pure states are associated with the stable minima of this functional, many other stationary points correspond to metastable or unstable states that do not represent equilibrium phases. The TAP approach allows computation of the complexity (the number of metastable states) as a function of free energy and temperature, and it provides an ordering of states by their free-energy values. It is particularly powerful for analyzing the metastable landscape, marginally stable states, and dynamical arrest (or threshold states) in mean-field spin glass models such as the SK and p-spin variants. Rigorous advances have clarified the conditions under which TAP descriptions are exact in the thermodynamic limit and how the ensemble of TAP solutions encodes the hierarchical structure predicted by Parisi's replica symmetry breaking (RSB) theory

TAP is essentially a mean-field construction: while exact in the infinite-range limit and often asymptotically controlled for dense graphs, it does not by itself resolve finite-dimensional short-range physics. Convergence of iterative algorithms to find TAP solutions is subtle; many TAP solutions exist, some are marginally stable, and naive iteration may pick up only a subset (often the marginal states). Moreover, deriving TAP beyond leading orders requires care (Plefka expansion) and the interpretation of TAP minima as thermodynamic pure states requires matching with replica/cavity analyses. Recent rigorous works have clarified which TAP solutions correspond to pure states and how TAP complexity matches Parisi predictions in p-spin and mixed models.[128,129]

## 4.4 The Cavity Method

The cavity method[130] provides an alternative, physically intuitive route to mean-field solutions without relying on the formal replica $n \to 0$ analytic continuation. The method analyzes the effect of adding (or removing) a single spin, the "cavity spin", to a large system and tracks how the local field distribution seen by that spin derives from the distribution of fields in the $N-1$ system. In locally tree-like graphs (or in the infinite-connectivity limit treated appropriately) the cavity recursion becomes exact in the thermodynamic limit. Self-consistency equations can be formulated for the probability distribution of cavity fields, and solving these equations provides access to both thermodynamic properties and local observables. Under the assumption of replica symmetry, these recursions reduce to the belief-propagation (BP) or Bethe equations. However, when the configuration space fractures into many mutually inaccessible clusters of metastable states (separated by large energy barriers) this signals the onset of clustering, and one must generalize to distributional recursions via population dynamics and incorporate one-step or full replica symmetry breaking (RSB) within the cavity formalism. It is also important to note that BP relies on the assumption of a locally tree-like interaction graph; in systems characterized by many short loops or strong correlations, this assumption fails, and BP becomes only an approximate inference method, requiring corrections from higher-order or RSB-based treatments.[131] For CSPs and diluted spin glasses, the cavity method yields powerful predictions at zero and finite temperature, notably including threshold values, entropies of clusters, and algorithmic consequences. The cavity derivation of TAP equations also provides a transparent route to Onsager corrections: cavity fields give the intuitive origin of the TAP reaction term. Mézard, Parisi and Virasoro's development of the cavity method and its relationship to Parisi RSB is a central reference, and later algorithmic incarnations (belief propagation, survey propagation) implement these ideas computationally.[132]

The cavity method is constructive and often more transparent than replica algebra: for sparse (finite-connectivity) random graphs it provides testable distributional equations and supports algorithm design (BP, message passing, survey propagation). It has been used with great success in combinatorial optimization (random K-SAT, graph coloring), coding theory, and inference problems in statistical learning. The method also naturally extends to one-step RSB (1RSB) and full-RSB schemes for problems where solution clustering is relevant; survey propagation (SP) is a prominent algorithmic output of 1RSB cavity ideas that proved decisive in solving hard SAT instances near the satisfiability threshold.[133] The cavity method relies on assumptions about locally tree-like structure or on the smallness of loops; on finite-dimensional lattices (short-range models like EA) those assumptions fail and the cavity equations are only approximate. When many hierarchical levels of RSB are present, the cavity formalism becomes technically involved (functional order parameters). For finite-dimensional problems, the cavity predictions must therefore be used with caution and checked against numeric or rigorous results.

### 4.5 Alternative Theoretical Pictures and Renormalization Ideas

For finite-dimensional short-range spin glasses, alternative conceptual frameworks are important. The droplet/scaling theory (Fisher & Huse, Bray & Moore and others) views the low-temperature physics as dominated by compact droplet excitations with scale-dependent free-energy cost, predicting essentially only a pair of thermodynamic pure states and specific scaling of barriers and correlation lengths.[134,82] The metastate approach (Newman & Stein, Aizenman-Wehr) formalizes how multiple competing thermodynamic states can appear in disordered finite systems and how chaotic size dependence and sample-to-sample fluctuations are described by probability measures on infinite-volume states. These approaches are central to the debate over whether Parisi-type (many-state) RSB survives in finite dimensions or whether a droplet-like picture describes the true low-temperature phase; the controversy remains an active research area.[135]

### 4.6 Numerical and Algorithmic Methods

Because many analytical approaches are approximate or limited to mean-field settings, numerical approaches play a major role. Monte Carlo methods, in particular parallel tempering / replica-exchange Monte Carlo, are widely used to equilibrate glassy systems and measure thermodynamic quantities and overlap distributions. Exchange Monte Carlo[136] significantly improves sampling in rugged landscapes by allowing replicas at different temperatures to swap configurations. Exact ground-state computations, population-dynamics methods for cavity distributions, belief-propagation algorithms for sparse models, and message-passing generalizations such as survey propagation all provide computational tools to study both physical spin glasses and optimization problems inspired by them. Finite-size scaling analyses of numeric test competing theories (RSB vs droplet) in low dimensions.[137]

In Table 1 we summarize the main spin glass models and methods we've discussed so far (EA model, SK model, Parisi's replica symmetry breaking, TAP equations, cavity method, etc.), with concise pros/cons.

## 5 Spin Glass Excitations

Understanding the low-energy excitations of spin glasses is essential for describing their thermodynamics, dynamics, and stability. In finite-dimensional short-range systems, several distinct types of excitations have been identified within theoretical frameworks and numerical studies. These excitations differ in geometry, energy scaling, and physical interpretation, and together they constitute the modern understanding of the spin glass energy landscape.

### 5.1 Droplet Excitations

In the droplet theory, the fundamental excitations are compact droplets with radius $R$ (clusters of flipped spins). A key tenet of the droplet model is that the energy cost of creating a droplet of radius $R$, scales as a power of $R$, specifically as $\Delta F(R) \sim R^{\theta}$ where $\theta$ (the stiffness exponent) is positive in sufficiently high dimension so that the spin glass state is stable.[138] The physics of the spin glass phase, according to this model, is dominated by thermally activated droplets, which are those with free energies less than or on the order of the thermal energy ($\Delta F \leq k_{\mathrm{B}}T$). This implies that large-scale excitations are energetically unfavorable. The model also describes the fractal dimension of the interface or "domain wall" of these droplets, which is predicted to be less than the spatial dimension of the system.[139] These excitations are crucial for understanding the dynamic properties of spin glasses, such as the logarithmic decay of autocorrelations with time, which arises from the large activation barriers for the creation and annihilation of these droplets.[140,134]

## 5.2 Valley Excitations

In stark contrast to the localized droplet model, the RSB theory, particularly in the mean-field description of spin glasses, posits a far more complex picture of excitations rooted in a rugged energy landscape with a vast number of metastable states, often called "valleys".[141] This framework suggests that the low-temperature phase is not a single ground state but is fractured into a multitude of inequivalent pure states.[142] Excitations in this valley picture are not merely localized spin flips but correspond to transitions between these different thermodynamic states. Interfaces between pure states are predicted to be space-filling.[‡‡,138]

A key prediction of the full RSB solution is the existence of "soft" or "massless" excitations; it is possible to rearrange a large, even system-spanning, fraction of the spins with an arbitrarily small energy cost. [143,144] This implies that, unlike the droplet model where the energy of an excitation grows with its size, the RSB landscape allows for large-scale rearrangements without a significant energy penalty, a feature intimately linked to the ultrametric organization of states within the theory.[145]

## 5.3 Fractal Cluster Excitations

Beyond the standard droplet and RSB pictures, numerical studies have identified non-compact fractal excitations with anomalously small energy costs.[146] A key finding is that the energy of these fractal clusters does not increase with their size; the associated exponent is slightly negative. This is in stark contrast to the predictions of the droplet model, where the energy of compact clusters is expected to grow with their size. The concept of fractal clusters is not entirely new to the field. A "fractal cluster model" has been used to study the dynamical susceptibility of spin glasses above their critical temperature, yielding critical exponents in good agreement with experimental data.[147] This model has also been applied to analyze the relaxation behavior of spin glasses, successfully describing fundamental features of their magnetic relaxation.[148]

## 5.4 Domain-Wall Excitations

In the study of spin glasses, domain-wall excitations are a fundamental concept used to probe the stability and structure of the ordered phase.[149] A domain wall is a large-scale excitation that can be created computationally by comparing the ground-state energies of a system under different boundary conditions, such as periodic versus antiperiodic.[150] The energy cost of introducing this wall, known as the domain-wall energy, is a cornerstone of the droplet theory and its scaling with system size $L$, described by the power law $\Delta E \sim L^{\theta}$ defines the crucial stiffness exponent $\theta$.[151]

The geometric properties of these domain walls are also central to the theoretical debate, particularly their fractal dimension. Extensive numerical studies have shown that these interfaces are not smooth but are fractal, with a length that scales as $L^{d_s}$.[152] The value of this fractal dimension provides a key distinction between competing theories: the droplet model predicts a fractal dimension less than the spatial dimension

‡‡ In the droplet picture, excitations are compact clusters of spins whose boundaries are fractal but not space-filling; that is, the surface dimension satisfies $d_s < d$, so the boundary occupies an asymptotically negligible fraction of the system volume. In contrast, in hierarchical or valley excitations derived from the replica-symmetry-breaking scenario, the excitation interfaces become space-filling ($d_s = d$), forming extended, system-spanning rearrangements. This distinction (compact droplets versus space-filling valleys) lies at the core of the differing predictions of the two theories for the geometry and scaling of low-energy spin glass excitations.[140,138]

($d_s < d$), meaning the interface is not space-filling. Conversely, the replica symmetry breaking picture suggests space-filling interfaces. Numerical evidence for dimensions up to five has found that $d_f$ is indeed less than the space dimension, lending support to the droplet picture in these regimes.[153] The precise values of both $\theta$ and $d_f$ can show subtle variations depending on the distribution of the random couplings which points to a rich and complex scaling behavior.[154]

Figure X presents a schematic illustration of the four major excitation models in spin glasses, providing a clearer and more intuitive visualization of their distinct characteristics.

## 5.5 Role of Chaos in Spin Glass Excitations

A distinct and profound form of excitation in spin glasses arises from their chaotic nature, where the equilibrium spin configuration is exquisitely sensitive to minute changes in external parameters like temperature or the quenched bond strengths.[155] This phenomenon, known as "temperature chaos" or "bond chaos," implies that an infinitesimally small perturbation can lead to a complete and dramatic reorganization of the ground state spins over large length scales.[156,157] This global rearrangement can be viewed as a system-spanning excitation, fundamentally different from the localized droplet picture. The existence of chaos is a key prediction of the replica symmetry breaking theory, where the complex energy landscape allows for a reshuffling of the infinite number of metastable states with a slight change in parameters.[158] The droplet-scaling picture also accommodates chaos, predicting that the spin orientations at large separations become sensitive to small temperature changes. Numerical simulations and theoretical arguments have established scaling relations for this sensitivity, defining a "chaos length" beyond which the spin configuration decorrelates.[159] Studies have shown that both temperature- and bond-induced chaos can be described by universal scaling exponents, suggesting a deep, common underlying principle.[160] This extreme sensitivity has profound implications, impacting everything from the system's memory and aging effects to the computational hardness of finding spin glass ground states.

Ultimately, no single model fully captures the multifaceted nature of spin glass behavior. To bridge the gap between these compelling theoretical constructs and the physical world, it is essential to examine the materials that exhibit these fascinating properties. In the next section, we will therefore introduce different canonical spin glass materials to better understand how these theoretical concepts manifest in real systems.

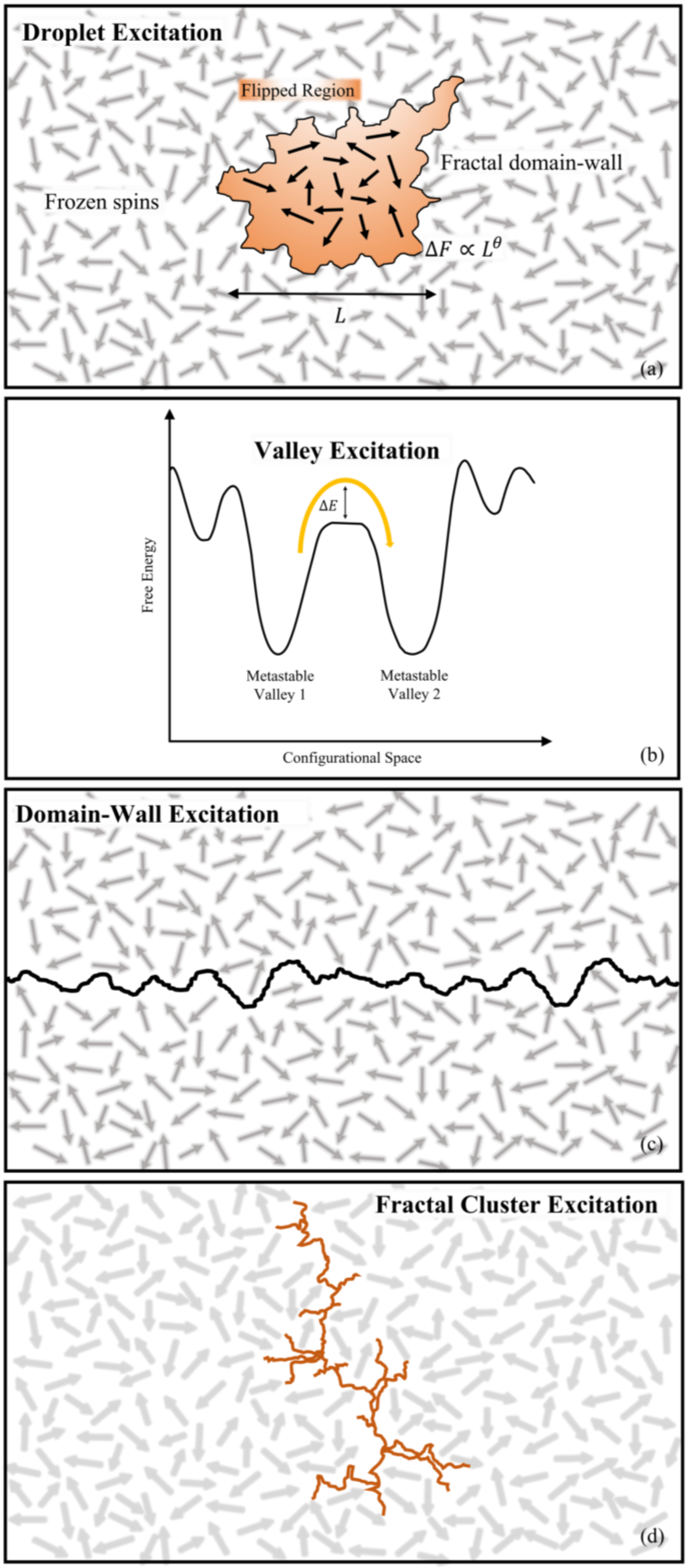

Figure 14. Schematic illustration of the principal excitation modes in spin glasses: (a) droplet excitations, representing compact rearrangements of spins with minimal surface area; (b) valley excitations, corresponding to transitions between nearby metastable energy minima; (c) domain-wall excitations, showing extended boundaries separating regions of different spin orientations; and (d) fractal excitations, characterized by irregular, scale-dependent spin rearrangements that reflect the complex energy landscape of spin glass systems.

Table 1 .Main models and analytical approaches in spin glass theory

| Model / Method | Description | Advantages | Limitations | Refs |
|---|---|---|---|---|
| EA | Short-range Ising spin glass model on a lattice with random nearest-neighbor couplings $J_{ij}$ . Characterized by the EA order parameter $q_{\mathrm{EA}}$ , defined via time average $\langle S_i \rangle^2$or configurational average over disorder. | Captures real materials with finite-range interactions; physically intuitive; foundation for more complex theories. | Difficult to solve analytically in finite dimensions; relies heavily on numerical simulations; unclear lower critical dimension. | 24 |
| SK | Infinite-range Ising spins with random couplings $J_{ij}$ drawn from a Gaussian distribution; solved using replica trick. | Exactly solvable in thermodynamic limit; provides deep insight into glassy order; starting point for RSB. | Replica symmetric solution is unstable at low $T$; physical interpretation less direct for real short-range systems. | 25 |
| RSB | Hierarchical solution to SK model introducing infinite levels of symmetry breaking in replica space. Order parameter becomes a distribution $P(q)$. | Yields stable free energy; explains ultrametricity and nontrivial overlap structure (though ultrametricity in finite dimensions remains debated); matches experiments and simulations for mean-field glasses. | Mathematically complex; direct experimental verification of ultrametricity remains subtle. | 26 |
| TAP | Mean-field self-consistent equations for local magnetizations including Onsager reaction term; equivalent to SK free energy minima. | Provides local, site-resolved description; connects thermodynamics with metastable states; useful for dynamics; Computes the complexity as a function of free energy and temperature, providing an ordering of states by free-energy values. | Many solutions (metastable states) make analysis complex; convergence issues near spin glass transition. | 126 |
| Cavity | Alternative to replicas: study effect of adding a new spin to an $N$-spin system, leading to self-consistent equations for cavity fields. | Physically transparent; avoids replica formalism; adaptable to diluted graphs and constraint satisfaction problems. | For finite dimensions, still challenging; needs assumptions on distribution of local fields. | 132 |
| Numerical Simulation | Computational techniques to study EA and related models in finite dimensions. | Directly applicable to realistic short-range systems; can handle disorder exactly. | Limited by system size and equilibration times; glassy dynamics makes convergence slow. | 136 |

# 6 Spin Glasses in Different Conducting Regimes

## 6.1 Insulating Spin Glasses

Unlike their metallic counterparts, insulating spin glasses do not rely on conduction electrons to mediate interactions between magnetic ions. Instead, interactions occur directly between the ions themselves. The interplay of ferromagnetic and antiferromagnetic interactions leads to a frustrated system, making it difficult for spins to find a low-energy configuration. Below a certain temperature (glass transition temperature), the system freezes into a disordered state with no long-range magnetic order. Studying insulating spin glasses is challenging due to the complexity of their interactions and the experimental difficulties associated with measuring their properties at low temperatures. However, understanding these systems is crucial for providing insights into the nature of disordered systems and phase transitions, developing new materials with tailored magnetic properties, and potential applications in spintronic devices and magnetic memory. To effectively model an insulating spin glass using the Heisenberg Hamiltonian, a foundational understanding of spin glass behavior is essential. A spin glass is characterized by long-range correlations between atomic magnetic moments that persist at extremely low temperatures. While this definition aligns with the core concept introduced by Sherrington,[161] it is generally considered more applicable to real-world systems. However, experimentally confirming these long-range correlations in actual spin glasses remains a significant challenge.[162]

One of the earliest studies on insulating spin glasses was conducted based on research involving $Eu_xSr_{1-x}S$.[163] This research investigates the magnetic properties of insulating $Eu_xSr_{1-x}S$ compounds. It finds evidence of spin glass behavior for specific concentrations of europium (Eu) due to short-range interactions between atoms. The researchers found that these insulating spin glasses exhibit characteristics similar to metallic spin glasses, including a freezing temperature below which long-range magnetic order vanishes, the presence of short-range magnetic ordering, and slow magnetization relaxation. They propose that the presence of competing ferromagnetic and antiferromagnetic interactions, rather than the RKKY interaction found in metals, is responsible for the spin glass behavior in these insulating compounds.[164]

Figure 15a presents the differential magnetic mass susceptibility measured under low magnetic field and low frequency conditions. They found that the magnetic susceptibility increases with the concentration of Eu and peaks at a specific temperature. This peak referred to as the $T_f$. Below this temperature, the material lacks long-range magnetic order, a fact confirmed by neutron diffraction experiments where it's a sign of spin glass phase as mentioned before.

Another research is based on magnetic susceptibility and magnetization measurements on amorphous manganese aluminosilicate ($Al_2Mn_3Si_3O_{12}$). Their findings suggest that this material exhibits spin glass behavior, characterized by strong antiferromagnetic interactions. Supporting evidence includes the rounding of the susceptibility peak under weak magnetic fields, magnetization curves similar to those of metallic spin glasses, and a weak dependence of the freezing temperature on measurement frequency. These characteristics distinguish the material from a system of small magnetic particles.[165] The authors measured the magnetic susceptibility of the material in a weak magnetic field and observed a typical spin glass behavior (as can be seen in Figure 15b): a rounded peak at the $T_f$. As the applied magnetic field increased, the temperature at which the susceptibility reached its maximum value decreased. This behavior is similar to what has been observed in other spin glass alloys, such as AuFe.

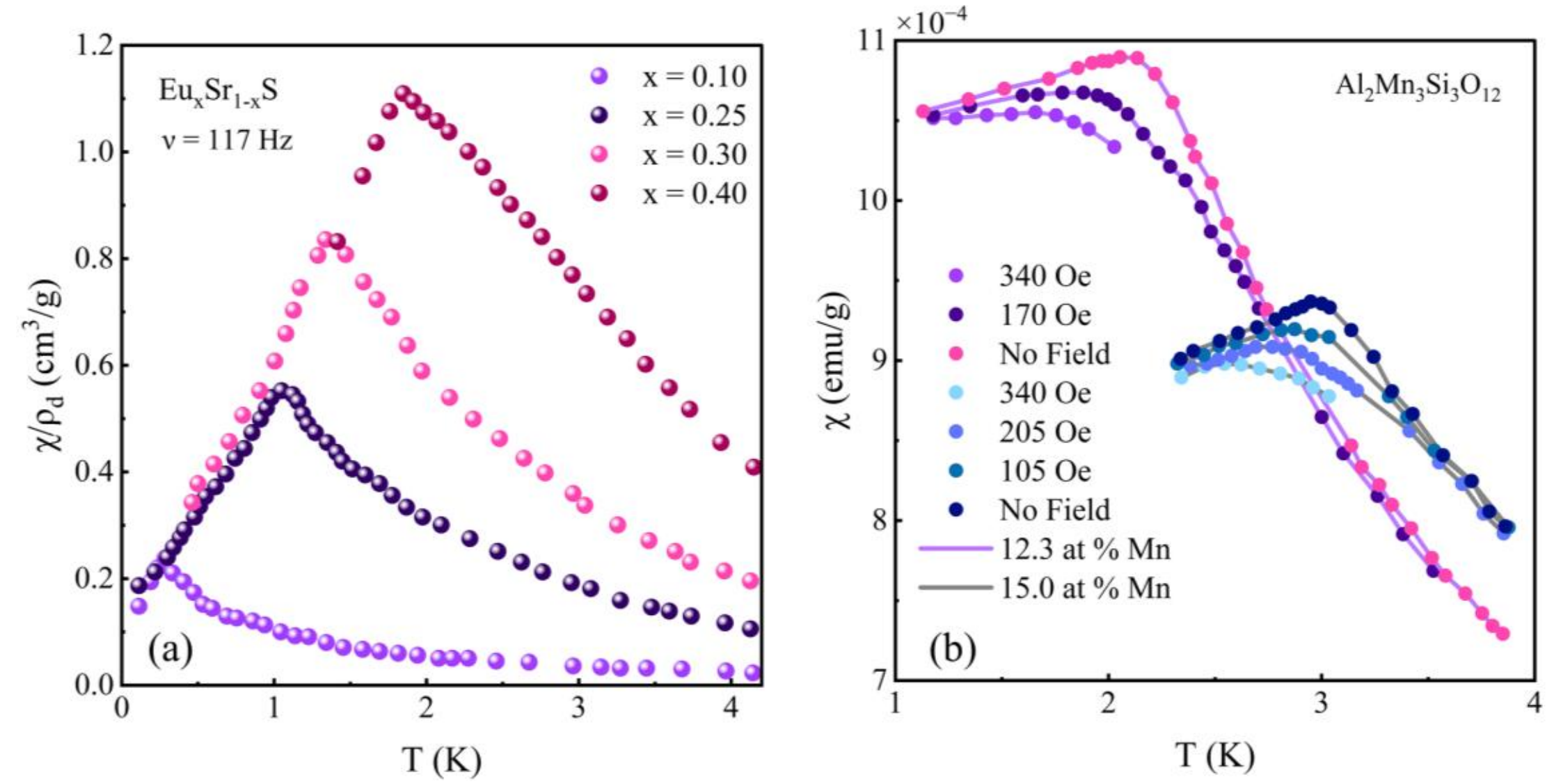

Figure 15. Differential magnetic susceptibility of (a) $Eu_xSr_{1-x}S$ compounds, showing an increase with Eu concentration and a peak at the spin glass freezing temperature $T_f$, below which long-range magnetic order is absent, and (b) amorphous $Al_2Mn_3Si_3O_{12}$, exhibiting typical spin glass behavior with a rounded susceptibility peak at $T_f$. In both cases, increasing the applied magnetic field lowers the peak temperature, reflecting the characteristic response of spin glass systems. Figures reconstructed based on data extracted from refs. [164] and [165] respectively.

### 6.1.1 Manganites

Insulating manganites have emerged as a significant focus within the field of spin glass research. he crystal lattice of manganites exhibiting a perovskite-like structure is presented in Figure 16. While early investigations of these materials primarily centered on their giant magnetoresistance (GMR) behavior,[166] , subsequent studies have revealed complex glassy magnetic dynamics and disorder-driven localization effects. Empirically, the magnetoresistance in these systems often follows a scaling relation $\Delta\rho/\rho = H^2/(T+T_0)^2$, where $T_0$ is a material-dependent constant. This form highlights the interplay between magnetic field, temperature, and spin disorder[8,167], providing a phenomenological bridge between experimental data and theoretical models of spin-glass-like magnetotransport.

Further studies have discovered their potential for thermoelectric applications. Key factors influencing a material's thermoelectric efficiency include a high atomic density, substantial atomic mass, and a robust phonon scattering mechanism.[168] The perovskite structure of manganites, when combined with rare earth and alkali metal elements, offers a promising framework for achieving these criteria. Additionally, a narrow bandgap is essential for optimizing the Seebeck coefficient, a crucial thermoelectric parameter. Sagar et al. conducted a pioneering study on Gd-Sr-MnO3, a representative manganite compound. Their research revealed a pronounced peak in thermopower around 40 K, with an optimal doping level of 0.5 yielding a record-breaking value of 35 mV/K at the time (Figure 17b). Concurrently, magnetization measurements indicated a phase transition from a paramagnetic to a spin glass state at this critical temperature. Further analysis of thermopower data, demonstrated a negative Seebeck coefficient at room temperature, characteristic of electron-dominated conduction. Upon cooling, this coefficient gradually decreases before reversing polarity to positive values near 70 K.

---

[8] The relation does not contain an explicit spin disorder term, however in magnetic materials, temperature is directly linked to the degree of spin disorder. At absolute zero ($T = 0$), spins would ideally be perfectly ordered in a ferromagnetic state. As temperature increases, thermal energy introduces fluctuations, causing individual spins to deviate from the aligned state. Also, the term $T_0$ is a crucial parameter that represents a characteristic temperature related to the intrinsic spin interactions within the material. It can be seen as a measure of the energy scale of spin correlations or frustrations that exist even in the absence of thermal agitation.[167]

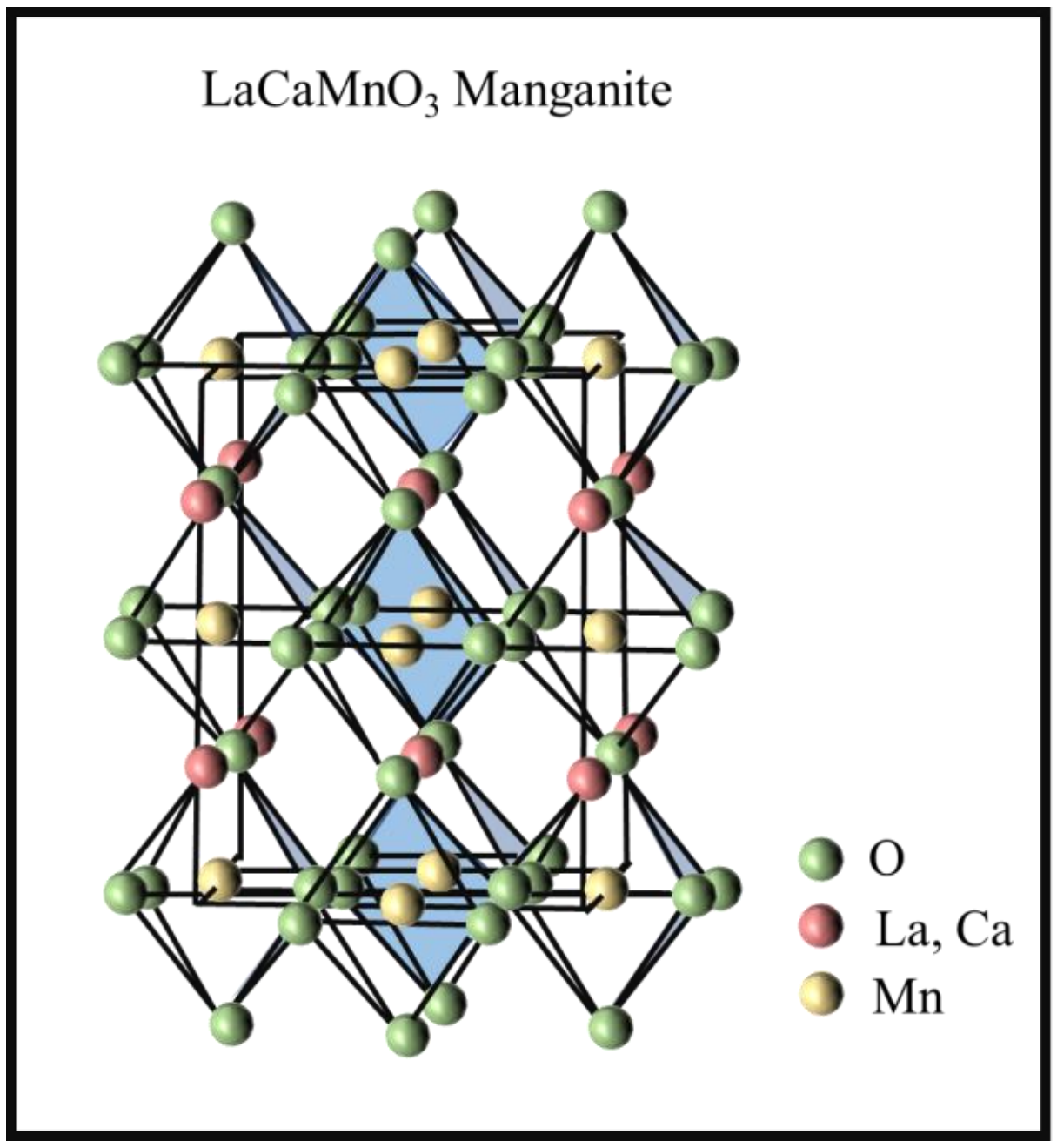

Figure 16. Structural model of the $La_{1-x}Ca_xMnO_3$ manganite, illustrating the three-dimensional network of corner-sharing $MnO_6$ octahedra (yellow Mn at the center, green O at the vertices). The A-site cations La/Ca (pink) occupy the voids between octahedra, producing a distorted perovskite lattice whose symmetry depends sensitively on ionic size mismatch, Mn-O-Mn bond angles, and octahedral tilting. These distortions play a central role in governing the electronic bandwidth, Jahn-Teller activity, double-exchange interactions, and hence the rich magnetic and transport properties of manganites.

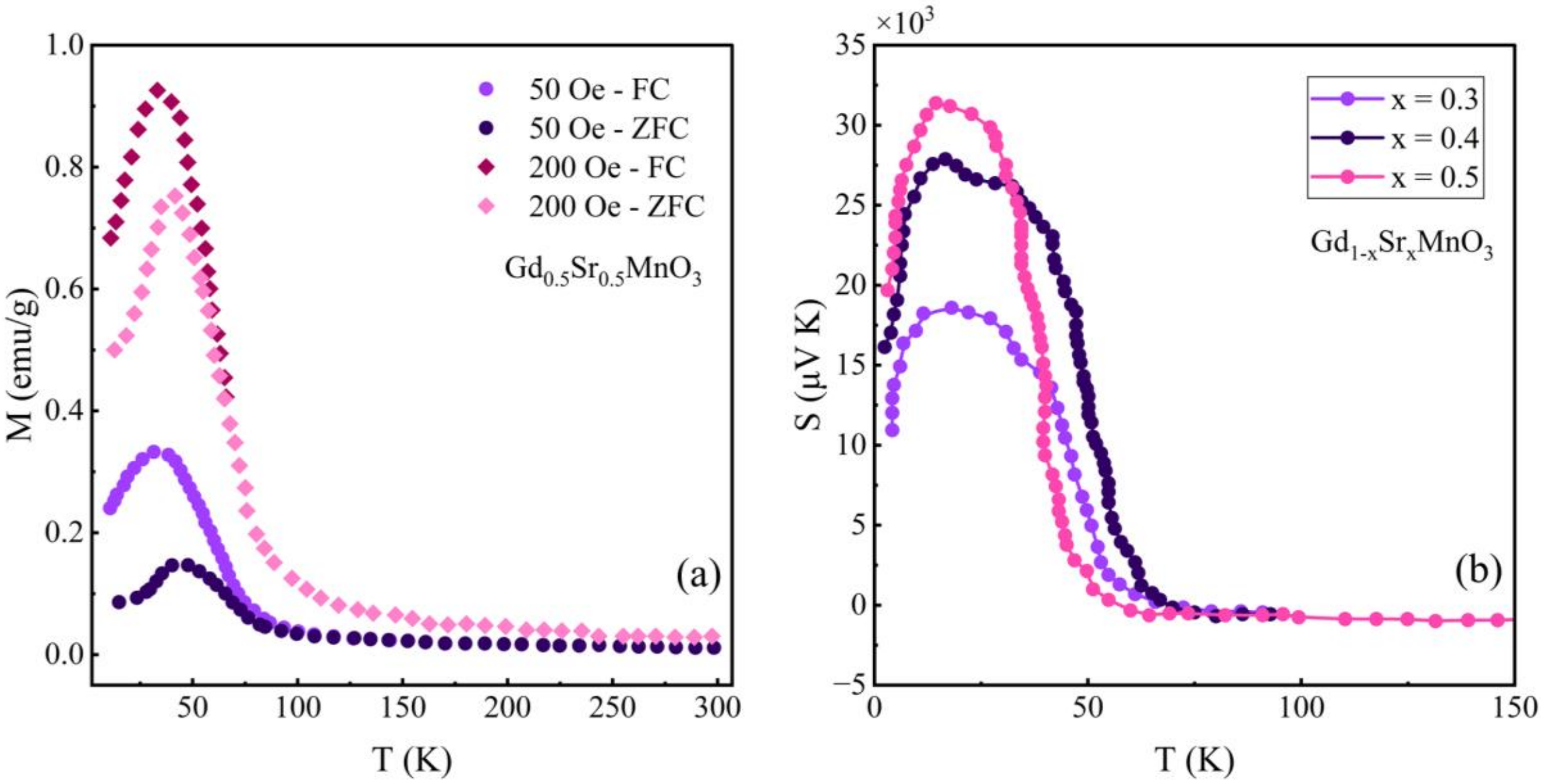

Figure 17. (a) $Gd_{0.5}Sr_{0.5}MnO_3$ FC and ZFC magnetization over 50 Oe and 200 Oe. (b) Thermopower of $Gd_{1-x}Sr_xMnO_3$ ($x = 0.3, 0.4, 0.5$) showing a pronounced thermopower peak (~35 mV/K) near 40 K, coinciding with a paramagnetic to spin glass transition. Figures reconstructed based on data extracted from Ref. [168].

The paper posits that the observed colossal thermopower may originate from phonon drag, magnon drag, charge ordering, or the formation of a spin glass phase. The FC and ZFC magnetization curves, previously discussed, are presented in Figure 17a for fields of 50 and 200 Oe with a doping level of $x = 0.5$. The samples display irreversible magnetization at low temperatures under a 50 Oe field. A clear separation between FC and ZFC magnetization is observed at 125 K, which shifts to 70 K when the field is increased to 200 Oersted. This bifurcation is a characteristic signature of spin glass behavior. The observed data suggest that the sample exhibits spin-glass-like properties below the bifurcation temperature (or irreversibility temperature, where ZFC-FC curves separate from each other), $T_{irr}$.[169] Typically, the coexistence of ferromagnetic and antiferromagnetic interactions within a material can result in a spin glass state, as evidenced by the divergence between FC and ZFC magnetization. Additionally, the absence of magnetization saturation at low temperatures further supports the classification of the system as a spin glass.[170]

The electrical resistivity of GdSrMnO compounds as a function of temperature and doping concentration has been also investigated. As temperature decreases, the resistivity of these materials generally increases, with higher Sr concentrations leading to lower resistivity values. The temperature dependence of resistivity for $EuSrMnO_3$, which exhibits behavior similar to that of $GdSrMnO_3$, is presented in Figure 34Figure 18. These materials exhibit high resistance characteristics typical of semiconductors or insulators. This behavior is attributed to lattice mismatches within the material and its spin glass nature.[166,171,172] Different models to explain the observed resistivity behavior in different temperature ranges has been explored:

- Thermal activation model[173,174]: Applicable at higher temperatures, it assumes that charge carriers overcome an energy barrier (activation energy) to contribute to conduction.[175,176] This model is commonly used for materials with magnetic charge carriers, posits that resistance increases exponentially with decreasing temperature as $\rho = \rho_0 \exp(E_A/k_B T)$, where $T$ is the absolute temperature, $\rho_0$ the value of resistivity at infinite temperature, $E_A$ the activation energy and $k_B$ the Boltzmann's constant. Based on the thermal activation model, a linear relationship between the natural logarithm of resistivity, $\ln(\rho)$, and the inverse of temperature, $1/T$, is anticipated. However, as shown in Figure 18b, the experimental data deviate from this expected linear behavior.
- Small polaron hopping (SPH) model[177,178]: Suitable for intermediate temperatures, it describes the hopping of charge carriers (polarons) between localized sites. This model suggests that resistance is directly proportional to temperature and exponentially related to an activation energy associated with charge carrier movement as $\rho = \rho_0 T \exp(E_A/k_B T)$. According to the small polaron hopping model, a linear relationship is anticipated between the natural logarithm of resistivity divided by temperature, $\ln(\rho/T)$, and $1/T$. This linear behavior is indeed observed within the temperature range of 200-300 K when fitting the experimental data to this model (Figure 18c).
- Variable range hopping (VRH) model:[179,180] Relevant for lower temperatures, it considers hopping of carriers between localized states over larger distances.[181,182] This model proposes a power-law relationship between resistance and temperature, influenced by factors such as the density of states at the Fermi level and carrier localization. The relation can be expressed as $\rho = \rho_0 \exp(T_0/T)^{1/4}$ with $T_0 = 18\alpha^3/(k_B N(E_F))$ where $\alpha$ is the electron wave function decay constant, and $N(E_F)$ is density of states at Fermi level. Based on the VRH model, a linear relationship is predicted between $\ln(\rho)$ and the inverse quarter power of temperature, $T^{-1/4}$. Experimental data in the 100-200 K range exhibit this expected linear behavior when fitted to the VRH model (Figure 18d).

Dysprosium manganite (DySrMnO)[183,184,176] is another manganite compound exhibiting spin-glass-like characteristics. The disordered arrangement of ions due to size differences between rare earth elements within the crystal structure contributes to this spin glass behavior. Given the magnetic ordering observed at low temperatures in these manganites, the high thermopower values are often attributed to charge ordering.

A consistent trend is observed across all samples: thermopower initially increases as temperature decreases, reaching a peak value ($T_{\text{peak}}$) before declining. This behavior is characteristic of many manganite materials.

Another report on $LaSrCoO_4$ reveals a colossal Seebeck coefficient, reaching approximately 15 mV/K in the low-temperature regime, as shown in Figure 19a.[186]

Europium manganite ($EuSrMnO_3$)[170] also demonstrates spin glass properties below 50 K. This behavior arises from the coexistence of ferromagnetic and antiferromagnetic phases within an antiferromagnetic matrix. Similar to dysprosium manganite, europium manganite exhibits a large thermopower at low temperatures, which is suggested to originate from either charge ordering or phonon drag mechanisms. Figure 19b illustrates the temperature dependence of the thermopower for various $Eu_{1-x}Sr_xMnO_3$ compositions. A clear trend emerges: as temperature decreases, the thermopower increases, indicating a reduction in metallic character. Additionally, the thermopower decreases as strontium (Sr) concentration increases. The $Eu_{0.8}Sr_{0.2}MnO_3$ and $Eu_{0.7}Sr_{0.3}MnO_3$ samples exhibit positive thermopower values across the entire temperature range, suggesting hole-dominated conduction. In contrast, $Eu_{0.5}Sr_{0.5}MnO_3$ and $Eu_{0.6}Sr_{0.4}MnO_3$ samples display a transition from negative to positive thermopower values with decreasing temperature, indicating a change in dominant charge carriers from electrons to holes.

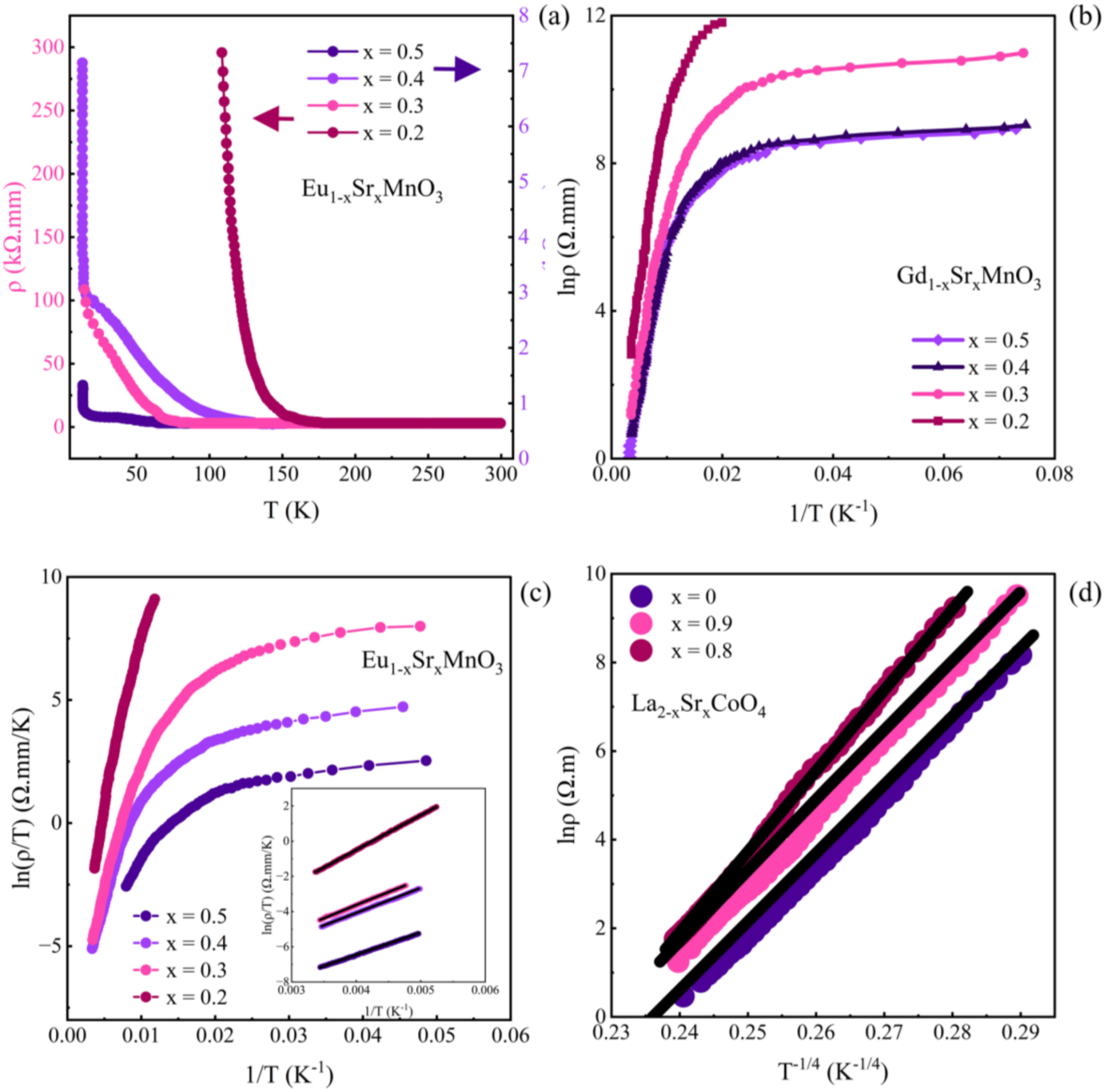

Figure 18. (a) Temperature dependence of electrical resistivity of $Eu_{1-x}Sr_xMnO_3$ ($0.2 \leq x \leq 0.5$) samples. (b) Natural logarithm of resistivity, $\ln(\rho)$, as a function of $1/T$ for $Gd_{1-x}Sr_xMnO_3$. (c) $\ln(\rho/T)$ as a function of $1/T$ for $Eu_{1-x}Sr_xMnO_3$. Inset: Fit of experimental data to the small polaron hopping (SPH) model (the solid lines). (d) $\ln(\rho)$ as a function of inverse temperature to the power of one-quarter, $T^{-1/4}$ for $La_{2-x}Sr_xCoO_4$. Solid lines are fit of experimental data to the variable range hopping (VRH) model. Figures reconstructed based on data extracted from refs. [170], [185], [170] and [186] respectively.

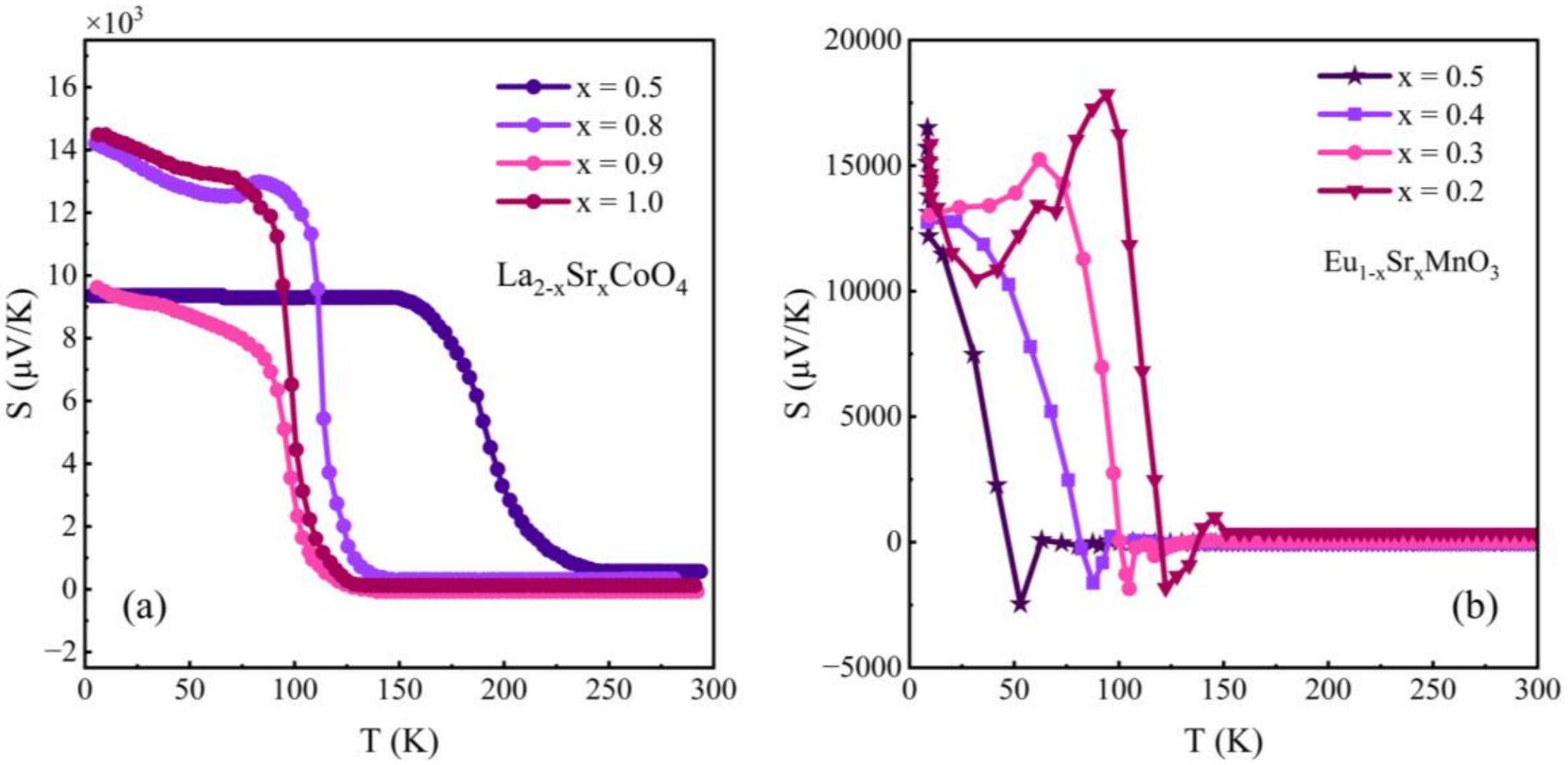

Figure 19. Temperature-dependent thermopower of (a) $La_{2-x}Sr_xCoO_4$, showing a colossal Seebeck coefficient (~15 mV/K) at low temperatures, and (b) $Eu_{1-x}Sr_xMnO_3$ compositions, highlighting spin glass behavior below 50 K. Thermopower increases at low temperatures and varies with Sr concentration, reflecting changes in metallic character and dominant charge carriers. Figures reconstructed based on data extracted from refs. [186] and [170] respectively.

Joy et al. investigated the impact of particle size on the colossal thermopower of charge-ordered, low-band Gd-Sr manganites, discovering that smaller particles exhibit a larger thermopower. Significantly, they linked this thermopower peak to the formation of a spin glass state. These manganites showcase a stable charge ordering arrangement where $Mn^{3+}$ and $Mn^{4+}$ ions occupy specific lattice sites, leading to the coexistence of ferromagnetic regions within an antiferromagnetic matrix. The competition between these opposing magnetic phases results in a spin glass state, characterized by magnetic frustration and the absence of long-range magnetic order. This spin glass formation is proposed as the origin of the colossal thermopower observed in these materials.[169]

Similar investigations have been conducted on $La_{0.5}Ca_{0.5}MnO_3$[187]. Figure 20a presents the temperature dependence of thermopower for this material, revealing a prominent minimum. To elucidate the underlying mechanisms governing thermopower in different temperature regimes, three distinct models have been proposed. At temperatures below the transition temperature ($T < 50$ K), the observed minimum is attributed to the presence of a spin glass phase. This interpretation is supported by fitting experimental data to a modified "Mandal's equation"[9,188] that includes an additional quadratic temperature term. This term, previously introduced by Fischer[189] to describe the low-temperature thermopower of metallic spin glasses under specific conditions, is crucial for capturing the observed minimum. Figure 20 (b-d) illustrates the excellent agreement between the experimental data and the fitted curves for various compositions.

---

[9] Mandal's equation is defined as: $S = S_0 + S_{3/2}T^{3/2} + S_2T^2 + S_4T^4$. $S_0$ is a constant for truncating in low temperature. $S_{3/2}T^{3/2}$ is magnon drag and $S_4T^4$ is for spin wave contribution. $T^2$ term adds the spin glass effect without which the peak in thermopower cannot be explained.

In the intermediate temperature range, where charge ordering effects persist, a phonon drag term[10] is incorporated into the Mandal's equation to account for the electron-phonon interactions. This modified-model provides an accurate description of the thermopower behavior in this region. Finally, at higher temperatures, Mott's equation[11,180,190,191,192] is employed to describe the thermopower, considering the dominant role of polaronic effects. In summary, the temperature dependence of thermopower in this material can be understood by considering the interplay between different physical mechanisms, including spin glass formation, electron-phonon interactions, and polaronic effects. The proposed models provide a comprehensive framework for interpreting experimental data and offer insights into the electronic properties of the material.

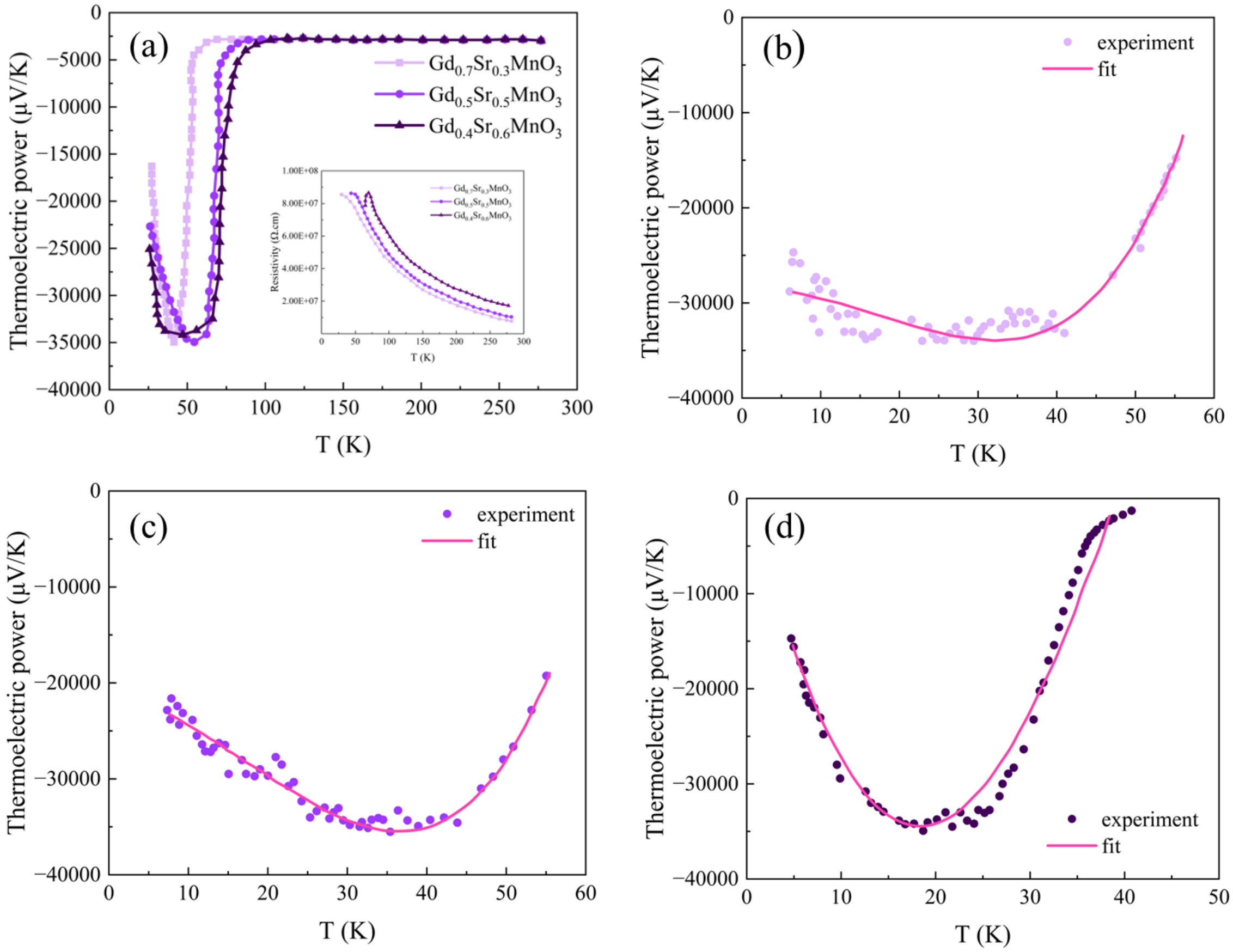

Figure 20. Temperature dependence of thermopower ($S$) for $Gd_{1-x}Sr_xMnO_3$ ($x = 0.3, 0.5, 0.6$) from 5 to 300 K. The inset displays the corresponding resistivity as a function of temperature. Magnified views of thermopower within the charge-ordered (CO) spin glass region for (b) $Gd_{0.7}Sr_{0.3}MnO_3$, (c) $Gd_{0.5}Sr_{0.5}MnO_3$, and (d) $Gd_{0.4}Sr_{0.6}MnO_3$. Solid lines represent fits to the equation [9]. Figures reconstructed based on data extracted from Ref. [169].

Table 2. Manganite materials and their freezing temperature.

| Compound | $T_f$ (K) | Refs. |
|---|---|---|

[10] $S = S_0 + S_{3/2}T^{3/2} + S_3T^3 + S_4T^4$ with $S_3T^3$ phonon drag term.

[11] Mott's equation $S = -\frac{k_B}{e}(\frac{\Delta}{k_BT} + B)$. Δ is activation energy, that is positive for the electron and negative for the hole. Activation energy is the energy required to jump over barriers between two neighboring sites. $B$ is a constant related to the entropy of the carriers.[180]

| | | |
|---|---|---|
| $Gd_2CoMnO_6$ | < 112 | [193] |
| $Gd_{1-x}Sr_xMnO_3$ | < 50 | [134] |
| $Dy_{1-x}Sr_xMnO_3$ | 32-35 | [176] |
| $Eu_{1-x}Sr_xMnO_3$ | < 50 | [184] |
| $La_{1-x}Ca_xMnO_3$ | < 120 | [187] |
| $La_{1-x}Sr_xCoO_4$ | 12 | [186] |

Several studies have reported spin glass behavior in perovskite manganites. For instance, DC magnetization and AC susceptibility measurements on $Gd_2CoMnO_6$ revealed a sharp spin glass transition at a remarkably high temperature of 112 K, following a paramagnetic to ferromagnetic phase transition.[193] The observed spin glass freezing temperature in this material is notably higher than typically reported values. This elevated transition temperature offers potential advantages for developing thermoelectric materials that can operate efficiently at higher temperatures.

Insulating spin glasses typically display large Seebeck coefficients, a desirable property for thermoelectric materials. However, their extremely high electrical resistivity severely limits their thermoelectric performance. Table 2 provides a concise summary of manganite compounds along with their corresponding freezing temperatures for quick reference.

### 6.1.2 Perovskite material

Magnetism in $ABO_3$ (Figure 21) and double-perovskite structures can often be understood via superexchange interactions, which hinge on linear cation-anion-cation alignments following the Goodenough-Kanamori-Anderson (GKA) rules. These rules explain the tendency toward ferromagnetic or antiferromagnetic coupling based on orbital orientations and bond angles. In the double perovskite $CaFeTi_2O_6$, high-field magnetization and specific heat measurements reveal no long-range magnetic order; instead, a spin-glass-like state appears below ~ 5.5 K at zero field. This state is gradually suppressed with increasing magnetic field, giving way to possible quantum spin-liquid like behavior, as evidenced by low-temperature specific heat following a power-law $\sim T^{1.6}$. Remarkably, the frustration in this material does not arise from geometry or competing exchange pathways, but likely from orbital modulation of exchange interactions.[194]

A variety of perovskite and double-perovskite materials exhibit spin glass or cluster glass behavior at low temperatures. These phenomena typically originate from cation disorder, site inversion, or competing ferro-/antiferromagnetic interactions. Below we provide a structured table (Table 3) that compacts key findings: the approximate freezing temperature, origin of glassy behavior, and references.

Wu et al. systematically studied $Pr_2CoFeO_6$ and Sr-doped derivatives as potential oxide thermoelectrics. The undoped double perovskite exhibits a large positive Seebeck coefficient ($\approx 600\ \mu V \cdot K^{-1}$ at 300 K) together with intrinsically low lattice thermal conductivity. However, its electrical conductivity is relatively poor, limiting the power factor. Sr substitution increases hole-carrier density by shifting the Fermi level deeper into the valence band, which enhances conductivity but reduces the magnitude of $S$, illustrating the typical thermopower-conductivity trade-off. In combination with hierarchical micro/nanostructuring that suppresses lattice thermal conductivity to $\sim 0.58\ W \cdot m^{-1} \cdot K^{-1}$, these modifications yield a peak ZT of $\sim 0.05$. The temperature-dependent transport properties are summarized in Figure 22: conductivity rises strongly with Sr content (Figure 22a), the Seebeck coefficient decreases with doping though remains sizable in the parent phase (Figure 22b), and the resulting power factor reaches ∼30 μW·m$^{-1}$·K$^{-2}$ near 600–700 K (Figure 22c).[195]

Oba and colleagues performed a comprehensive experimental study on substitutional tuning of the double-perovskite oxide $Sr_2FeMoO_6$, aiming to optimize its thermoelectric performance through targeted cation modifications. Three substitution schemes were explored[196]:

- B-site V substitution: In the system $Sr_2FeMo_{1-x}V_xO_6$, increasing V content led to improved bulk density, likely due to better sinterability,
- A-site Ba doping: Within $Sr_{2-y}Ba_yFeMo_{0.8}V_{0.2}O_6$, introducing Ba enhanced density markedly and significantly increased the electrical conductivity, particularly at $y = 0.6$.
- B-site Mn doping (with Ba co-substitution): The most compelling results came from $Sr_{1.4}Ba_{0.6}Fe_{1-z}Mn_zMo_{0.8}V_{0.2}O_6$. Increasing Mn content elevated the Seebeck coefficient markedly. The cumulative effect of these substitutions led to a power factor ($S^2\sigma$) peak of approximately 83.2 μW K$^{-2}$ m$^{-1}$ at 700 °C for the composition $Sr_{1.4}Ba_{0.6}Fe_{0.8}Mn_{0.2}Mo_{0.8}V_{0.2}O_6$, a substantial enhancement, quantitatively underscoring the promise of this substitution strategy.

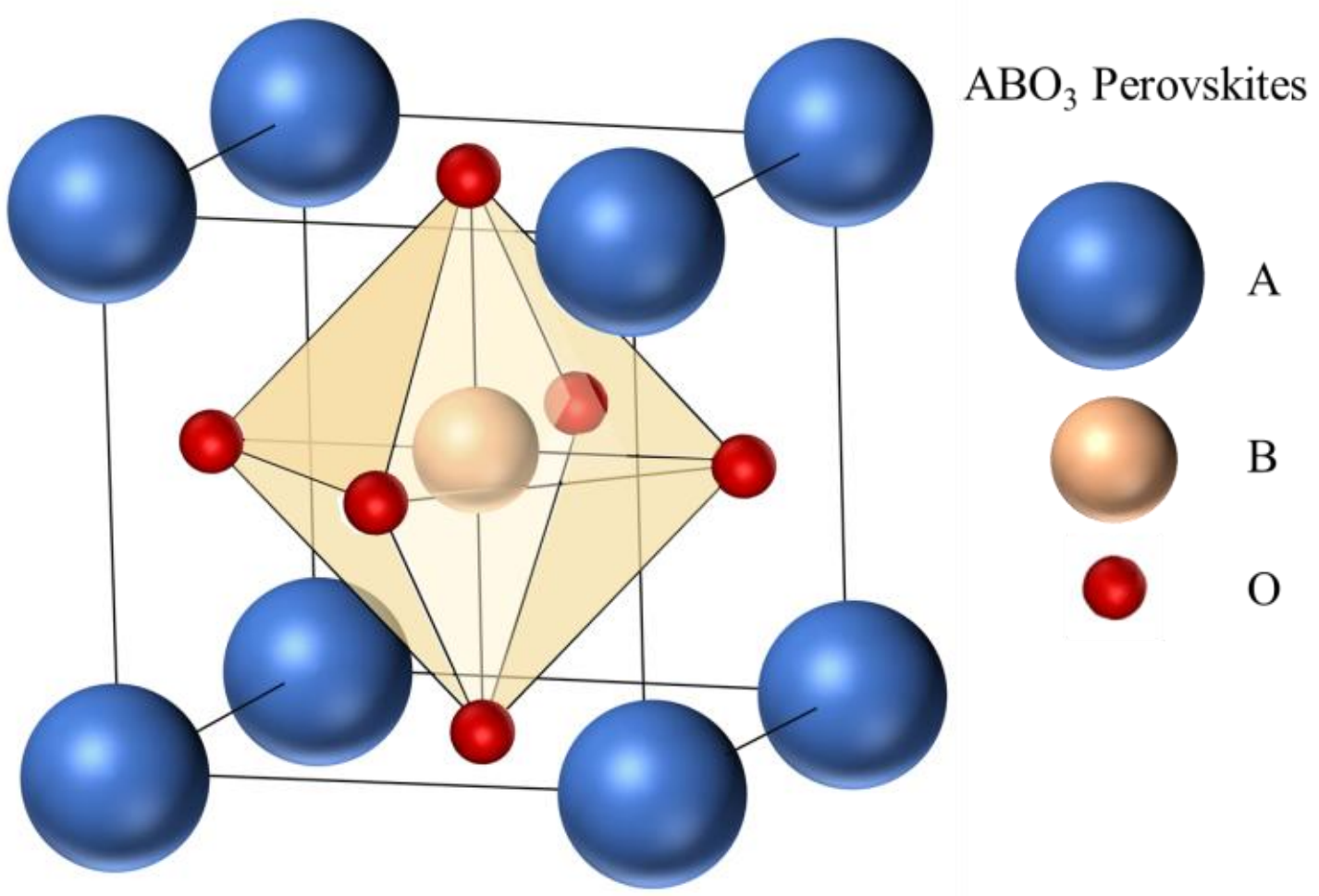

Figure 21. Idealized $ABO_3$ perovskite unit cell, highlighting the geometric arrangement of the large A-site cation (blue) at the cube corners, the smaller B-site cation (orange) at the body center, and oxygen ions (red) forming an octahedral cage around the B-site. This simple archetype provides a reference for understanding structural distortions in real perovskites, including the cooperative tilting and deformation of $BO_6$ octahedra, which strongly influence the functional behavior of complex oxides such as manganites.

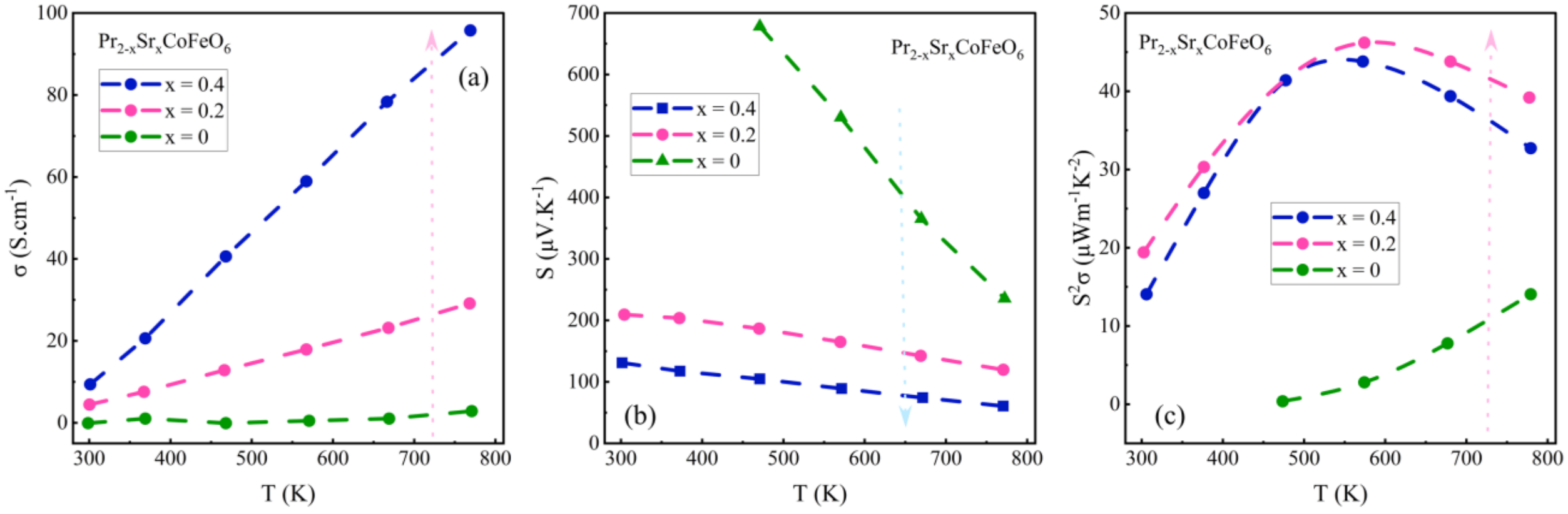

Figure 22. Thermoelectric properties of $Pr_{2-x}Sr_xCoFeO_6$ $(x = 0, 0.2, 0.4)$. (a) Electrical conductivity versus temperature, showing enhanced $\sigma$ with increasing Sr content. (b) Temperature dependence of Seebeck coefficient, highlighting the large positive $S$ in the pristine phase and its systematic reduction upon Sr doping. (c) Temperature dependence of the thermoelectric power factor ($S^2\sigma$), which peaks around 600–700 K, reflecting the optimal compromise between electrical conductivity and thermopower: $\sigma$ increases with temperature and Sr doping, while $S$ gradually decreases. Their competing evolution produces a temperature range where $S^2\sigma$ is maximized. Figures reconstructed based on data extracted from Ref. [195].

### 6.1.3 Cuprates

The spin glass state in high-$T_c$ cuprates, especially in $La_{2-x}Sr_xCuO_4$ (Figure 23), emerges when hole doping disrupts the long-range antiferromagnetic (AF) order of the $CuO_2$ planes, producing a frozen, disordered spin state in the range $0.02 \leq x \leq 0.05$. In the phenomenological dipolar-frustration model, holes localize near Sr dopants, generating local dipoles that frustrate the AF background.[197,198] Stripe order provides an alternative lens: in the charged-soliton model, doped holes form quasi-one-dimensional charge stripes that act as domain walls for spin correlations, fostering SG behavior.[199] Extensions involving spin-disordered "hedgehog" solitons and gauge-invariant effective Lagrangians reveal a quasi-two-dimensional chiral SG phase at low doping, driven by the interplay between massless gauge fields ($A\mu^3$) and spin chirality induced by mobile holes. Experimentally, in $La_{2-x}Sr_xCu_{0.95}Zn_{0.05}O_4$, the SG freezing temperature $T_f$ decreases with x and vanishes at a quantum critical point ($x_c \approx 0.19$).[200] Similar SG suppression occurs in Ni-doped $La_{1.85}Sr_{0.15}CuO_4$, where Ni's magnetic moment enhances spin disorder and destroys superconductivity.[201]

Magnetic phase diagram of $La_{2-x}Sr_xCuO_4$ (Figure 24) reveals a continuous evolution of magnetic behavior as holes are introduced into the $CuO_2$ planes.[202] At very low doping (x $\lesssim$ 0.01–0.015), the system hosts long-range antiferromagnetic (AF) order. As Sr substitution increases, AF order is destabilized, but the doped holes do not immediately form a static spin glass state. Instead, μSR detects an intermediate boundary labeled SF, which marks the onset of spin freezing of the doped holes. Below this line, the local fields created by hole-induced magnetic moments slow down sufficiently to appear quasi-static on the μSR timescale. Upon further cooling or increasing doping, these frozen hole moments evolve into the fully developed spin glass state, characterized by static, randomly oriented local moments and spatially inhomogeneous magnetism. Thus, the AF → SF → SG progression reflects a dopant-driven crossover from long-range magnetic order, through the initial freezing of hole moments, to a fully frozen glassy configuration.

Muon spin relaxation (μSR) and AC susceptibility measurements place the SG phase between AF and superconducting regions in the phase diagram, sometimes coexisting with superconductivity in nanoscale domains whose extent is tuned by the incommensurability, proportional to hole concentration.[203] The SG state in these cuprates has also been described as a cluster spin glass and, at very low doping, as a "charge glass" linked to glassy charge dynamics.[204]

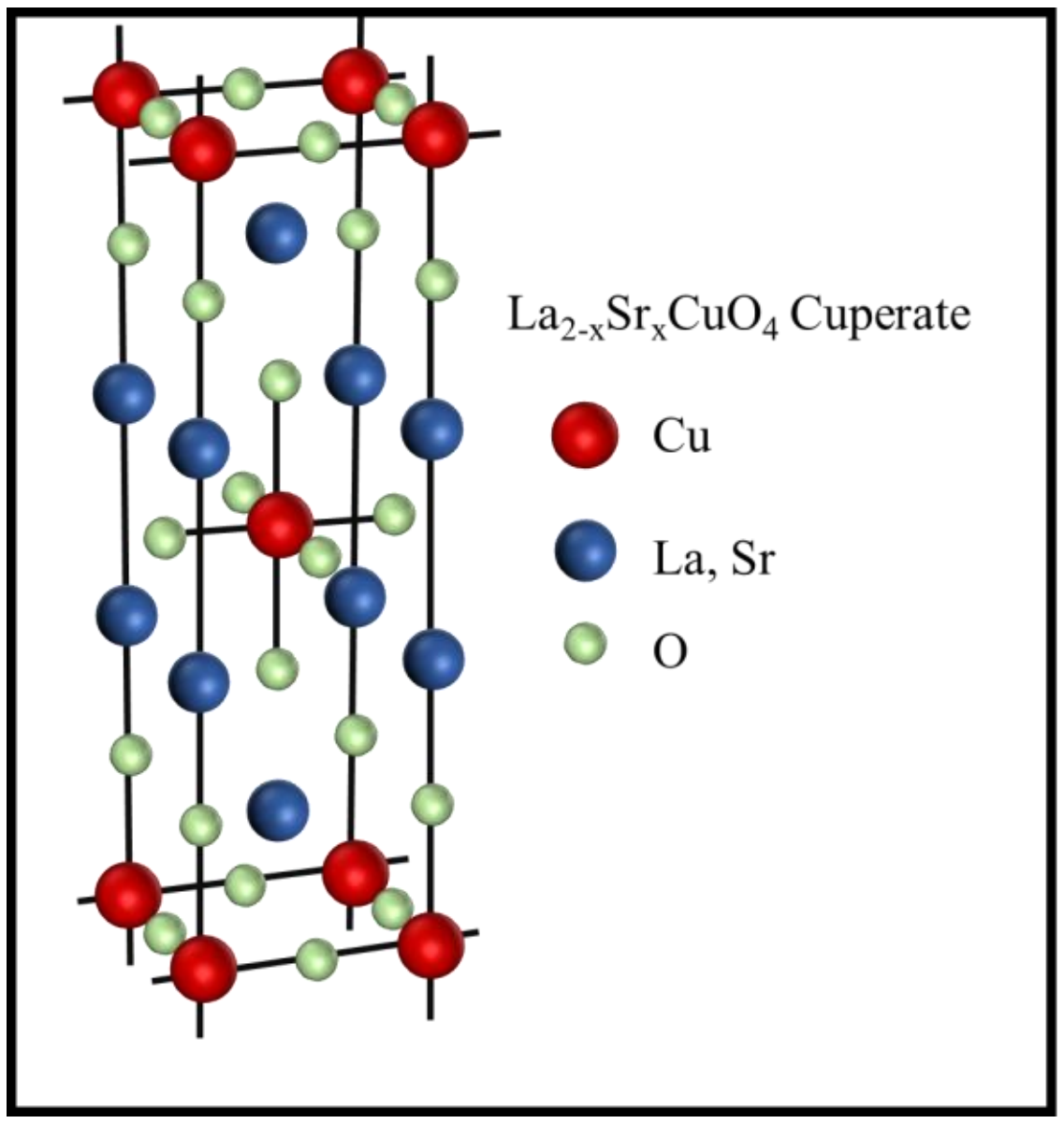

Figure 23. The layered perovskite-derived structure of $La_{2-x}Sr_xCuO_4$, a canonical high-temperature superconducting cuprate. The $CuO_2$ planes, formed by square-planar coordinated Cu ions (red) and in-plane oxygen (light green), provide the electronically active layers where superconductivity emerges. La/Sr cations (blue) reside between the planes and control hole doping, lattice distortions, and the delicate balance between antiferromagnetism and superconductivity characteristic of cuprate materials.

Collectively, these studies demonstrate that SG order in high-$T_c$ cuprates is a multifaceted phenomenon arising from the competition of doping-induced frustration, emergent stripe correlations, gauge-field mediated chirality, and the proximity of superconductivity. In Table 4, we present a selection of cuprate-based compounds that exhibit spin glass behavior, along with their spin-freezing temperatures and the key magnetic characteristics or mechanisms responsible for the emergence of the spin glass phase.

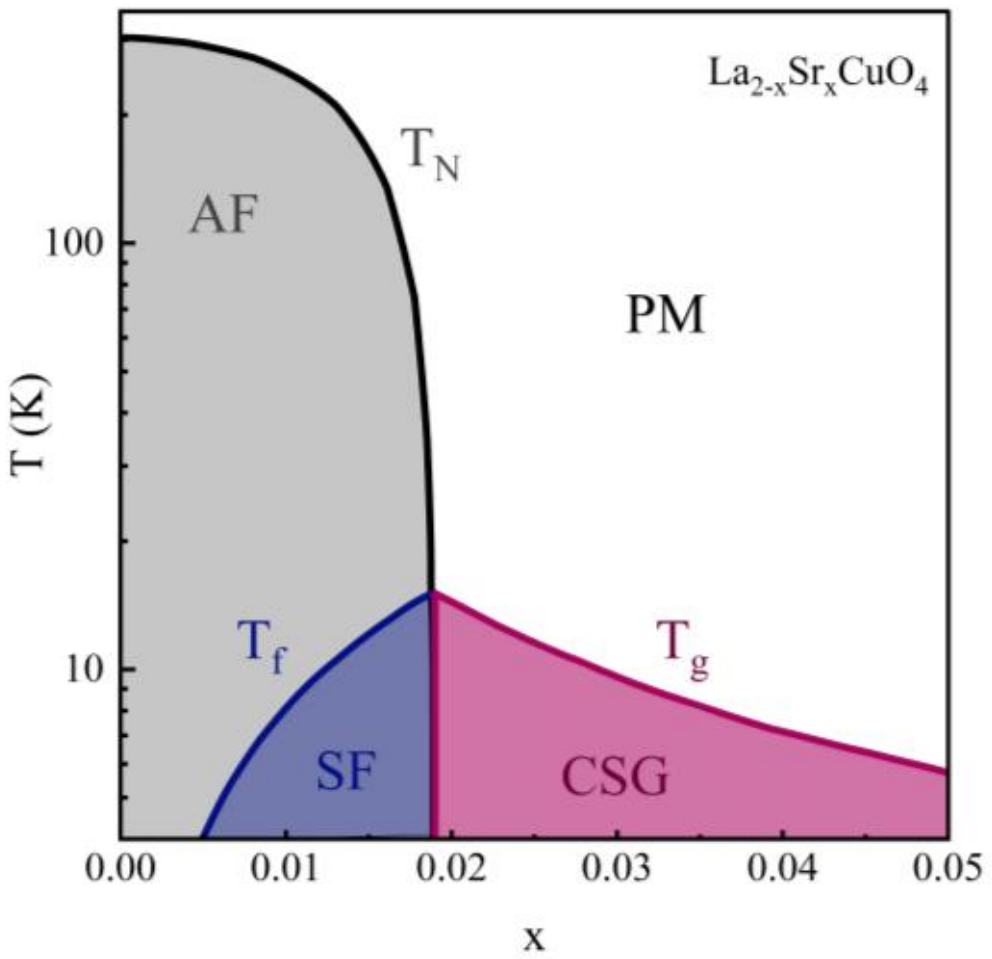

Figure 24. Magnetic phase diagram of lightly doped $La_{2-x}Sr_xCuO_4$. The plot illustrates how the characteristic magnetic energy scales evolve with increasing Sr concentration: the Néel temperature marking antiferromagnetic order ($T_N$), the temperature at which doped-hole moments begin to freeze ($T_f$, corresponding to the SF line), and the temperature where the system enters the fully developed spin glass state ($T_g$). All three boundaries are shown as functions of the hole-doping level x. superconductive phase starts at $x \approx 0.05$. Figure reconstructed based on data extracted from Ref. [202].

The Seebeck coefficient ($S$) in infinite-layer nickelate films exhibits distinct behaviors influenced by disorder and quantum criticality. In disordered $NdNiO_2$ films, $S/T$ remains nearly constant, indicating metallic conduction. Conversely, cleaner films display a logarithmic divergence in $S/T$ at low temperatures, followed by a pronounced "hump" near 25 K. This transition suggests a shift from band-structure-dominated[12,205] to quantum-critical-dominated[13,206] transport.[207]

In $La_{2-x}Sr_xNiO_{4+\delta}$, the Seebeck coefficient is positive, indicating hole-dominated conduction. At high temperatures, the temperature dependence of the Seebeck coefficient can be explained by a metal-like band conduction model, suggesting that the electron or hole is itinerant.[208]

### 6.1.4 Magnetic structures with pyrochlore lattice

The pyrochlore lattice, with the general formula $A_2B_2O_7$ (Figure 25), is one of the most prominent geometrically frustrated magnetic structures. Its framework is composed of two interpenetrating face-centered cubic (FCC) sublattices, each capable of hosting magnetic ions. These sublattices form corner-sharing tetrahedral three-dimensional analogs of triangular units, which prevent all antiferromagnetically interacting spins from simultaneously satisfying their exchange interactions. This intrinsic frustration can stabilize a wide variety of magnetic ground states, including spin glass phases, spin ices, and cooperative paramagnets. The magnetic behavior depends on the specific ions occupying the A and B sites, their spin-orbit coupling, and the balance between exchange and dipolar interactions. Substitutions, vacancies, and structural disorder can further complicate the energy landscape, leading to multiple nearly degenerate states. In contrast to canonical spin glasses, some pyrochlore SGs exhibit gradual spin freezing, retaining dynamic correlations down to very low temperatures. Studies on $Y_2Mo_2O_7$, $Tb_2Ti_2O_7$, and $Gd_2Ti_2O_7$ highlight the structural uniformity yet magnetic diversity of pyrochlores: from classical spin glass freezing to persistent spin dynamics and partial long-range ordering. Other systems such as $Y_2Mn_2O_7$, $Ho_2Mn_2O_7$, and $Yb_2Mn_2O_7$ reveal even more intricate phenomena, including short-range correlations, spin canting, and intermediate-range magnetic order[14]. Overall, pyrochlore oxides provide an exceptional platform for investigating the interplay between frustration, disorder, and exotic magnetic states.[209]

Table 5 lists several pyrochlore-based compounds that display spin glass behavior, highlighting their characteristic freezing temperatures along with the key magnetic interactions or structural factors that drive the formation of the spin glass phase.

---

[12] In conventional metals, transport is governed by Fermi-liquid quasiparticles whose behavior is determined primarily by the band structure and weakly temperature-dependent scattering rates. The Seebeck coefficient follows the Mott relation: $S/T \approx \frac{\pi^2 k_B^2}{3e} \frac{d \ln \sigma(\varepsilon)}{d\varepsilon} \Big|_{\varepsilon_F}$ and remains nearly constant at low temperature.[205]

[13] Near a quantum critical point (QCP), strong fluctuations of the order parameter dominate the low-temperature dynamics. These fluctuations destroy long-lived quasiparticles, make the scattering rate strongly energy and temperature dependent and increase the entropy per charge carrier, leading to non-Fermi-liquid behavior. A universal prediction of QCP theory is that $S/T \sim -\ln T$ which has been observed in heavy-fermion QCPs, cuprates, pnictides, and in clean $NdNiO_2$ films. or a more comprehensive insight.[206]

[14] In this context, intermediate-range refers specifically to the spatial extent of magnetic correlations. It means that the spins do not form true long-range magnetic order extending across the entire crystal, but they also do not remain purely short-range correlated. Instead, clusters or domains of correlated spins extend over tens to hundreds of lattice spacings.

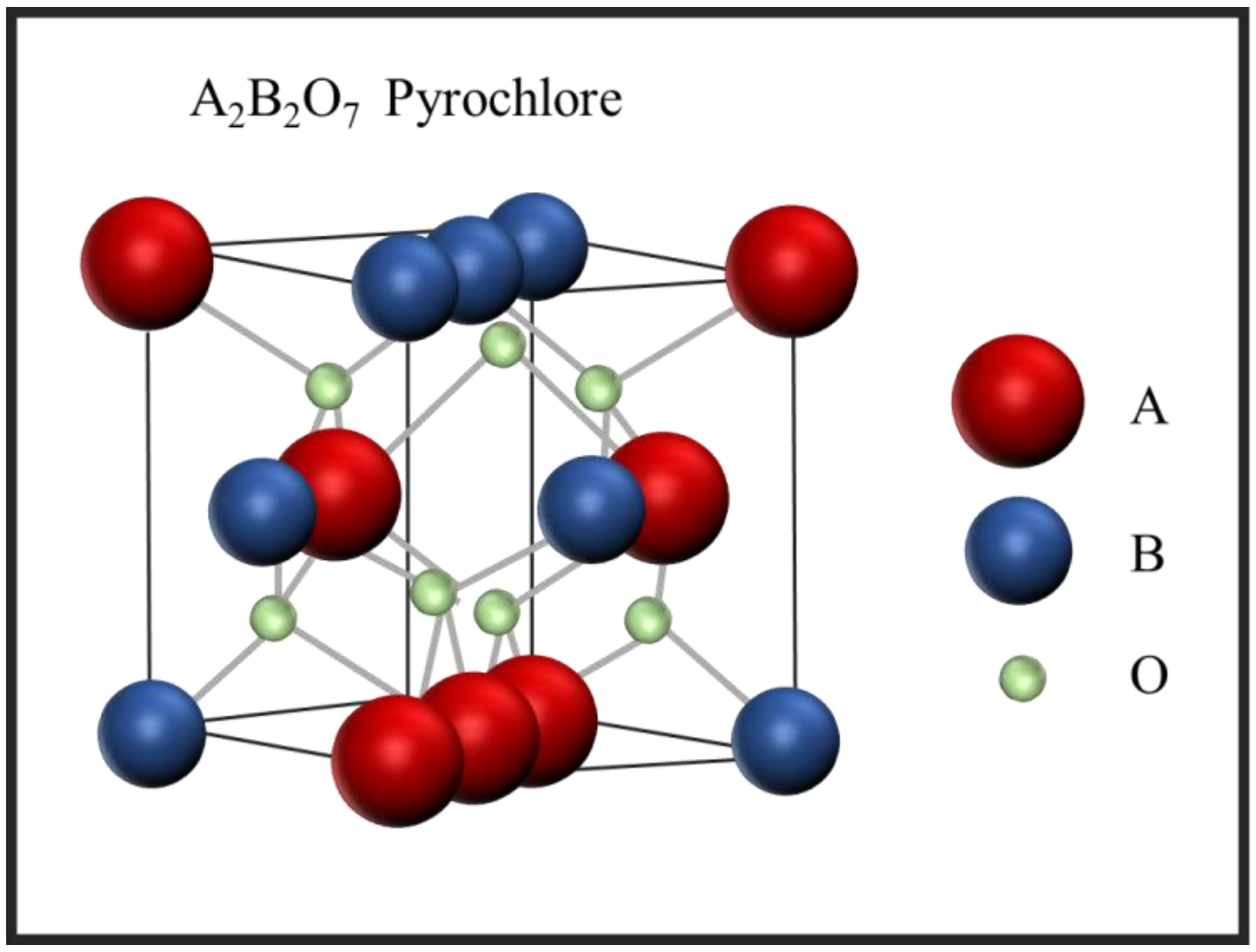

Figure 25. $A_2B_2O_7$ pyrochlore: The pyrochlore lattice consists of a three-dimensional network of corner-sharing $BO_6$ octahedra (blue) interpenetrated by AO′ polyhedra (red), with oxygen ions (green) completing the intricate architecture. The resulting geometry, featuring corner-sharing tetrahedral sublattices, gives rise to strong geometric frustration and exotic magnetic behavior such as spin-ice, spin-liquid, and glassy states, depending on the choice of A- and B-site cations.

### 6.1.5 Spinel oxides

Spinel oxides, with general formula $A^{2+}B_2^{3+}O_4$ (Figure 26), form a rich playground for geometrical frustration due to their dual-site lattice architecture, A-site cations in tetrahedral coordination form a diamond lattice, while B-site cations in octahedral coordination form a pyrochlore sublattice. When magnetic moments occupy either or both sublattices, their interactions-mediated through $O^{2-}$ ions, yield competitive exchanges that can thwart conventional magnetic order. A key amplifier of frustration is site inversion, where some A- and B-site ions exchange places; this disorder disrupts regular superexchange pathways and fosters glassy magnetic states. For instance, in Cr-based spinels such as $ZnCr_2O_4$ and $CdCr_2O_4$, the geometrical frustration suppresses long-range Néel order down to well below their Curie-Weiss temperatures (~390 K and ~88 K respectively), instead resulting in a spin-Peierls like transition and a "spin-liquid" type regime with fluctuating, correlated spins down to ~12.5 K and ~7.8 K.[210]

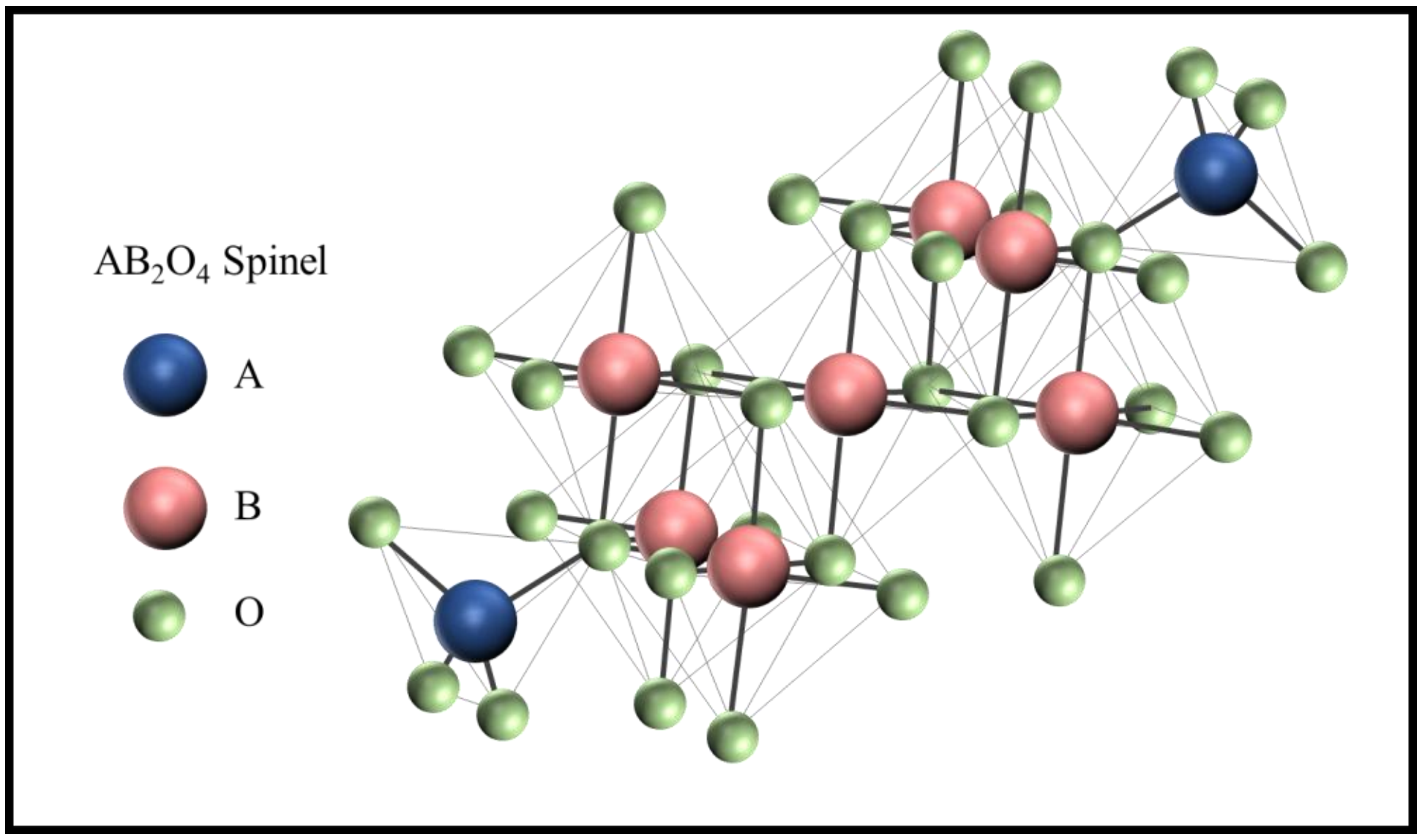

Figure 26. $AB_2O_4$ spinel: Structure of the spinel oxide, where A-site cations (blue) occupy tetrahedral interstices and B-site cations (pink) reside in octahedral sites within a close-packed oxygen framework (green). The arrangement of cations over these interstitial sites, and the possible presence of inversion or disorder, strongly influence the magnetic exchange pathways and electronic transport, making spinels a versatile family for studying magnetism, mixed valency, and frustration.

the A-site magnetic sublattice side, compounds like $FeSc_2S_4$ and $CoAl_2O_4$ exhibit strikingly suppressed magnetic order, even down to millikelvin, despite sizable antiferromagnetic interactions. In $FeSc_2S_4$, a combination of orbital degeneracy and strong spin-orbit coupling leads to a spin-orbital liquid, with neither magnetic nor orbital order observed down to 50 mK.[211] In $CoAl_2O_4$, theoretical models show that next-nearest-neighbor exchange ($J_2$) acting alongside nearest-neighbor $J_1$ on the diamond lattice can stabilize a highly degenerate spiral spin-liquid when $J_2/J_1 \gtrsim 1/8$. Subtle inversion disorder further disrupts this delicate balance, pushing the system into a spin glass state, experimentally observed as a freezing transition around 4.5 K that becomes pronounced above a critical inversion parameter $\eta \approx 0.10$. Together, these examples showcase how the confluence of lattice geometry, exchange competition, and disorder in spinels produces a spectrum of exotic magnetic ground states, from spin liquids to spin glasses.

Table 6 summarizes the freezing temperatures and underlying mechanisms responsible for spin glass behavior in a selection of spinel compounds, compiled from recent experimental studies.

### 6.1.6 Heavy fermion and non-magnetic disorder

Heavy-fermion (HF) materials, often containing 4f or 5f electrons, display enormously enhanced effective electron masses due to strong electronic correlations. These materials are fertile ground for exotic quantum phases, including unconventional superconductivity, Kondo lattice behavior, antiferromagnetic order, and, under suitable conditions, spin glass states.

The coexistence of RKKY interactions, Kondo screening, and random magnetic couplings in heavy-fermion Kondo-lattice systems can produce genuine spin glass transitions. The mean-field model by Theumann et al. captures this by introducing random long-range couplings among local moments over Kondo binding, leading to a finite-temperature spin glass phase that competes with the Kondo state[212]. Experimental realization of a cluster spin glass state is found in the skutterudite[15,213,214]-like system $Ce_3Co_4Sn_{13}$ when doped with Fe. For Fe concentrations above a critical value ($x \approx 0.1$), the system transitions from a dense Kondo lattice into a cluster spin glass phase, as evidenced by Mössbauer spectroscopy, magnetic relaxation, and distinct Mössbauer hyperfine features.[215]

A class of cerium-based intermetallics exhibits spin-glass-like behavior not due to f-electron disorder but rather due to non-magnetic atom disorder (NMAD) surrounding periodically arranged Ce or U sites. Notable examples include: $CePd_3B_{0.3}$: Displays a pronounced Schottky-like anomaly around $T_{\max} \approx 0.95$ K and an ac-susceptibility cusp at $T_f \approx 0.45$ K, accompanied by a substantial Sommerfeld coefficient ($\gamma \sim 240$ mJ/mol K$^2$). Here, about one-third of the B atoms randomly occupy body-centered sites, leading to fluctuating exchange interactions among Ce ions via RKKY coupling and resulting in spin glass freezing.[216] Other similar systems, such as $CeCu_{6.5}Al_{6.5}$ (random Cu-Al occupancy), $CePtGa_3$ (random Pt-Ga distribution), and mixed substitutions in $CeCu_4Al$, $CeCu_4Ga$, $CeNi_{1-x}Cu_xSn$, $CeRh_{2-x}Ni_xSi_2$, and $YbH_x$, are hypothesized to exhibit spin glass behavior stemming from NMAD rather than intrinsic f-electron complexity.

[15] "Skutterudites" are a family of intermetallic compounds characterized by a cage-like crystal structure derived from the mineral $CoAs_3$, typically written as $MX_3$ or, in their filled form, $R_xM_4X_{12}$, where $M$ is a transition metal and $X$ is a pnictogen. Their open framework contains large polyhedral cages capable of hosting "filler" atoms ($R$), whose low-frequency rattling modes strongly influence transport and magnetic properties. Filled skutterudites are known for heavy-fermion behavior, unconventional superconductivity, and low thermal conductivity arising from guest–host interactions.[213,214]

The heavy-fermion metallic compound CeFePO, non-superconducting and without long-range magnetic order, is extremely close to a ferromagnetic quantum critical point (QCP). Instead of reaching the QCP, however, it undergoes a spin-glass-like freezing below $T_f \approx 0.7$ K, with signatures in ac susceptibility (frequency-dependent peak) and μSR showing time-field scaling typical of cooperative glassy dynamics. This suppression of ferromagnetic quantum criticality is not disorder-driven but arises from critical magnetic fluctuations and anisotropy in a clean system.[217]

In Table 7, we summarize several heavy-fermion compounds and related f-electron systems that exhibit spin glass or cluster glass behavior. For each material, we list the reported spin-freezing temperature $T_f$, the physical mechanism attributed to the glassy behavior, and the corresponding reference.

Table 3. Spin glass/cluster glass behavior in perovskites

| Compounds | $T_f$ (K) | Mechanism / Reason for Glassy Behavior | Ref. |
|---|---|---|---|
| $Ba_4SbMn_3O_{12}$ | 11.5 | Ferromagnetic $Mn_3$ clusters freeze due to cluster-cluster interaction and cation order/dis-order | 218 |
| $Sm_2CuMn(MnTi_3)O_{12}$ | 7 | A-site columnar-ordered quadruple perovskite shows dielectric and spin glass behavior | 219 |
| $Gd_2CrMnO_6$ | 90 | B-site disorder & competing exchange interactions drive spin-glass-like and Griffiths-like phases | 220 |
| $CaMnFeTaO_6$ | 20 | Superparamagnetic ferrimagnetic clusters freeze to form cluster glass due to Fe/Mn/Ta disorder | 221 |
| $LaBiCaMn_2O_7$ | 76 | Spin glass due to frustration and competition short-range FM and AFM states | 222 |
| $SrLaGaRuO_6$ | 50 | Shows spin glass transition duo to isolated spins coupling via mobile $J_{\text{eff}} = 1$ excitons | 223 |
| $Ho_2CoMnO_6$ | 31 | Reentrant spin glass behavior in nanorod form driven by frustration in double perovskite | 224 |
| $CaCu_3Mn_2Ir_2O_{12}$ | 49 | Mn/Ir disorder at B-site leads to competing AFM/FM interactions; glassy arrest fits dy-namic slowing model | 225 |
| $Y_2MnGa(Mn_{4-x}Ga_x)O_{12}$ | $x = 0,1 \rightarrow 26$<br>$x = 2 \rightarrow 17$<br>$x = 3 \rightarrow 11$ | A-site ordered quadruple perovskite; spin-glass-like properties arise from Ga substitution at B-site | 226 |
| $SrTi_{0.5}Mn_{0.5}O_3$ | 14 | B-site disorder (Ti/Mn) and competing interactions produce cluster spin glass; ac & memory effects confirmed | 227 |
| $CaMn_3(Fe_3Mn)O_{12}$ | 14 | Spin glass due to the randomly distributed $Fe^{3+}$ and $Mn^{4+}$ at B site | 228 |
| $La_2CrNiO_6$ | 15 | Disorder induces ferrimagnetism, exchange bias, and spin glass behavior | 229 |
| $SrLaFeCoO_6$ | 75 | Mixed valency/disorder creates magnetic glass with magnetoresistance | 230 |
| $Sr_2FeTiO_6$ | 16 | Relaxor ferroelectric / spin glass coexistence due to disorder | 231 |
| $Pr_2CoFeO_6$ | 34 | Anti-site disorder drives reentrant cluster glass and Griffiths phases | 232 |
| $Nd_{0.5-x}Pr_xSr_{0.5}MnO_3$ | $x = 0.5 \rightarrow 250$ | Spin-glass-like state at low field related to charge ordering | 233 |
| $Sr_2YReO6$ | 12 | Shows a spin glass, due to spin-orbit coupling, crystal fields, and exchange interactions | 234 |
| $Ba_2In_{2-x}Co_xO_5$<br>$(0.70 \leq x \leq 1.70)$ | $x = 0.7 \rightarrow 10$<br>$x = 1.7 \rightarrow 30$ | Competing FM and AFM interactions cause frustration and spin glass state | 235 |
| $Nd_{0.67}Sr0.33Mn_{1-x}Fe_xO_3$ | 80 | Cation substitution induces CMR and spin glass behavior | 236 |

Table 4. Spin glass and magnetic properties of selected cuprate-type compounds

| Compounds | $T_f$ (K) | Mechanism / Reason for Glassy Behavior | Ref. |
|---|---|---|---|
| $LaNiO_2$ | 11.8 | Frequency-dependent AC susceptibility peaks, memory effects – interpreted as a spin glass state arising from disorder/complex magnetic interactions in bulk infinite-layer nickelate. | 237 |
| $PrNiO_2$ | 6.7 | Same phenomenology as $LaNiO_2$ (frequency shift, memory) – universal spin glass behavior across $RNiO_2$ family; likely due to structural/chemical disorder and mixed magnetic interactions. | 237 |
| $NdNiO_2$ | 4.8 | Same as above – frequency dependence + memory effects; bulk nickelates show spin glass rather than long-range AFM | 237 |
| $La_2CuO_4$ lightly doped (Li, Sr) (LSCO, low x) | - | Impedance spectroscopy & magnetic probes reveal spatially segregated dynamic charge domains; holes/dopants cause dipolar frustration of the AF background – leads to cluster/spin glass behavior in lightly doped samples. The paper does not explicitly state $T_f$ for LSCO $x = 0.03$ but shows frequency dependent peaks (~5 K) in dielectric loss and ac susceptibility that shift to lower temperatures as frequency is decreased. This is indicative of spin-glass-like dynamics. | 238 |
| $Pr_4Ni_3O_8$ | 68 | Complex spin-freezing process: mixed-valence Ni, short-range correlated regions produce two relaxation channels (fast + slow); distinct magnetic memory and glassy behavior driven by short-range correlations. | 239 |
| $RuSr_2Ln_{1.5}Ce_{0.5}Cu_2O_{10-\delta}$<br>$RuSr_2LnCu_2O_{8-\delta}$<br>Ln=Y, Dy, Ho | Dy →87.4<br>Ho →103.3<br>Y→101.5 | Rutheno-cuprates show coexistence / competition of magnetic order (Ru sublattice) and superconductivity; oxygen nonstoichiometry, mixed valence and disorder produce glassy or glass-like magnetic signatures depending on Ln and δ. | 240 |
| $Gd_3Ba_2Mn_2Cu_2O_{12}$ | 2.5, 4.8, 9 | Manganocuprate with multiple magnetic sublattices (Gd, Mn, Cu) shows complex magnetic anomalies (cluster-like freezing due to competing interactions among sublattices). | 241 |
| weakly doped cuprates | - | Theoretical works show that spin twists, domain walls (stripes), and frustrating bonds lead naturally to a cluster spin glass phase in weakly doped cuprates; spin textures (twists) and domain walls pin clusters producing glassy freezing. | 242 |
| $La_{2-x}Sr_xNiO_{4+\delta}$<br>$La_{1.6-x}Nd_{0.4}Sr_xCuO_4$ | - | Neutron/μSR studies show stripe correlations (charge + spin), and static/disordered stripe regions produce slow spin dynamics and glassy freezing in certain doping windows. The paper does not explicitly state a $T_f$, however, it reports static spin stripe order emerging below ~70K | 243 |

Table 5. Pyrochlore compounds exhibiting spin glass or related behavior

| Compounds | $T_f$ (K) | Mechanism / Reason for Glassy Behavior | Ref. |
|---|---|---|---|
| $Lu_2Mo_2O_7$ | 16 | Spin glass freezing arises from strong geometric frustration of S=1 Mo moments combined with disorder/quenched randomness in exchange pathways; chemically tuning (oxynitriding) reduces the effective moment and/or modifies exchange, stabilizing a quantum-disordered (quantum-spin-liquid like) ground state instead of glassy freezing | 244 |
| $Ca_2Ru_2O_7$ | 25 | Small itinerant Ru moments and competing exchange interactions in a metallic pyrochlore produce frustrated magnetic interactions and a negative nonlinear susceptibility; conduction-electron damping and quenched disorder lead to a non-canonical (metallic) spin glass transition | 245 |
| $(Bi_{1.88}Fe_{0.12})(Fe_{1.42}Te_{0.58})O_{6.87}$ | 20 | High Fe content and chemical/oxygen nonstoichiometry produce competing exchange paths and local structural/displacive disorder; these produce short-range correlated magnetic clusters whose freezing manifests as a spin glass transition. | 246 |
| $Tb_2Mo_2O_7$ | 24 | Even in the absence of chemical disorder, geometric frustration and the formation of short-range ferromagnetic clusters (with a broad distribution of relaxation times) lead to unconventional glassy freezing; the glassiness is intrinsic and arises from frustrated cluster dynamics rather than site disorder. | 247 |
| $A_2Sb_2O_7$ (A=Mn, Co, Ni) | 41 | Spin glass transitions occur without measurable chemical disorder; the origin is geometric frustration amplified when local moment size is reduced (increasing frustration index), driving the inability to form a single ordered ground state and producing glassy freezing of spins. | 248 |
| $Ga_2Mn_{2-x}Cr_xO_7$ | 12 | Cr substitution introduces bond disorder and competition between AFM and FM interactions; this produces a Griffiths-like regime (clusters of locally correlated spins) that evolves into cluster spin glass behavior as substitution increases. | 249 |
| $Y_2Ir_{2-x}Cr_xO7$ | x = 0→130<br>x = 0.05→46<br>x = 0.1→48<br>x = 0.2→50 | Isoelectronic Cr substitution does not change global structure but enhances local magnetic inhomogeneity and rare-region effects; the system develops a Griffiths-like phase that precedes or coexists with cluster glass behavior – i.e., substitution-induced local moments and competing interactions create glassy dynamics without wholesale structural disorder. | 250 |
| $Bi_{2-x}Fe_x(FeSb)O_7$ x = 0.1,0.2,0.3 | < 14 | Displacive A-site disorder (off-center Bi displacements), oxygen vacancies and mixed-valence transition metals create strong local lattice distortions and inhomogeneous exchange; these structural irregularities produce random fields and bond disorder that promote glassy magnetic states. | 251 |
| $A_2Mo_2O_7$ (A = Gd, Dy, Ho, Er) | Y → 22<br>Dy → 24<br>Ho → 21<br>Er → 18 | Local lattice distortions and inhomogeneous bonding create site-selective changes in exchange pathways and local anisotropies. These microscopic distortions increase the distribution of exchange energies and relaxation times, favoring cluster formation and spin glass freezing even when global crystallographic order is retained. | 252 |

Table 6. Spin-glass and magnetic properties of selected spinel compounds.

| Compounds | $T_f$ (K) | Mechanism / Reason for Glassy Behavior | Ref. |
|---|---|---|---|
| $ZnCoTiO_4$ | < 14 | Cation disorder and competing FM/AFM interactions due to mixed $Co^{2+}/Zn^{2+}/Ti^{4+}$ site occupation on A and B sites | 253 |
| $Co_{2-x}Zn_xSnO_4$ ($0 \leq x \leq 1$) | ~10-15 | Substitution-driven disorder modifies superexchange pathways, leading to frustration and spin glass state | 254 |
| $Mn[Cr_{0.5}Ga_{1.5}]S_4$ | 7.9 | Competing AFM interactions on $Cr^{3+}/Mn^{2+}$ sublattices and dilution by nonmagnetic $Ga^{3+}$ ions | 255 |
| Ti-doped $CuMn_2O_4$ | ~ 25 | $Ti^{4+}$ substitution for Mn disrupts double exchange and enhances frustration, producing spin glass state | 256 |
| $Zn_{0.8-x}Ni_xCu_{0.2}Fe_2O_4$ ($0 \leq x \leq 0.28$) | 26 | $Ni^{2+}$ substitution alters cation distribution and increases competing interactions | 257 |
| $LiCoMnO_4$ | 7 | Geometrical frustration and antisite disorder between $Li^+$, $Co^{2+}$, and $Mn^{3+}$ ions | 258 |
| $Ni_{1-x}Cd_xCr_2O_4$ ($x \leq 0.30$) | 22 | $Cd^{2+}$ substitution dilutes B-site $Cr^{3+}$ network, increasing disorder and frustration | 259 |
| $MnCr_2O_4$ & Co $Cr_2O_4$ | ~ 50 & ~ 98 | Mixed B-site occupation ($Cr^{3+}$, $Co^{2+}$, $Mn^{3+}$) yields exchange competition | 260 |
| $ZnCrMnO_4$ | 26 | Strong frustration from mixed B-site cations and possible cation disorder | 261 |
| Re-doped $CoFe_2O_4$ | ~ 200 | $Re^{5+}$ doping induces structural distortion and cation redistribution, enhancing frustration | 262 |
| $CuCrTiS_4$ | 7.91 | Disorder in $Cr^{3+}$ sublattice and competing exchange interactions | 263 |
| Mn-doped zinc spinel | 5.8 | Magnetic $Mn^{2+}$ doping in nonmagnetic zinc spinel lattice produces exchange disorder | 264 |
| $ZnTiCoO_4$ | 12.9 | Dilution of magnetic $Co^{2+}$ network by nonmagnetic $Zn^{2+}/Ti^{4+}$ ions create cluster spin glass state | 265 |
| $Cu_{1-x}Ag_xCrSnS_4$ | 16 | Ag substitution on Cu site changes local environment and increases frustration | 266 |
| $CoAl_2O_4$ | 4.5 | Anti-site disorder between $Co^{2+}$ and $Al^{3+}$ on A/B sites in frustrated diamond lattice | 267 |

Table 7. Heavy-fermion & related materials exhibiting spin glass phases

| Compounds | $T_f$ (K) | Mechanism / Reason for Glassy Behavior | Ref. |
|---|---|---|---|
| $Ce_2AgIn_3$ | 1.86 | RKKY-mediated random interactions amid heavy-fermion matrix; ac susceptibility shows cusp, divergence in nonlinear $\chi_{\mathrm{dc}}$ , no anomaly in $C$ or $\rho$ supports spin glass with HF behavior. | 268 |
| $CeNi_2Sn_2$ | 3 | Heavy-fermion background with random moments; ZFC-FC divergence, hysteresis, slow relaxation at ∼2 K, strong time dependence (canonical spin glass superimposed on HF behavior). | 269 |
| $CeRh_2Si_2$ | 1 | Structural (random-type) and magnetic (random-bond type) instabilities might lead to the formation of magnetic clusters[16] and spin glass phase. | 270 |
| $CeT_4M$ (T=Ni,Cu; M=Al,Ga,Mn) | - | Transition between HF and spin glass or magnetic order depending on composition; chemical substitution induces disorder leading to spin glass in some variants. | 271 |
| Zn-doped $LiV_2O_4$ | 2.6 | Disordering the spin background destroys the heavy fermion state and results spin glass state. | 272 |
| $PrIr_2B_2$ | 36 & 3.5 | Spin glass behavior attributed to 4f-electron magnetic frustration in HF background. | 273 |
| $U(Ga_{0.95}Mn_{0.05})_3$ | 40.9 | Mn doping in heavy-fermion U-Ga matrix produces cluster spin glass freezing from disorder and RKKY modulations. | 274 |
| $Sn_{1-x}U_xPt_3$ | 0.85-1.17 | Spin glass state likely results from magnetic clustering due to compositional inhomogeneity. | 275 |
| $UNi_4B$ | 9 | Spin glass state arises from disorder (mixture of two crystal structures) along with frustration of magnetic moments. | 276 |
| $(U_{0.25}Pt_{0.75})_{1-x}Si_x$ | x = 0.20 → 2.3<br>x = 0.25 → 2.9 | A-site disorder leads to HF spin glass freezing in amorphous alloys. | 277 |
| $URh_2Ge_2$ | 9.38 | Local structural disorder leads to random bonds and Ising-like HF spin glass state with non-Fermi-liquid behavior coexisting at low T. | 278 |

[16] "Magnetic clusters" refer to small regions in which spins develop short-range correlated behavior due to disorder, even though the system as a whole lacks long-range magnetic order.[23]

| | | | |
|---|---|---|---|
| $PrAu_2Si_2$ | 3 | Canonical spin glass in well-localized Pr moments in structurally ordered lattice, without intentional disorder–intrinsic excitonic frustration. | 279 |
| $SmCu_6$ | 3.8 | Origins unclear but amorphization seems to suppress long-range magnetic order found in crystalline $SmCu_6$ may lead to frustration from random stacking and variations in magnetic interactions. | 280 |
| $Pr_2CuIn_3$ | 5.4 | Site disorder and resulting frustrated, competing interactions in these nonmagnetic atom disordered compounds. | 281 |
| $Dy_2PdSi_3$ | 2.6 | Frustration effects due to the statistically disordered distribution of nonmagnetic Pd/Si atoms. This disordered structure leads to random, competing exchange interactions between the magnetic Dy atoms. | 282 |

## 6.2 Metallic spin glasses

In metallic spin glasses, localized magnetic moments interact indirectly through exchange mechanisms mediated by conduction electrons. These interactions, typically described by the RKKY model, can be either ferromagnetic or antiferromagnetic, with their competition giving rise to frustration and glassy magnetic states.[283] A canonical approach to these systems involves the introduction of dilute magnetic ions, such as manganese (Mn), into a non-magnetic metallic host (e.g., Cu, Ag, or Au). The resulting magnetic behavior depends sensitively on the impurity concentration, producing a sequence of regimes schematically summarized in Figure 27. At extremely low concentrations (≈tens of ppm, Figure 27a), individual Mn moments behave as isolated impurities. In this dilute limit, the Kondo effect dominates: conduction electrons screen the impurity moments below a characteristic Kondo temperature, suppressing magnetic ordering and leading to apparent non-magnetic behavior.[284] As impurity concentration rises above $\sim 0.1$ at.% (Figure 27b), indirect RKKY interactions between impurities become comparable to Kondo screening, leading to a spin glass transition. The freezing temperature $T_f$ scales with the impurity concentration $c$, often following a power law such as $T_f \propto c$ at low $c$, changing to $c^{2/3}$ for higher $c$.[285,286]

At intermediate impurity concentrations, beyond the dilute Kondo regime but still below percolation ($\sim 1 - 5$ at.% , Figure 27c), magnetic ions begin forming correlated clusters. These "superspins"[287] produce a cluster spin glass state, marked by frequency-dependent susceptibility peaks, memory effects, and slow magnetic relaxation, as reported for systems such as $Cr_{0.5}Fe_{0.5}Ga$.[288] Short-range correlations and dipolar interactions become increasingly relevant in this regime.

When the impurity concentration approaches the percolation threshold ($\approx 5 - 10$ at.%, Figure 27d), clusters coalesce into larger ferromagnetic-like domains within a frustrated background, giving rise to so-called giant-cluster spin glass behavior.[289] This regime shows slow dynamics, micromagnetic features, and signatures of both glassy freezing and partial long-range correlations. Finally, above the percolation threshold ($> 10 - 15$ at.%, Figure 27e), inhomogeneous long-range order emerges. Here, direct exchange interactions between neighboring magnetic ions dominate, producing ferromagnetic or antiferromagnetic order depending on composition, although residual disorder still broadens the magnetic transition and frustrates perfect long-range coherence. Thus, the progression from isolated Kondo impurities through RKKY-driven spin glass freezing, cluster glass formation, and eventually percolative long-range order[290] reflects the concentration-tuned competition of exchange interactions in metallic spin glasses.[291]

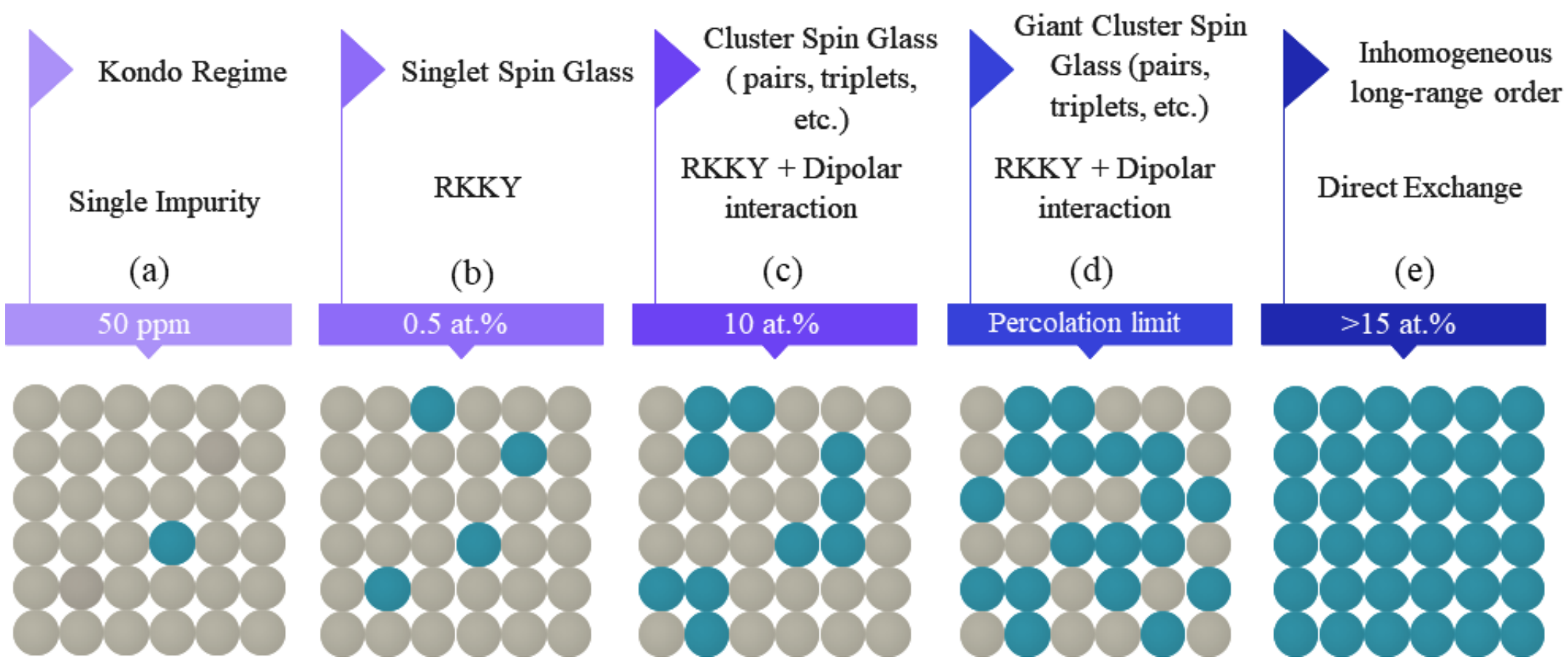

Figure 27. Metallic spin glasses: depiction of the influence of rising concentrations of magnetic ions within a non-magnetic metallic framework, inspired by Coey and Mydosh. (a) Kondo regime in minimal trace of magnetic ions, (b) adequate magnetic ion concentration leading to RKKY interactions and signs of spin glass behavior; (c) indicates the emerging significance of short-range correlations, the prominence of dipolar interactions, and the formation of a cluster glass; in (d) the magnetic ion concentration reaches a level where magnetic percolation, long-distance magnetic alignment, and direct exchange are evident; and in (e) Inhomogeneous long-range order and related direct exchange interaction.

As mentioned earlier among the first metallic alloys discovered to display spin glass properties were noble-metal systems doped with dilute transition-metal impurities, such as AuMn, AuCr, AgMn, and AuFe. Their unusual electronic transport properties, particularly in resistivity and thermopower, have long been attributed to the underlying spin glass ordering and the competition between Kondo screening and RKKY exchange. To observe spin glass behavior clearly, an alloy system must satisfy two essential criteria: (i) the magnetic impurities should be highly soluble in the metallic host to ensure uniform distribution of local moments, and (ii) the Kondo temperature $T_{\mathrm{K}}$ of the system should be low ($T_{\mathrm{K}} < 5$ K), so that collective freezing rather than complete screening governs the low-temperature physics.

The concentration dependence of impurity resistivity in these alloys is summarized in Figure 28 a-c. At low concentrations ($\sim 1$ at.% impurities, Figure 28a), the resistivity rises more steeply than expected from a simple linear temperature dependence. AuCr displays a sharp resistance peak, indicating strong magnetic scattering, while AuMn and AgMn show broader, less pronounced maxima. AuFe, in contrast, exhibits only a weak resistance anomaly, suggesting a comparatively subdued role of magnetic correlations at this dilution. As the impurity concentration increases to $\sim 5$ at.% (Figure 28b), the resistance maximum shifts to higher temperatures, reflecting the enhanced role of impurity-impurity interactions in controlling spin dynamics. Interestingly, the residual resistivity at $T \rightarrow 0$ is higher for AuFe and AuCr than for AuMn and AgMn, consistent with stronger impurity scattering in Fe- and Cr-based alloys. At even higher concentrations (8–11 at.%, Figure 28c), the resistivity shows a more pronounced linear rise with temperature up to the ordering temperature, followed by a resistance peak associated with the onset of collective magnetic freezing. The sharpness of this peak, particularly in AuCr, points to fragile antiferromagnetic clusters that disintegrate more readily than ferromagnetically correlated ones.[292,293]

Complementary insights come from thermopower measurements (Figure 28d). For $\sim 5$ at.% impurity concentration, AuFe exhibits a strongly negative thermopower across the full measured temperature range, while AuCr shows a weaker signal that changes sign from negative to positive near 10 K. AuMn, combining features of both, displays a pronounced negative peak but also undergoes a sign inversion at low temperatures. These observations are consistent with earlier reports in AgMn and CuMn, where low-temperature thermopower sign reversals were also observed, but absent in alloys like AuFe, CuFe, CuCo, and CuMo. Theoretical interpretation of these results has been provided by Matho and Béal-Monod, who proposed that resonant scattering of conduction electrons from impurity pairs undergoing Zeeman splitting near $k_{\mathrm{B}}T$ can generate a distinct contribution to thermopower. In cases where impurity-impurity interactions are predominantly antiferromagnetic (e.g., Cr-based systems), this resonant contribution drives the sign change in $S(T)$. Conversely, when ferromagnetic correlations dominate (e.g., Fe-based alloys), the thermopower remains negative down to low temperatures. These trends align with susceptibility measurements that indicate predominantly ferromagnetic couplings between Fe impurities, antiferromagnetic interactions between Cr impurities, and mixed interactions for Mn, depending on local atomic environment.

Taken together, the concentration- and temperature-dependent resistivity and thermopower of AuMn, AuCr, AgMn, and AuFe alloys (Figure 28 a–d) illustrate the complex interplay between Kondo physics, RKKY-driven spin glass freezing, and impurity-cluster formation. These transport anomalies provide some of the earliest experimental evidence for the rich variety of spin glass behavior in metallic hosts.

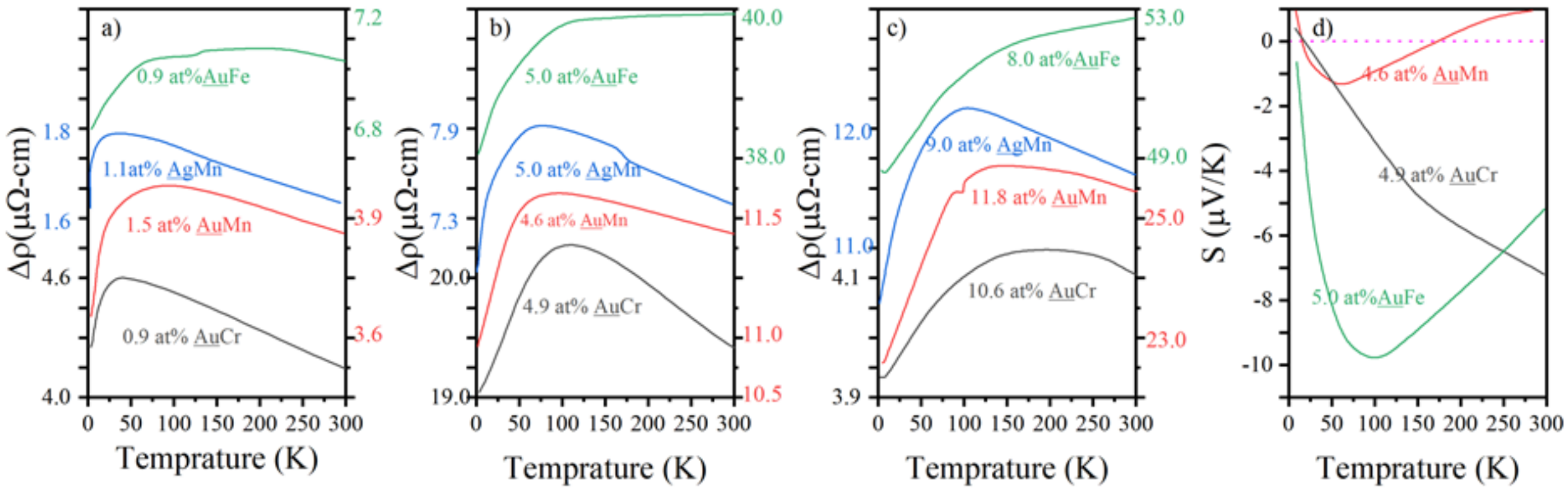

Figure 28. Temperature-dependent impurity resistivity ($\Delta\rho$) and thermopower ($S$) in representative metallic spin glass alloys. (a) Low impurity concentration (~1 at.%): all alloys show an enhanced resistivity increase with decreasing temperature. AuCr exhibits a sharp resistance peak, while AuMn and AgMn display broader maxima and AuFe only a weak anomaly. (b) Intermediate concentration (~5 at.%): resistance peaks shift to higher temperatures with increasing concentration, highlighting the growing role of impurity-impurity interactions. (c) High concentrations (8–11 at.%): resistivity retains a nearly linear temperature dependence up to the ordering temperature, followed by distinct resistance maxima associated with cluster formation; the sharp peak in AuCr points to fragile antiferromagnetic clusters. (d) Thermopower for ~5 at.% alloys: AuFe shows a large negative peak that persists to low temperatures; AuCr exhibits no clear maximum but undergoes a sign change from negative to positive near 10 K; AuMn combines both features, with a pronounced negative peak and a low-temperature sign inversion. Figures reconstructed based on data extracted from Ref. [293].

The family of quadruple perovskites $CaCu_3B_2Ir_2O_{121}$ (B = Mn, Fe, Co, Ni) has been widely investigated for its crystal structure, electronic band topology, and magnetic ground states. These materials display contrasting behaviors depending on the B-site cation. For B = Mn and Fe, spin glass states are observed, while B = Co and Ni stabilize type-I ferrimagnetic ordering. Their transport characteristics also differ: $CaCu_3Mn_2Ir_2O_{12}$ behaves as a semiconductor, $CaCu_3Fe_2Ir_2O_{12}$ and $CaCu_3Co_2Ir_2O_{12}$ are metallic, and $CaCu_3Ni_2Ir_2O_{12}$ exhibits semi-metallic character. The magnetic susceptibility of this series follows the Curie–Weiss law at high temperatures. For B = Mn and Fe, the susceptibility increases initially before dropping with further heating, consistent with a spin glass state. By contrast, B = Co and Ni compounds exhibit steadily decreasing susceptibility, characteristic of ferrimagnetic (FiM) ordering. Transition temperatures ($T_f$ or $T_c$) can be determined by analyzing the derivative of the magnetization with respect to temperature ($d\chi/dT$). The anomalies occur at ~62 K, 75 K, 250 K, and 275 K for B = Mn, Fe, Co, and Ni, respectively.[294]

Beyond perovskites, metallic glasses are an important class of amorphous systems known to host spin glass states. Owing to their large magnetocaloric effect (MCE), they are regarded as promising candidates for magnetic refrigeration.[295] Two primary categories of magnetocaloric metallic glasses include transition metal (TM)-based and rare-earth (RE)-based alloys. Figure 29a illustrates the temperature-dependent magnetization curves for $RE_{55}Co_{17.5}Al_{27.5}$ metallic glasses with RE = Gd[296], Tb[297], and Dy. The FC and ZFC curves, measured between 5 and 200 K in a 200 Oe applied field, reveal a significant reduction in magnetization upon heating, indicative of a ferromagnetic to paramagnetic transition. Analysis of the ZFC curves for $Tb_{55}Co_{17.5}Al_{27.5}$ and $Dy_{55}Co_{17.5}Al_{27.5}$ reveals a distinct cusp, a hallmark of spin glass behavior previously observed in other Tb- and Dy-based metallic glasses. The cusp temperature, $T_f$, is determined to be 37 K and 23 K for $Tb_{55}Co_{17.5}Al_{27.5}$ and $Dy_{55}Co_{17.5}Al_{27.5}$, respectively.[298]

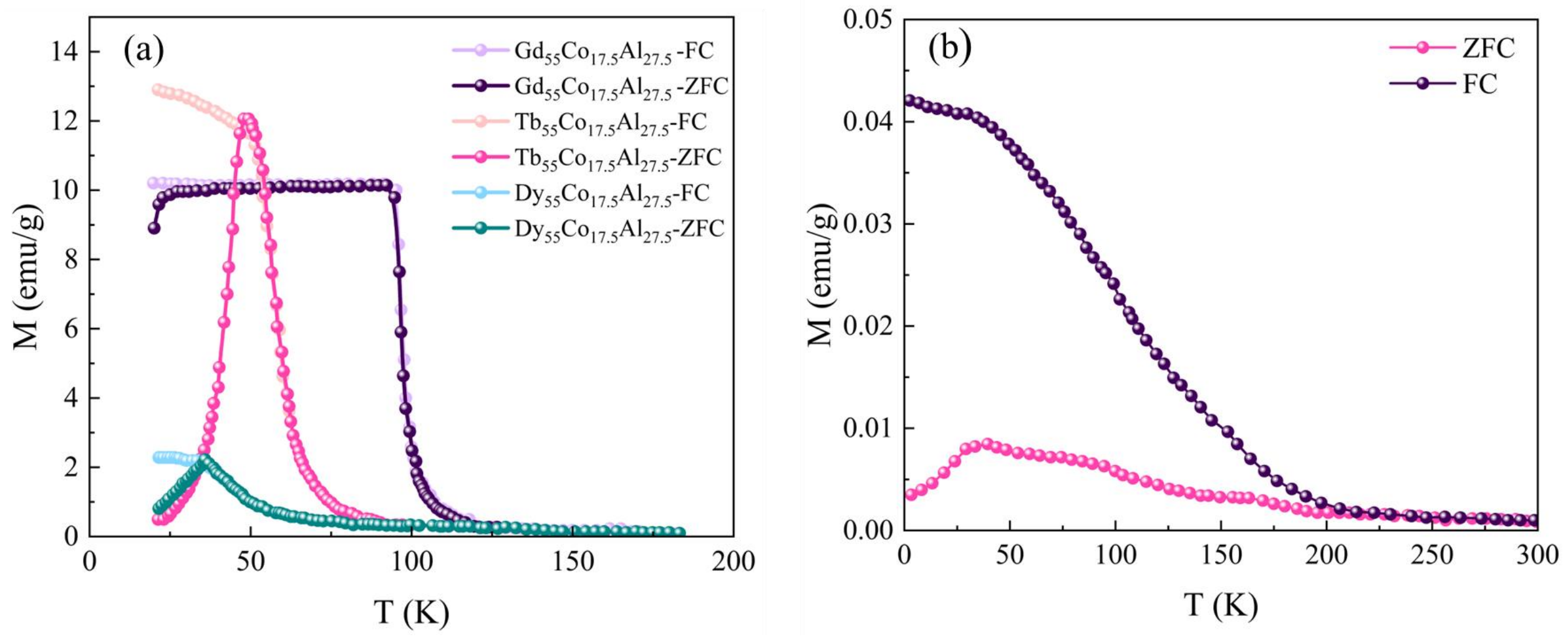

Figure 29. (a) Temperature dependence of field-cooled magnetization curves for $RE_{55}Co_{17.5}Al_{27.5}$ metallic glasses with RE = Gd, Tb, and Dy under a 200 Oe (0.02 T) magnetic field. (b) Temperature-dependent magnetization curves for $La_2CrNiO_6$ powders measured in ZFC and FC modes under a 100 Oe magnetic field. Figures reconstructed based on data extracted from refs. [295] and [299] respectively.

Another intriguing example is the double perovskite $La_2CrNiO_6$, predicted to be a half-metallic ferrimagnet.[299] Its negative Curie–Weiss temperature (-35 K) indicates dominant antiferromagnetic (AFM) correlations, while its Curie temperature of 106 K is consistent with ferrimagnetism. Upon cooling from the paramagnetic phase, $La_2CrNiO_6$ powders exhibit a spin-glass-like transition. This arises from competition between AFM interactions ($Cr^{4+}$-O-$Cr^{4+}$, $Ni^{2+}$-O-$Ni^{2+}$, and $Ni^{2+}$-O-$Cr^{3+}$) and FM couplings ($Ni^{2+}$-O- $Cr^{4+}$, $Ni^{3+}$-O-$Cr^{3+}$). AC susceptibility measurements (Figure 29b) confirm this glassy state: the ZFC curve shows a broad cusp near 40 K, consistent with the freezing of magnetic clusters. Below this temperature, spins freeze into a disordered configuration, producing canonical spin glass dynamics.

Thin films also provide fertile ground for reentrant spin glass phenomena. An FeAs layer grown on $LaAlO_3$ (100) substrates displays metallic transport, strong thermoelectric response, and a reentrant spin glass state.[300] Magnetization versus temperature curves recorded under 500 Oe (Figure 30a) reveal a sharp peak at ∼50 K in both FC and ZFC branches. This feature reflects the interplay of ferromagnetic and antiferromagnetic correlations characteristic of reentrant spin glasses. Thermopower measurements (Figure 30b) complement this picture: $S$ remains negative between 30–370 K, with a broad minimum of -11 μV/K near 100 K, surpassing values in bulk FeAs. At $T < 30$ K and $T > 370$ K, the thermopower becomes positive, indicating complex electronic scattering associated with magnetic fluctuations.

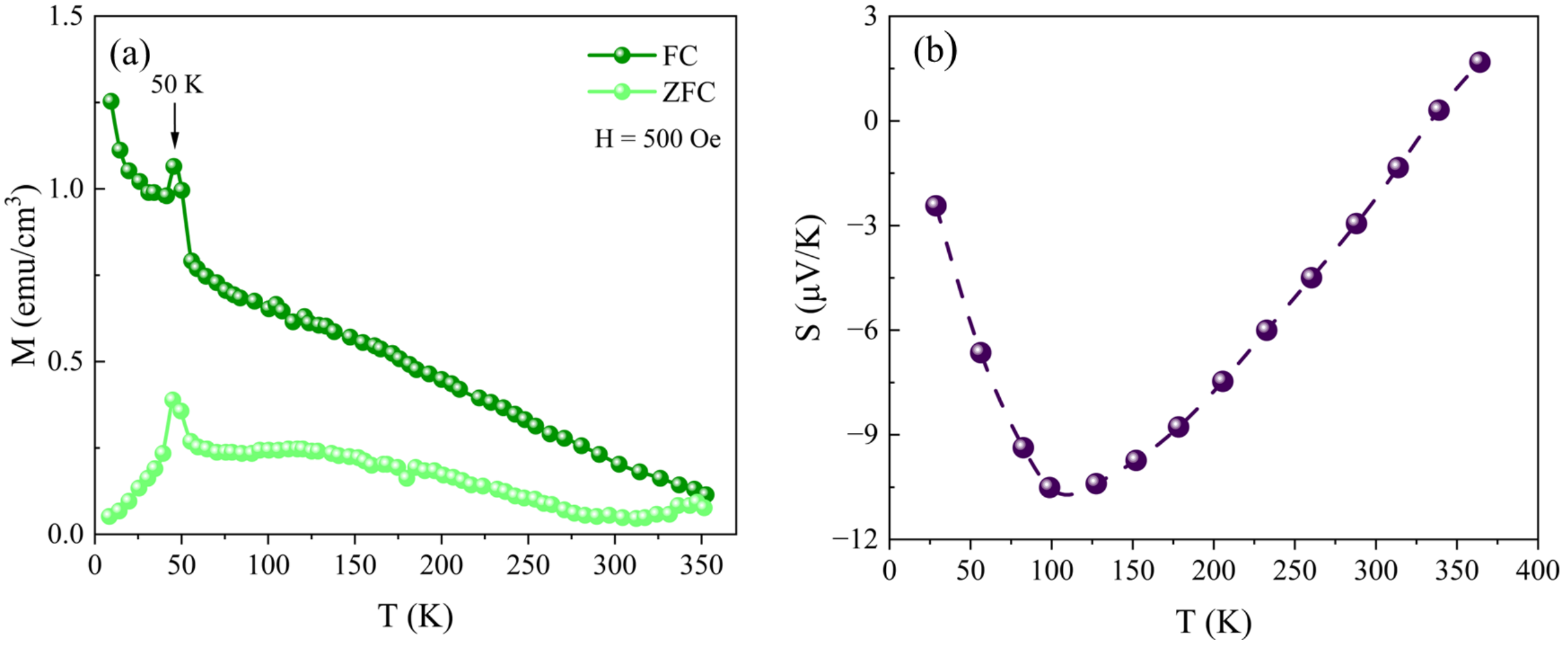

Figure 30. Magnetic and thermoelectric properties of an FeAs thin film on $LaAlO_3$ (100) substrates. (a) Magnetization versus temperature under 500 Oe shows a sharp peak at ∼50 K in both FC and ZFC branches, reflecting reentrant spin-glass behavior arising from competing ferromagnetic and antiferromagnetic correlations. (b) Thermopower measurements reveal a broad minimum (∼ − 11 μV/K) near 100 K, with $S$ remaining negative from 30–370 K and turning positive outside this range, indicating complex electronic scattering linked to magnetic fluctuations. Figure reconstructed based on data extracted from Ref. [300].

Spin glass physics is also evident in amorphous transition-metal-metalloid alloys such as $(A_wB_{1-w})_{75}P_{16}B_6Al_2$ (A, B = Fe, Ni; Co, Ni; Fe, Mn).[301] At low concentrations (w), these alloys undergo a para-magnetic-spin-glass (PM-SG) transition upon cooling, while higher concentrations exhibit a paramagnetic-ferromagnetic (PM-FM) transition followed by a FM- SG crossover. AC susceptibility $\chi_{ac}(T)$ measure-ments show sharp peaks at the corresponding transition temperatures, with the lower-temperature cusp marking the spin glass freezing point. For $(Fe_xNi_{1-x})_{75}P_{16}B_6Al_2$ with $x \leq 0.16$, the peak corresponds to a direct PM-SG transition (Figure 31a). In Co-Ni-based alloys with $y \leq 0.32$, cusps in $\chi_{ac}(T)$ similarly iden-tify PM- SG transitions (Figure 31b). For Fe-Ni alloys with $0.40 \leq z \leq 0.90$, successive PM- FM and FM-SG transitions are observed, and the corresponding phase diagram is presented in Figure 31c. Here, $T_f$ denotes the FM-SG transition temperature, while $T_{SG}$ refers to the PM-SG transition temperature.

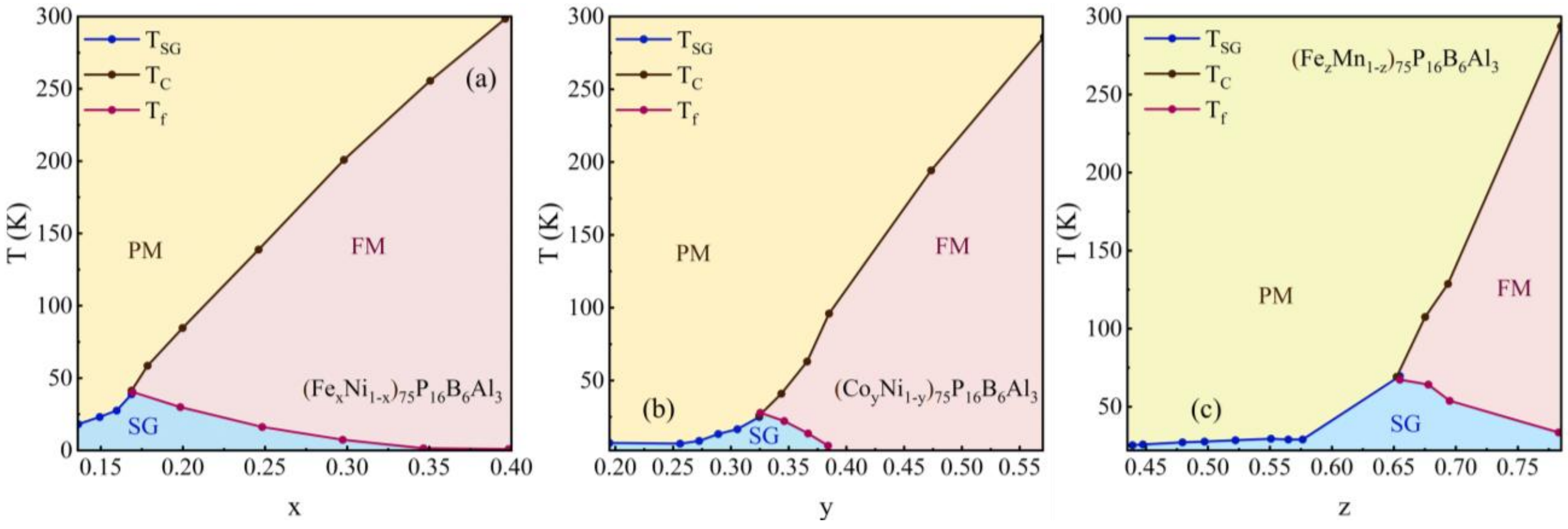

Figure 31. Magnetic phase diagrams of amorphous transition-metal–metalloid alloys: (a) Fe–Ni, (b) Co–Ni, and (c) Fe–Mn series. The diagrams illustrate PM–SG, PM–FM, and FM–SG regions, with $T_{SG}$ and $T_f$ denoting the spin-glass and ferromagnetic–spin-glass transition temperatures, respectively. Figure reconstructed based on data extracted from Ref. [301].

In Table 8, we have summarized the metallic spin glasses discussed, including their freezing temperatures, for improved accessibility.

Table 8. Metallic spin glasses and their FM-SG and PM-SG freezing temperatures ($T_f$ and $T_{SG}$)

| Compounds | | $T_f$ or $T_{SG}$ (K) | Refs. |
|---|---|---|---|
| AuFe | | 0.9 at. % → 8.5<br>1.0 at. % → 9.1<br>1.4 at. % → 12.8<br>5.0 at. % →22<br>8.0 at. % →32 | 293-302 |
| CuMn | | 1.1 at. % → 10.8<br>3 at. % → 20.0<br>5 at. % → 27.4 | 293-302 |
| AgMn | | 1.1 at. % → 5<br>5.0 at. % → 19<br>9.0 at. % →30 | 293 |
| AuCr | | 0.9 at. % → 15<br>4.9 at. % → 50<br>10.6 at. % → 80 | 293 |
| AuMn | | 1.5 at. % → 5.5<br>4.6 at. % → 15<br>11.8 at. % → 25 | 293 |
| CuMn (slow cooled) | | 0.23 at. % → 2.85<br>0.46 at. % → 5.00<br>1.46 at. % → 12.40 | 111 |
| CuMn (quenched) | | 0.57 at. % → 6.00<br>0.70 at. % → 7.65<br>0.94 at. % → 9.40<br>2.00 at. % → 15.50<br>6.30 at. % → 32.3 | 111 |
| $Gd_{0.37}Al_{0.63}$ | | 15.8-16 | 303 |
| $Tb_{55}Co_{17.5}Al_{27.5}$ | | 37 | 298 |
| $Dy_{55}Co_{17.5}Al_{27.5}$ | | 23 | 298 |
| $La_2CrNiO_6$ | | 290 | 299 |
| FeAs Film on $LaAlO_3$(100) | | 50 | 300 |
| $CaCu_3Mn_2Ir_2O_{12}$ | | 49 | 294 |
| $CaCu_3Fe_2Ir_2O_{12}$ | | 75 | 16 |
| $(Fe_xNi_{1-x})_{75}P_{16}B_6A_{13}$ | PM-SG | x = 0.14 → 20<br>x = 0.15 → 24<br>x = 0.16 → 31 | 301 |
| | FM-SG | x = 0.17 → 43<br>x = 0.18 → 40<br>x = 0.20 → 33<br>x = 0.25 → 18<br>x = 0.30 → 8<br>x = 0.35 → 2<br>x = 0.40 → ≈ 0 | |
| $(Co_yNi_{1-y})_{75}P_{16}B_6A_{13}$ | PM-SG | x = 0.20 → 6.5 | 301 |
| | | x = 0.26 → 7.5<br>x = 0.28 → 9.5<br>x = 0.30 → 12.5<br>x = 0.32 → 14.5 | |
| | FM-SG | x = 0.34 → 23.5<br>x = 0.36 → 21<br>x = 0.38 → 13<br>x = 0.40 → 4 | |
| $(Fe_zMn_{1-z})_{75}P_{16}B_6A_{13}$ | PM-SG | x = 0.40 → 26<br>x = 0.45 → 26.5<br>x = 0.50 → 27<br>x = 0.55 → 29<br>x = 0.60 → 42 | 301 |
| | FM-SG | x = 0.65 → 63<br>x = 0.70 → 54<br>x = 0.80 → 34<br>x = 0.85 → < 5 | |

### 6.2.1 Heusler spin glasses

Another broad family of intermetallic compounds that has attracted attention in magnetism research is the Heusler alloys, which can be broadly categorized into two types: (i) full Heusler alloys with the idealized $X_2YZ$ stoichiometry (Figure 32), where X and Y are transition metals and Z is a main-group element, and (ii) half Heusler alloys with the general formula XYZ. Magnetic Heusler alloys are typically rich in 3d transition elements and exhibit a wide range of magnetic ground states, including ferromagnetism, antiferromagnetism, and spin glass phases of the reentrant type.[304,305,306]

Because these materials are usually excellent conductors, the RKKY exchange interaction dominates their magnetism, leading to competing ferro and antiferromagnetic couplings between magnetic sublattices. Moreover, antisite disorder (site exchanges between atoms) is frequently observed and plays a critical role in stabilizing spin glass states. A prominent example is the half-Heusler compound IrMnGa, crystallizing in space group *F-43m*.

Kroder et al. used neutron diffraction and magnetic measurements to establish its canonical spin glass behavior. They observed:

- absence of magnetic Bragg peaks in neutron diffraction, ruling out long-range order,
- bifurcation of magnetization curves below the freezing temperature $T_f = 74$ K,
- a sharp cusp in the ac susceptibility shifting with frequency,
- exchange bias effects under field cooling,
- pronounced magnetic aftereffect and memory effects.[307]

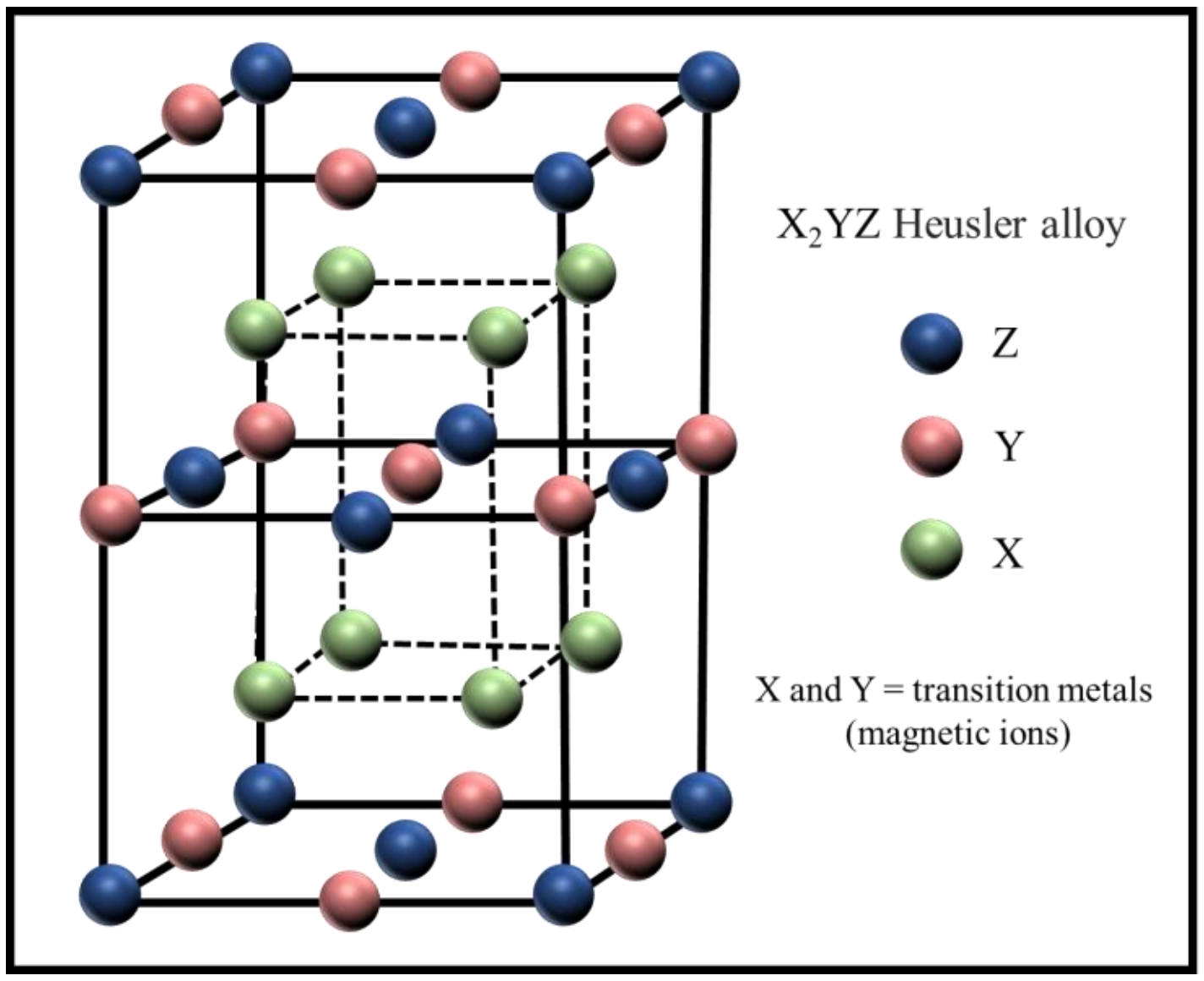

Figure 32. $X_2YZ$ Heusler alloy: The full-Heusler structure, consisting of four interpenetrating fcc sublattices, with X (green), Y (pink), and Z (blue) atoms occupying specific Wyckoff positions. The precise arrangement of these elements controls the degree of hybridization between atomic orbitals, leading to remarkable functionalities such as half-metallicity, high spin polarization, magnetocaloric behavior, and tunable magnetic anisotropy. Heusler phases thus serve as a versatile platform for spintronic, thermoelectric, and multifunctional materials research.

Analysis via both the Vogel-Fulcher law and critical scaling confirmed that IrMnGa behaves as a canonical spin glass. The absence of any anomaly in the specific heat at $T_f$ further supports this conclusion. The noncollinear spin arrangements in Heusler compounds are widely believed to arise from frustration between competing exchange interactions, which suppress a unique magnetic ground state and instead stabilize a disordered glassy configuration. These frustrated arrangements can also produce exotic phenomena such as the topological Hall effect, as reported in several $D_{2d}$ Heusler systems.[308] Noncollinear textures including anti-skyrmions and elliptical Bloch skyrmions have been observed in both centrosymmetric and non-centrosymmetric Heusler compounds, further underscoring the diversity of magnetic states accessible in this family.[309]

In a different context, Gupta et al. investigated the quaternary equiatomic Heusler alloy NiRuMnSn, where Ru (a 4d element) substitutes Fe (a 3d element). Structural refinements using X-ray diffraction (XRD), extended X-ray absorption fine structure (EXAFS), and neutron diffraction revealed partial site disorder, with Ni and Ru statistically sharing the 4c and 4d Wyckoff positions.[17] Magnetic measurements showed a FM-PM transition near 214 K, and below ~60 K the emergence of a spin glass phase was evidenced by ac susceptibility, neutron depolarization[18], and magnetization bifurcation. The discrepancy between the theoretical and experimental magnetic moments was attributed to Ni/Ru site disorder and low-temperature spin glass freezing. Notably, this system represents a rare case among the nearly one thousand reported Heusler alloys that exhibit clear spin glass behavior. Interestingly, while isoelectronic substitution of 3d by 4d elements is typically thought to improve structural order and enhance spin polarization, the results of Gupta *et al.* show that the essentially nonmagnetic character of Ru (whose 4d electrons tend to remain in a low-spin, weakly magnetic state) suppresses long-range spin polarization and instead promotes reentrant spin glass ordering. This highlights the delicate interplay between structural order, electronic configuration, and magnetic frustration in Heusler alloys.[310]

[17] In the tetragonal crystal structure (space group I4/mmm or a similar variant), Wyckoff positions such as 4c and 4d refer to symmetry-defined atomic sites characterized by specific fractional coordinates and site symmetries. These labels indicate crystallographically distinct locations where atoms may occupy with full or partial occupancy.

[18] Neutron depolarization is a technique in which a beam of polarized neutrons passes through a magnetic material. If the sample contains magnetic domains or frozen, randomly oriented local moments (such as those found in spin glasses) the internal magnetic fields cause the neutron spins to precess differently along their paths. This leads to a reduction ("depolarization") in the polarization of the outgoing neutron beam. By measuring the amount of depolarization, one can detect the onset of magnetic inhomogeneity, domain formation, or spin freezing, making this method particularly sensitive to the emergence of a spin glass state or other non-uniform magnetic phases.

Other metallic systems including Heusler alloys that exhibit spin glass ordering, with their freezing temperature and the primary reason for spin glass formation is given in Table 9.

Table 9. Heusler alloys exhibiting spin glass phases and their properties.

| Compounds | $T_f$ (K) | Mechanism / Reason for Glassy Behavior | Ref. |
|---|---|---|---|
| $Bi(Fe_{0.95}Co_{0.05})O_3$ | 260 & 100 | Competition between ferromagnetic and antiferromagnetic exchange at cluster boundaries | 311 |
| $Cu_2VAl$ | 45 | Structural disorder leading to frustrated exchange interactions | 312 |
| IrMnGa | 74 | Site disorder from Mn/Ga mixing causing frustrated magnetic couplings | 307 |
| $Mn_{50}Ni_{40}Ga_{10}$ | 70 | Formation of magnetic clusters with competing local interactions | 313 |
| MnCuNiSn | 150 | Chemical disorder (Ni/Cu) giving rise to exchange frustration | 314 |
| $Ni_{50}Mn_{42}In_3Sb_5$ | 94-108 | Disorder-driven frustration linked with exchange bias phenomena | 315 |
| $Ru_{2-x}Fe_xCrSi$ | 45-70 | Substitutional disorder and FM/AFM competition on Ru/Fe sites | 316 |
| $Mn_{1.5}Fe_{1.5}Al$ | 34.5 | Chemical randomness promoting canonical spin glass freezing | 317 |
| $Ni_2Mn_{1.36}In_{0.64}$ | 200 | Off-stoichiometry predicted to enhance frustrated Mn–Mn interactions | 318 |
| $Ni_{2.01}Mn_{1.58}Sn_{0.41}$ | 151 | Due to the composition disorder between Mn and Sn atoms in the martensite phase, magnetic clusters arise in the alloy and freeze into spin glass state | 319 |
| $Ni_{50}Mn_{37}Sb_{13-x}Al_x$ | 100-150 | Lattice contraction tuning magnetic exchange and suppressing reentrant spin glass | 320 |
| $Mn_2Ni_{1.4}Ga_{0.6}$ | 300 | Competition between short-range ferromagnetism and glassy dynamics | 321 |
| $Ru_2Cr_{1-x}Si_x$ | 15 | Si substitution inducing frustration of Ru-Cr exchange interactions | 322 |
| NiFeGa | 27.2 | Antiphase boundaries introducing local frustration and cluster spin glass behavior | 323 |
| $Mn_{1.89}Pt_{0.98}Ga_{1.12}$ | 123 | Coexistence of reentrant spin glass with topological Hall response due to frustrated Mn sublattice | 324 |
| $Mn_2Ni_{1.6}Sn_{0.4}$ | 117 | Competing ferromagnetic martensitic phase and glassy freezing | 325 |
| $Mn_{50}Ni_{38}Al_{12}$ | 130 | Superspin glass state from exchange-biased nanoscale clusters | 326 |
| $Ni_{47}Mn_{36}Cr_4Sn_{13}$ | 146.5-155.6 | Martensitic transformation producing coexisting FM and spin glass states | 327 |
| Si-doped $Ni_{50}Mn_{36}Sn_{14}$ | 136.5 | Site disorder coupled with martensitic instability | 328 |
| Ni-Mn (general) | 170 | Disordered Mn sublattice leading to glassy magnetic transitions | 329 |
| Au–Cr | 90-100 | Antiferromagnetic clusters embedded in metallic matrix | 330 |
| Au–Fe | 50 | Finite Fe-rich magnetic clusters close to percolation threshold | 331 |
| AlN | - | Defect-induced magnetism creating superspin glass behavior | 332 |
| $Tm_2Cu_2In$ | - | Intrinsic frustration from Shastry–Sutherland lattice geometry | 333 |

## 6.3 Semiconducting Spin Glasses

### 6.3.1 Chalcogenides

Monochalcogenides (e.g., sulfides, selenides, tellurides) represent a unique class of semiconducting spin glasses where structural tunability, dimensionality, and competing exchange pathways drive exotic low-temperature magnetism. Their layered architectures, stabilized by weak van der Waals bonding, allow both chemical substitution and exfoliation into quasi-2D forms, which has recently intensified interest due to their potential integration in electronic and spintronic devices.[23] Structural defects, non-stoichiometry, and aliovalent doping are key in generating local magnetic moments and frustration, with consequences ranging from superparamagnetic blocking to cooperative glassy freezing.

A prototypical system is $Eu_xSr_{1-x}S$, often regarded as a diluted Heisenberg spin glass.[334] Here, the competition between ferromagnetic nearest-neighbor exchange $J_1$ and antiferromagnetic next-nearest neighbor exchange $J_2$ ($J_2/J_1 \approx -0.5$) drives frustration. Spin glass freezing occurs below $\sim$3 K in the Eu concentration window $0.13 \leq x \leq 0.63$, bounded by a percolation threshold at $x \approx 0.13$. Below this limit, isolated Eu ions form superparamagnetic clusters without collective glassy dynamics, underscoring how finite connectivity is essential for cooperative freezing.[335]

Raman scattering experiments have further revealed short-range spin correlations deep into the glassy phase.[336] By contrast, EuCa solid solutions fail to exhibit spin glass ordering, suggesting that the balance of exchange interactions disfavors frustration. Interestingly, in $Eu_xSr_{1-x}SO_4$, glassy freezing is observed across the full composition range, though only at very low temperatures ($\sim$1 K), again emphasizing the sensitivity of these systems to subtle structural and chemical details.[337]

A microscopic explanation has been attributed to the interplay between direct f-d overlap and indirect superexchange. For nearest-neighbor Eu-Eu pairs, hybridization between the Eu 4f and excited $5d–t_{2g}$ states stabilizes a ferromagnetic $J_1$ . In contrast, second-neighbor interactions proceed via the ligand's p-orbitals, generating an antiferromagnetic $J_2$. The delicate balance between these two competing channels gives rise to frustration and, depending on concentration and lattice environment, stabilizes either a spin glass or a conventional ordered phase.[338] Compounds such as ErSrTe and ErSrSe also fall into this family, showing spin glass freezing due to similar competing exchanges, whereas ErSrO, dominated by ferromagnetic $J_1$ and $J_2$, instead stabilizes long-range ferromagnetism up to percolation.

Beyond rare-earth chalcogenides, transition-metal chalcogenides provide another rich playground where subtle structural effects control the magnetic ground state. The $AFeMCh_2$ (Ch = S, Se, Te) family in particular shows a remarkable crossover from long-range AFM to spin glass behavior depending on lattice parameters. Compounds with larger in-plane lattice constants $a$, such as $KFeAgTe_2$ or $KFeCuTe_2$, stabilize AFM order, while more compressed lattices (e.g., $KFeAgS_2$) favor glassy freezing. This correlation suggests that compression in the ab-plane broadens the Fe-3d band overlap with chalcogen p-states, destabilizing AFM order in favor of frustrated spin correlations.[339,340,341,342]

Interestingly, the interlayer distance (half of lattice parameter $c$) appears to play a negligible role compared to the in-plane parameter $a$. Compounds such as $KFeAgSe_2$ and $RbFeLiSe_2$ sit at the boundary between AFM and spin glass states, making them useful model systems for studying this crossover. This family thus illustrates a broader principle: structural control of magnetism. Across $AFeMCh_2$, the transition from Te → Se → S systematically tunes reactivity and exchange competition. In parallel, alkali-metal substitution (K → Rb → Na) further modifies lattice dimensions and the magnetic ground state.[343]

These observations highlight how spin glasses in semiconducting chalcogenides emerge not simply from random site disorder, but from a subtle interplay of lattice compression, orbital hybridization, and exchange frustration–an interplay that remains a fertile ground for both theoretical modeling and experimental discovery.

### 6.3.2 Dilute magnetic semiconductor

Diluted magnetic semiconductors (DMSs) of the form $A_{1-x}Mn_xB^{VI}$ (A = Hg, Cd, Zn; $B^{VI}$ = S, Se, Te) exhibit spin glass characteristics. Table 1 of [47] provides a comprehensive list of these ternary alloys, including their crystal structures and composition ranges. Low-field magnetic susceptibility measurements of $A_{1-x}Mn_xB^{VI}$ DMS reveal a well-defined cusp at the spin glass transition temperature ($T_f$), indicating a phase transition. These observations led to an understanding of how $T_f$ varies as a function of x. Irreversible phenomena are evident in magnetization experiments below $T_f$, as demonstrated by the difference in susceptibility data between FC and ZFC measurements. This divergence further supports the presence of a spin glass state in these materials. The spin glass behavior of DMS has usually been linked to frustration of antiferromagnetic interactions between Mn ions caused by the lattice structure of the $A_{1-x}Mn_xB_{VI}$ alloys[47].

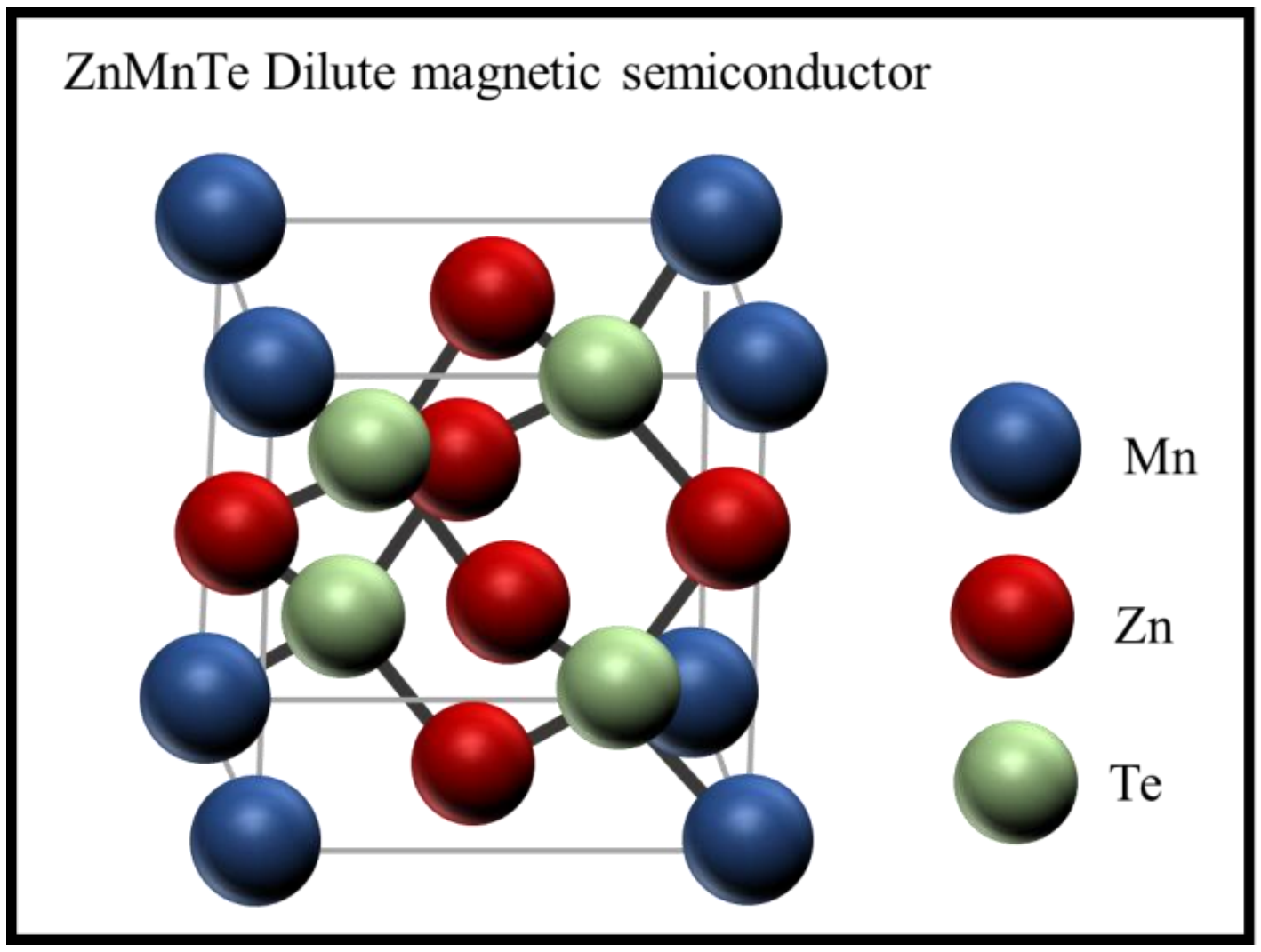

Figure 33. $Zn_{1-x}Mn_xTe$ dilute magnetic semiconductor: The zinc-blende-type structure of ZnMnTe, showing the substitution of a fraction of Zn cations (red) by magnetic Mn ions (blue) within the tetrahedrally coordinated Te anion framework (green). The incorporation of Mn introduces localized magnetic moments into an otherwise nonmagnetic semiconductor, enabling strong carrier-spin exchange interactions. This coupling gives rise to tunable magneto-optical effects, giant Zeeman splitting, and the possibility of engineering spin-dependent electronic states.

Experimental studies on SnMnTe and PbSnMnTe, two related semi-magnetic semiconductors, revealed distinct magnetic properties. RKKY indirect exchange interactions mediated by free carriers lead to the emergence of both ferromagnetic and spin glass phases. In $Sn_{1-x}Gd_xTe$ samples with $x < 0.05$, a prominent cusp in the magnetic susceptibility was observed at $T = 1.8$ K, suggesting the formation of a spin glass phase. This phase transition is attributed to an increase in spin-spin interactions. For $Sn_{1-x}Mn_xTe$, the magnetic behavior depends on carrier concentration. At relatively low carrier concentrations ($p = 5 - 7 \times 10^{20}$ $cm^{-1}$), the system exhibits ferromagnetism at liquid helium temperatures. As carrier concentration increases ($p = 9 - 12 \times 10^{20}$ $cm^{-3}$), the ferromagnetic nature weakens, leading to a reentrant spin glass phase. At even higher carrier concentrations ($p = 20 - 23 \times 10^{20}$ $cm^{-3}$), ferromagnetism vanishes, and a pure spin glass phase is observed. Figure 2 of [344] illustrates the real component of the AC magnetic susceptibility of the studied samples, revealing a clear cusp at $x = 0.04$ and 0.06, indicative of a spin glass transition. For $x = 0.08$, two distinct maxima are observed: one at $T = 10.8$ K, corresponding to a paramagnetic-to-ferromagnetic transition, and another at $T = 2.3$ K, suggesting a FM-SG transition. This behavior is characteristic of a reentrant spin glass phase. At $x = 0.10$, the sample exhibits a ferromagnetic transition at $T =$ 15.5 K without further transitions at higher temperatures. These findings demonstrate a correlation between manganese concentration and the occurrence of the spin glass state, with higher manganese concentrations requiring larger carrier concentrations to induce the spin glass transition.[344]

$Ga_{1-x}Mn_xS$, a layered III-VI diluted magnetic semiconductor, exhibits spin glass behavior near the transition temperature. In a single-crystalline sample with $x = 0.09$, a distinct cusp separates the FC and ZFC magnetization curves. This cusp is indicative of a spin glass transition, as confirmed by nonlinear magnetization scaling analysis.[345]

The magnetic semiconductor $MnIn_2Se_4$ and its diluted alloy $Zn_{1-x}Mn_xIn_2Se_4$ ($x = 0.87$) were investigated for their magnetic and structural properties. Localized Mn ions in the layered rhombohedral structure undergo a spin-freezing transition below 3.5 K. Irreversible behavior in DC magnetization suggests a spin-glass-like transition at temperatures below 3.5 K for $MnIn_2Se_4$ and 2.9 K for $Zn_{1-x}Mn_xIn_2Se_4$. This conclusion is further supported by FC and ZFC magnetization measurements at low fields (20 Oe). Additionally, a spin glass transition was observed in $Zn_{1-x}Mn_xTe$ ($x = 0.51$), as evidenced by the cusp in the magnetization curve (Figure 34a). The estimated spin glass $T_f$ for this sample is 20.8 K[346].

Mn-doped $V_2VI_3$ semiconductors, known for their potential in room-temperature thermoelectric and thermomagnetic applications[347], also exhibit spin glass behavior. Four representative compounds were studied: $Bi_{2-x}Mn_xTe_3$, $Sb_{2-x}Mn_xTe_3$, $Bi_{2-x}Mn_xSe_3$, and $Sb_{2-x}Mn_xSe_3$. While $Bi_{1.96}Mn_{0.04}Te_3$ and $Sb_{1.97}Mn_{0.03}Te_3$ demonstrate ferromagnetic ordering[348], $Bi_{1.97}Mn_{0.03}Se_3$ exhibits spin glass behavior, as evidenced by ZFC and FC magnetization curves (Figure 34b). In contrast, $Sb_{1.96}Mn_{0.04}Se_3$ displays paramagnetic behavior[349].

lists additional semiconducting spin glasses along with their freezing temperatures for clarity and reference. While semiconducting spin glasses offer a unique platform for exploring the interplay between magnetic and electronic properties, their thermoelectric properties have been relatively less explored. This area presents a promising avenue for future research, with the potential to uncover novel materials and applications.

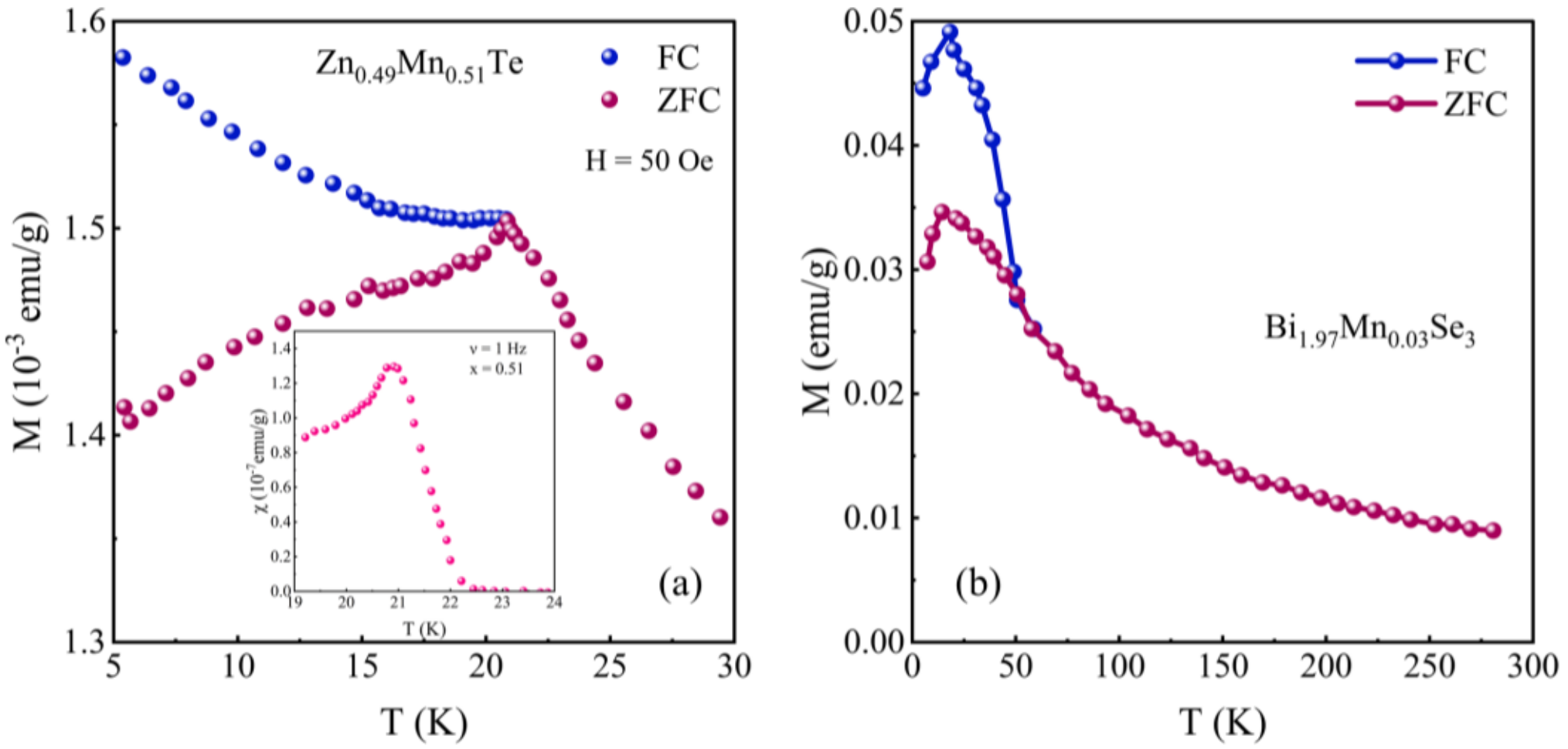

Figure 34. (a) ZFC and FC magnetization curves as a function of temperature for $Zn_{0.49}Mn_{0.51}Te$. The characteristic cusp in the ZFC curve at $T = 20.8 \pm 0.2$ K indicates a spin glass transition. Inset: Temperature dependence of $\chi''$ near the spin glass transition for $Zn_{0.49}Mn_{0.51}Te$. (b) Temperature-dependent ZFC and FC magnetization of a $Bi_{1.97}Mn_{0.03}Se_3$ single crystal exhibiting spin glass behavior. Figures reconstructed based on data extracted from refs. [346] and [349] respectively.

Table 10. DMS compounds that display spin glass behavior together with their main physical and magnetic properties.

| Compounds | $T_f$ (K) | Mechanism / Reason for Glassy Behavior | Ref. |
|---|---|---|---|
| $(Ba,K)(Zn,Mn)_2Sb_2$ | < 10 | Random distribution of Mn and competing exchange interactions | 350 |
| $(Ba,Na)(Zn,Mn)_2As_2$ | 12 | Asperomagnetic order due to carrier-mediated frustrated exchange | 351 |
| $CdCr_{2-2x}In_{2x}Se_4$ | < 120 | Site dilution and frustration in spinel lattice | 352 |
| $Cd_{1-x}Mn_xS$ | x = 0.3 → 2.5<br>x = 0.5→ 18 | Competing AFM/FM exchange in nanosized grains | 353 |
| $Dy_xGe_{1-x}$ ($x \leq 0.02$) | 10 | Dilute Dy moments with indirect RKKY-type interactions | 354 |
| $Zn_{1-x}Fe_xO$ | x = 0.01 → 23<br>x = 0.03 → 33 | Defect-induced magnetism and Fe clustering | 355 |
| $Ga_{1-x}Mn_xS$ | 11.2 | Impurity band formation and frustration of Mn-Mn interactions | 356 |
| $Ge_{1-x-y}(Sn_xMn_y)Te$ | 5.3 | Coexistence of ferromagnetism and competing Mn-Mn interactions | 357 |
| $Ge_{1-x-y}A_xMn_yTe$ (A=Pb, Sn, Eu, Cr,) | 5 & 90 | Magnetic clusters in IV–VI lattice with strong disorder | 358 |
| $Fe_{1-x}Zn_xPS_3$ | - | Magnetic dilution in van der Waals layered system | 359 |
| $Mg_{1-x}Mn_xTe$ | - | Diluted Mn moments and AFM competing exchange | 360 |
| Na(Zn,Mn)Sb | < 15 | Carrier-mediated Mn interactions leading to spin glass state | 361 |
| $Cd_{1-x}Mn_xTe$ | 6.45 | Random Mn distribution and AFM superexchange | 362 |
| $Pb_{1-x-y}Sn_xSc_yTe$ | < 100 | Strong magnetic disorder in IV–VI system | 363 |
| $Sn_{1-x-y}Mn_xCo_yFe_zO_2$ | y = 0 → 150 | High defect concentration and magnetic clustering | 364 |
| $Ga_{1-x}Mn_xS$ | 11.2 | Random Mn distribution in layered lattice | 365 |
| $Zn_{1-x}Mn_xTe$ | 20.8 | Frustrated Mn-Mn interactions | 366 |
| $Hg_{0.7}Mn_{0.3}Te$ | 8.4 | Random Mn distribution + dominant AFM superexchange in fcc lattice | 367 |
| $Bi_{2-x}Mn_xSe_3$ | ~ 30 | Mn clustering and disorder-induced frustrated exchange | 349 |
| $Sn_{1-x}Mn_xTe$ | 2.3 | Competition between AFM superexchange and carrier-mediated (RKKY) interactions | 344 |
| $Zn_{1-x}Mn_xIn_2Se_4$ | x = 1 → < 3.5<br>x = 0.87 → < 2.9 | AFM superexchange via Se and random Mn substitution ⟶ frustrated clusters | 346 |

## 6.4 Room-Temperature Spin Glasses

Spin glasses typically exhibit low freezing temperatures, which presents challenges for many practical applications. The limited availability of room-temperature spin glasses significantly hinders their integration into industrial devices. Therefore, the development of new materials with higher spin glass transition temperatures is a crucial area of research.

Zinc ferrite ($ZnFe_2O_4$) has been extensively studied for its diverse magnetic behaviors, ranging from ferromagnetism to super-paramagnetism and spin glass states. While spin glass behavior has been commonly observed in bulk $ZnFe_2O_4$ samples, thin film samples often exhibit cluster glass characteristics. $ZnFe_2O_4$ films grown on $MgAl_2O_4$ substrates at high deposition rates demonstrate spin-glass-like behavior, which shifts to lower temperatures when deposited on $Al_2O_3$ substrates (as shown in Figure 35 from ZFC and FC measurements).[368]

The magnetic semiconductor $Fe_{3-x}Ti_xO_4$ ($0.8 \leq x \leq 1.0$) exhibits spin glass behavior, as evidenced by the following observations: The ZFC and FC magnetization curves diverge at a characteristic temperature ($T^* \approx 380$ K). A clear cusp appears in the ZFC curve at 240 K when a 500 Oe magnetic field is applied and the $T_f$ shifts from 240 K to 230 K as the magnetic field increases from 500 Oe to 1000 Oe.[369]

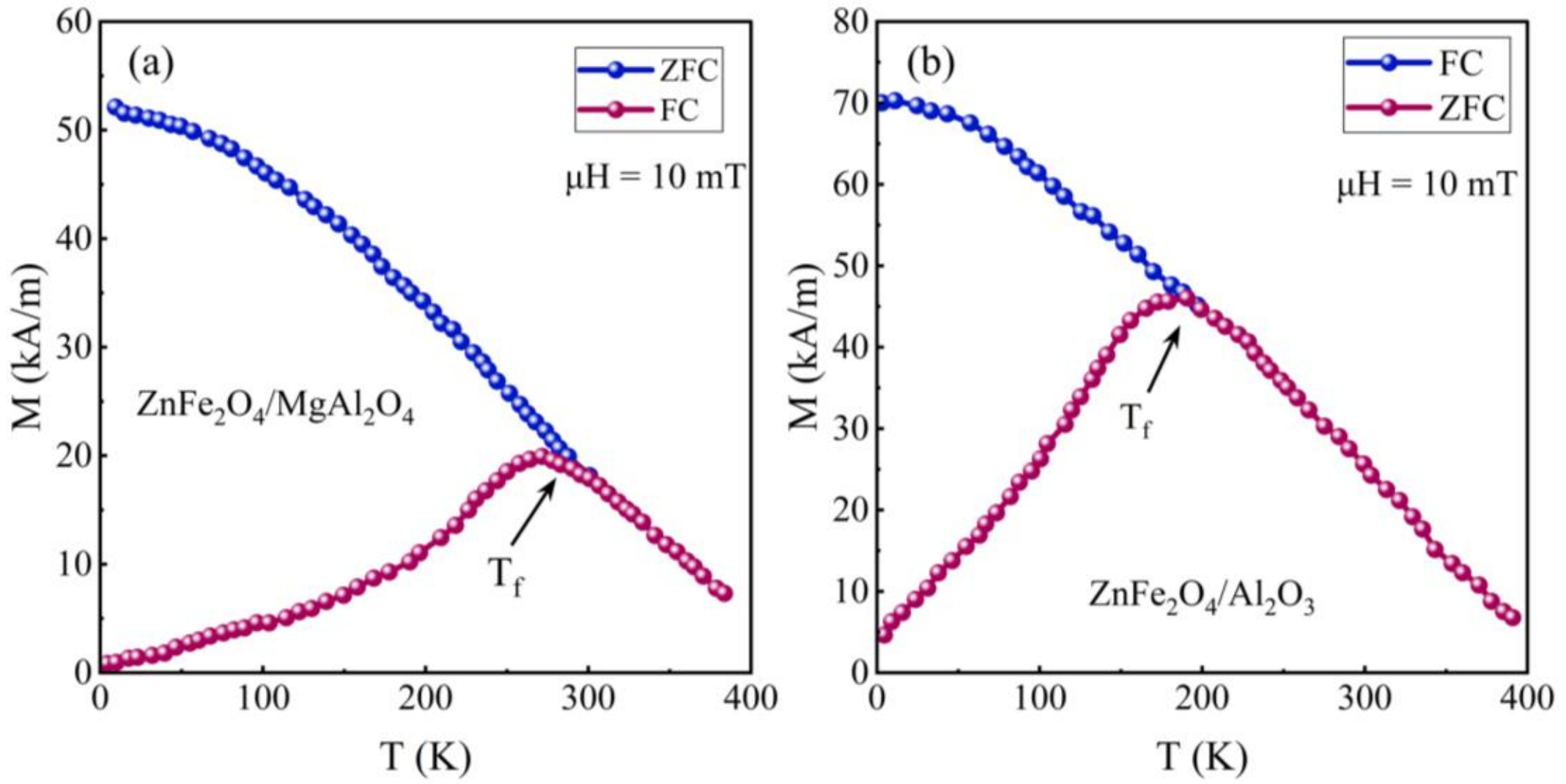

Figure 35. ZFC and FC magnetization of $ZnFe_2O_4$ thin films showing spin-glass-like behavior. Films grown on $MgAl_2O_4$ substrates display typical spin glass features, while deposition on $Al_2O_3$ shifts the freezing temperature to lower values, highlighting the influence of substrate and growth conditions on magnetic behavior. Figures reconstructed based on data extracted from Ref. [368].

Last report is on Fe-Co-Ni-Mn chemically complex alloys (CCAs) with varying Mn concentrations ($14 \leq x \leq 50$). Samples are of the form $(FeCoNi)_{100-x}Mn_x$. This analysis demonstrates the unique ability of Fe-Co-Ni-Mn CCAs to exhibit room-temperature spin glass behavior, making them promising candidates for future applications. Fe, Co and Ni are FM, whereas Mn is AFM. The interaction between these atoms yields to a frustration and formation of spin glass state which occurs in higher temperatures than a bulk spin glass. A clear bifurcation between ZFC and FC magnetization curves indicates spin glass formation. For $Mn_{40}$ and $Mn_{50}$ samples, $T_f$ exceeds room temperature varying with Mn concentration, with $T_f \approx 298$ K for $Mn_{40}$ and $T_f \approx 398$ K for $Mn_{50}$.[370]

Table 11. Some semiconducting spin glasses and their freezing temperature

| Compound | $T_f$ (K) | Refs. |
|---|---|---|
| $ZnFe_2O_4/MgAl_2O_4$ | 290 | 368 |
| $ZnFe_2O_4/Al_2O_3$ | 190 | 368 |
| $(FeCoNi)_{100-x}Mn_x$ | x = 25 → 103-109, x = 40 → 298, x = 50 → 398 | 370 |
| $La_2CoMnO_6$ | 220 | 371 |
| $Fe_{3-x}Ti_xO_4$ | 230 | 369 |

# 7 Conclusion

Spin glasses remain one of the clearest settings where microscopic competition (frustrated exchange, random fields, and quenched disorder) produces macroscopic complexity: an intricate free-energy topography with an extensive number of metastable states, broad distributions of relaxation times, and dynamics that are intrinsically history dependent. Across canonical models and real materials, a consistent physical picture emerges: glassy freezing is not adequately characterized by a single equilibrium observable, but instead by the coupled structure of (i) a low-temperature phase with nontrivial state organization, and (ii) nonequilibrium response governed by barriers that grow with length scale and lead to aging, memory, and rejuvenation. The continuing relevance of spin glasses is that they force a unified treatment of thermodynamics, kinetics, and disorder, precisely the ingredients that also underlie slow dynamics and emergent complexity in a wide range of correlated systems.

A central technical lesson is that "spin-glass order" is best interpreted as a hierarchy of correlated length and time scales rather than a sharp, universally accessible experimental signature. In mean-field settings, replica symmetry breaking and the full order-parameter distribution formalize this hierarchy; in finite-dimensional systems, droplet-like scaling and temperature/field chaos provide an alternative organization in terms of collective excitations and barrier scaling. A productive way forward is to treat these as complementary limiting descriptions and to focus on discriminating, experimentally accessible consequences: how relaxation times scale with temperature and length; whether rejuvenation reflects true temperature chaos or a separation of active length scales; and how weak fields reshape aging protocols, nonlinear susceptibility, and fluctuation–dissipation relations. This perspective naturally emphasizes quantitative comparisons across experiments, simulations, and theory using the same operational definitions of timescales, protocols, and finite-size effects.

On the materials side, the field is shifting from "existence proofs" of glassiness to controlled materials-informed mechanisms. In metallic alloys, insulating oxides, and geometrically frustrated hosts, glassiness can originate from different microscopic routes (random exchange, random anisotropy, site dilution, competing superexchange pathways, charge/orbital degrees of freedom, and disorder-induced random fields). The technical opportunity is to convert this diversity into leverage: by designing disorder in a controlled way (chemical substitution, strain, dimensional reduction, interface engineering, nanoparticle assemblies), one can tune the relative importance of frustration, anisotropy, and randomness and thereby test scaling predictions for freezing temperatures, critical exponents, barrier growth, and noise spectra. At the same time, the community benefits from more rigorous diagnostics for distinguishing spin glasses from nearby phenomenology such as cluster-glass behavior, superparamagnetic blocking, domain-wall pinning, and kinetically arrested first-order transitions. For room-temperature and reentrant systems in particular, high-impact progress will come from establishing a stringent, protocol-consistent evidentiary standard: simultaneous characterization of frequency-dependent susceptibility, aging and memory under controlled waiting times, nonlinear response, and field dependence, ideally accompanied by local probes (μSR/NMR, neutron scattering, coherent x-ray techniques) that can separate intrinsic glassiness from mesoscale clustering.

A parallel technical frontier is the integration of modern computational and inference tools into spin-glass physics without diluting physical interpretability. Machine-learning classifiers can be valuable, but their highest impact will come when they are paired with physically meaningful features (overlap distributions, susceptibilities, correlation lengths, barrier proxies, noise statistics) and when they make falsifiable predictions rather than simply labeling phases. Similarly, algorithmic analogies between spin glasses and optimization should be treated as bidirectional: spin-glass theory can inform which landscapes are intrinsically hard (metastability, clustering, chaotic sensitivity), while optimization tools can provide practical approximations for large-scale disorder realizations and help extract effective parameters from experimental data. A concrete goal for the next phase of the field is to standardize "analysis pipelines" that map experimental time series and frequency response into model-comparable objects: distributions of relaxation times, effective barrier scaling, correlation-length growth laws, and deviations from equilibrium fluctuation–dissipation behavior under well-defined protocols.

Several open problems stand out as both technically sharp and broadly consequential:

1. Which features of glassy freezing are genuinely universal across disorder realizations and interaction ranges, and which are controlled by microscopic details such as anisotropy, itinerancy, and coupling to lattice, orbital, or charge degrees of freedom? Progress likely requires systematic cross-material comparisons using identical protocols and carefully quantified disorder.
2. Aging and memory are commonly interpreted through phenomenological length-scale growth. Directly measuring or tightly constraining the evolving correlation length (and its dependence on temperature shifts and fields) remains a key challenge. Combining time-domain protocols with scattering and local probes is an important path forward.
3. The interplay between weak fields, aging, and putative phase boundaries (e.g., AT-like lines) remains unsettled in many experimentally relevant regimes. High-quality nonlinear susceptibility and noise measurements, analyzed with finite-size/time scaling, are critical for clarifying which aspects are equilibrium-like and which are purely dynamical crossovers.
4. Rather than treating cluster-glass behavior as a confounding artifact, it can be used as a tunable intermediate regime that tests how glassiness emerges as clusters couple and as disorder changes character (random exchange vs random anisotropy vs random fields). Developing quantitative criteria for this crossover would help unify disparate materials reports.
5. Quantum spin glasses, transverse-field realizations, and programmable Ising platforms offer controlled settings where dissipation, entanglement, and measurement back-action may reshape glassy phenomenology. Connecting classical aging concepts to quantum relaxation, decoherence, and non-thermal steady states is an emerging opportunity with high conceptual payoff.
6. Low-frequency noise (1/f-like spectra, intermittency, telegraph signals) provides a direct window into barrier-dominated dynamics and rare events. Establishing robust links between noise statistics and microscopic models could make noise a standard "bridge observable" between theory, simulation, and experiment.

Taken together, these directions point to a mature and still-expanding field. The highest-impact spin-glass research over the next few years is likely to be the work that (i) uses controlled materials design to tune the microscopic origin of disorder and frustration, (ii) employs protocol-consistent, multi-modal diagnostics to separate true glassiness from blocking and clustering, and (iii) translates nonequilibrium measurements into model-comparable quantities with clear scaling meaning. In that role, spin glasses continue to function as a unifying framework: they connect statistical mechanics, materials physics, and computation through a common language of landscapes, metastability, and slow dynamics, while continually testing the limits of how we define phases and transitions in complex, disordered matter.

# 8 Acknowledgement

This study was supported partially by NSF grant CBET-2110603.

# 9 References

[1] "Scientific Background for the Nobel Prize in Physics 2021" (PDF). Nobel Committee for Physics. 5 October 2021. Retrieved 22 February 2025.

[2] Mézard, Marc, Giorgio Parisi, and Miguel Angel Virasoro. Spin glass theory and beyond: An Introduction to the Replica Method and Its Applications. Vol. 9. World Scientific Publishing Company, 1987.

[3] Krempaský, Juraj, Gunther Springholz, Sunil Wilfred D'souza, Ondřej Caha, Martin Gmitra, Andreas Ney, C. A. F. Vaz et al. "Efficient magnetic switching in a correlated spin glass." *Nature Communications* 14, no. 1 (2023): 6127.

[4] Chen, S., Zhong, Y., Cai, J., Zhang, Z., Gao, F., Huo, S., ... & Jiang, J. (2024). High thermoelectric performance of GeTe-MnTe alloy driven by spin degree of freedom. *Materials Today Physics*, *43*, 101393.

[5] Charbonneau, Patrick, Enzo Marinari, Giorgio Parisi, Federico Ricci-tersenghi, Gabriele Sicuro, Francesco Zamponi, and Marc Mezard, eds. Spin glass theory and far beyond: replica symmetry breaking after 40 years. World Scientific, 2023.

[6] Gardner, Elizabeth, and Bernard Derrida. "Optimal storage properties of neural network models." *Journal of Physics A: Mathematical and general* 21, no. 1 (1988): 271.

[7] Kenning, Gregory G., Maxine Brandt, Ryan Brake, Morgan Hepler, and Daniel Tennant. "Observation of critical scaling in spin glasses below $T_c$ using thermoremanent magnetization." *Frontiers in Physics* 12 (2024): 1443298.

[8] Joy, Lija K., S. Shanmukharao Samatham, Senoy Thomas, V. Ganesan, Salim Al-Harthi, A. Liebig, Manfred Albrecht, and M. R. Anantharaman. "Colossal thermoelectric power in charge ordered lanthanum calcium manganites ($La_{0.5}Ca_{0.5}MnO_3$)." *Journal of Applied Physics* 116, no. 21 (2014).

[9] Sagar, S., V. Ganesan, P. A. Joy, Senoy Thomas, A. Liebig, Manfred Albrecht, and M. R. Anantharaman. "Colossal thermoelectric power in Gd-Sr manganites." *Europhysics Letters* 91, no. 1 (2010): 17008.

[10] Yu, Jihao, Weiwei Wu, Huaping Zhang, Ruiwen Shao, Fan Zhang, Hong Wang, Zian Li et al. "Robust spin glass state with exceptional thermal stability in a chemically complex alloy." Physical Review Materials 6, no. 9 (2022): L091401.

[11] Katzgraber, H. G., Hamze, F., Zhu, Z., Ochoa, A. J., & Munoz-Bauza, H. (2015). Seeking quantum speedup through spin glasses: The good, the bad, and the ugly. *Physical Review X*, *5*(3), 031026.

[12] King, Andrew D., Sei Suzuki, Jack Raymond, Alex Zucca, Trevor Lanting, Fabio Altomare, Andrew J. Berkley et al. "Coherent quantum annealing in a programmable 2,000 qubit Ising chain." *Nature Physics* 18, no. 11 (2022): 1324-1328.

[13] King, Andrew D., Jack Raymond, Trevor Lanting, Richard Harris, Alex Zucca, Fabio Altomare, Andrew J. Berkley et al. "Quantum critical dynamics in a 5,000-qubit programmable spin glass." *Nature* 617, no. 7959 (2023): 61-66.

[14] Das, A., & Chakrabarti, B. K. (2008). Colloquium: Quantum annealing and analog quantum computation. Reviews of Modern Physics, 80(3), 1061-1081.

[15] Brodoloni, L., and S. Pilati. "Zero-temperature Monte Carlo simulations of two-dimensional quantum spin glasses guided by neural network states." *Physical Review E* 110, no. 6 (2024): 065305.

[16] Placke, Benedikt, Tibor Rakovszky, Nikolas P. Breuckmann, and Vedika Khemani. "Topological Quantum Spin Glass Order and its realization in qLDPC codes." *arXiv preprint arXiv:2412.13248* (2024).

[17] Parisi, Giorgio. "Spin glasses and fragile glasses: Statics, dynamics, and complexity." *Proceedings of the National Academy of Sciences* 103, no. 21 (2006): 7948-7955.

[18] Alcantara, A. R., S. Barrett, D. Matev, I. Miotkowski, A. K. Ramdas, T. M. Pekarek, and J. T. Haraldsen. "Enhancement of the spin-glass transition temperature through pd-orbital hybridization in $Zn_{1-x}Mn_xTe$." *Physical Review B* 104, no. 10 (2021): 104423.

[19] F. Izadi, and R. Sepehrinia, “Magnetic proximity in a coupled ferromagnet–spin glass system”, Phys. Rev. B **107**, 094207 (2023)

[20] Ghio, Davide, Yatin Dandi, Florent Krzakala, and Lenka Zdeborová. "Sampling with flows, diffusion, and autoregressive neural networks from a spin-glass perspective." *Proceedings of the National Academy of Sciences* 121, no. 27 (2024): e2311810121.

[21] Sappington, Anna, and Vaibhav Mohanty. "Probabilistic genotype-phenotype maps reveal mutational robustness of RNA folding, spin glasses, and quantum circuits." *Physical Review Research* 7, no. 1 (2025): 013118.

[22] Mydosh, J. A. (1983). Spin-glasses. *Europhysics News*, *14*(12), 2-4.

[23] Mydosh, J. A. (1993). Spin glasses: an experimental introduction. CRC Press.

[24] Edwards, S. F., and Anderson, P. W. (1975). Theory of spin glasses. Journal of Physics F: Metal Physics, 5(5), 965.

[25] Sherrington, D., and Kirkpatrick, S. (1975). Solvable model of a spin-glass. Physical review letters, 35(26), 1792.

[26] Parisi, G. (1979). Infinite number of order parameters for spin glasses. Physical Review Letters, 43(23), 1754.

[27] Parisi, G. (1979). Toward a mean field theory for spin glasses. *Physics Letters A*, *73*(3), 203-205.

[28] Sherrington, D. (2007). Spin glasses: a perspective. In Spin glasses (pp. 45-62). Berlin, Heidelberg: Springer Berlin Heidelberg.

[29] Bolthausen, E., & Bovier, A. (Eds.). (2007). *Spin glasses*. Springer.

[30] Fischer, K. H., & Hertz, J. A. (1993). *Spin glasses* (No. 1). Cambridge university press.

[31] Nordblad, P. (2013). Competing interaction in magnets: the root of ordered disorder or only frustration?. *Physica Scripta*, *88*(5), 058301.

[32] Gyorffy, B. L., Pindor, A. J., Staunton, J., Stocks, G. M., & Winter, H. (1985). A first-principles theory of ferromagnetic phase transitions in metals. *Journal of Physics F: Metal Physics*, *15*(6), 1337.

[33] Feigel'man, M. V., & Tsvelik, A. M. (1979). Phase transition in a spin glass. *Zh. Eksp. Teor. Fiz.*, *77*(6), 2524-2538.

[34] Gunnarsson, K., Svedlindh, P., Andersson, J. O., Nordblad, P., Lundgren, L., Katori, H. A., & Ito, A. (1992). Magnetic behavior of a reentrant Ising spin glass. *Physical Review B*, *46*(13), 8227.

[35] Guchhait, S. (2022). Mesoscale Spin Glass Dynamics. Spectroscopy and Characterization of Nanomaterials and Novel Materials: Experiments, Modeling, Simulations, and Applications, 55-66.

[36] Hahlen, T. (2007). From quantum glass to addressable spin clusters in the model magnet $LiHo_xY_{1-x}F_4$.

[37] Aveline, I. M. S., & Holz, A. (1973). Ruderman-Kittel-Kasuya-Yosida (RKKY) interaction in alloys of $Gd_{1-x}Mg_x$ and $Gd_{1-x}Th_x$. *Cienc. Cult.(Sao Paulo)*, *25*(6), 46.

[38] Singh, K., Maignan, A., Simon, C., Hardy, V., Pachoud, E., & Martin, C. (2011). The spin glass delafossite $CuFe_{0.5}V_{0.5}O_2$: A dipolar glass?. *Journal of Physics: Condensed Matter*, *23*(12), 126005.

[39] Piatek, J. O., Dalla Piazza, B., Nikseresht, N., Tsyrulin, N., Živković, I., Krämer, K. W., ... & Rønnow, H. M. (2013). Phase diagram with an enhanced spin-glass region of the mixed Ising–X Y magnet $LiHo_xEr_{1-x}F_4$. *Physical Review B*, *88*(1), 014408.

[40] Yuan, F., Du, J., & Shen, B. (2012). Controllable spin-glass behavior and large magnetocaloric effect in Gd-Ni-Al bulk metallic glasses. *Applied Physics Letters*, *101*(3).

[41] Gatteschi, D., & Sessoli, R. (2004). Molecular nanomagnets: the first 10 years. *Journal of magnetism and magnetic materials*, *272*, 1030-1036.

[42] Mo, J., Chen, H., Chen, R., Jin, M., Liu, M., & Xia, Y. (2023). Nature of Spin-Glass Behavior of Cobalt-Doped Iron Disulfide Nanospheres Using the Monte Carlo Method. *The Journal of Physical Chemistry C*, *127*(3), 1475-1486.

[43] Gao, C., Tan, S., Wang, C., Sun, Y., Cao, R., Han, K., ... & Xing, F. (2020). Observation of spin-glass like be-havior in the layered oxyselenides $La_2O_3(Mn_{1-x}Co_x)_2Se_2$. RSC advances, 10(24), 14033-14039.

[44] Morgownik, A. F. J., & Mydosh, J. A. (1983). Analysis of the high-temperature spin-glass susceptibility: Determination of the local magnetic exchange. *Solid state communications*, *47*(5), 321-324.

[45] Reimers, J. N., Greedan, J. E., & Sato, M. (1988). The crystal structure of the spin-glass pyrochlore, $Y_2Mo_2O_7$. *Journal of Solid State Chemistry*, *72*(2), 390-394.

[46] Poddar, A., Bhowmik, R. N., Muthuselvam, I. P., & Das, N. (2009). Evidence of disorder induced magnetic spin glass phase in $Sr_2FeMoO_6$ double perovskite. *Journal of Applied Physics*, *106*(7).

[47] Furdyna, J. K. (1988). Diluted magnetic semiconductors. *Journal of Applied Physics*, *64*(4), R29-R64.

[48] Sherrington, D. (2014). A spin glass perspective on ferroic glasses. *physica status solidi (b)*, *251*(10), 1967-1981.

[49] da Silva Jr, A. F., Martins, A. S., & de Campos, M. F. (2019). Spin glass transition in AuFe, CuMn, AuMn, AgMn and AuCr systems. *Journal of Magnetism and Magnetic Materials*, *479*, 222-228.

[50] F. Izadi, and R. Sepehrinia, “Griffiths singularity in quasi-one-dimensional restricted $\pm J$ Ising spin glass”, Phys. Rev. E 108, 064117 (2023)

[51] Shi, Y., Nisoli, C., & Chern, G. W. (2021). Ice, glass, and solid phases in artificial spin systems with quenched disorder. *Applied Physics Letters*, *118*(12), 122407.

[52] Bramwell, S. T., & Gingras, M. J. (2001). Spin ice state in frustrated magnetic pyrochlore materials. *Science*, *294*(5546), 1495-1501.
[53] Snyder, J., Ueland, B. G., Slusky, J. S., Karunadasa, H., Cava, R. J., & Schiffer, P. (2004). Low-temperature spin freezing in the $Dy_2Ti_2O_7$ spin ice. *Physical Review B*, *69*(6), 064414.
[54] Goremychkin, E. A., Osborn, R., Rainford, B. D., Macaluso, R. T., Adroja, D. T., & Koza, M. (2008). Spin-glass order induced by dynamic frustration. *Nature Physics*, *4*(10), 766-770.
[55] Süllow, S., Nieuwenhuys, G. J., Menovsky, A. A., Mydosh, J. A., Mentink, S. A. M., Mason, T. E., & Buyers, W. J. L. (1997). Spin glass behavior in $URh_2Ge_2$. *Physical review letters*, *78*(2), 354.
[56] Anderson, P. W. (1978). The concept of frustration in spin glasses. *Journal of the Less Common Metals*, *62*, 291-294.
[57] Chandra, P., Coleman, P., & Ritchey, I. (1993). The anisotropic kagome antiferromagnet: a topological spin glass? *Journal de Physique I*, *3*(2), 591-610.
[58] Dutta, K., & Talukdar, D. (2022). Varying critical behaviour of stacked triangular lattice Ising antiferromagnets in the presence of spin–lattice coupling. *Journal of Magnetism and Magnetic Materials*, *556*, 169344.
[59] Dorogovtsev, S. N., Goltsev, A. V., & Mendes, J. F. (2008). Critical phenomena in complex networks. *Reviews of Modern Physics*, *80*(4), 1275-1335.
[60] Silveira, Alexandre, R. Erichsen Jr, and S. G. Magalhaes. "Geometrical frustration and cluster spin glass with random graphs." *Physical Review E* 103, no. 5 (2021): 052110.
[61] Okunev, V. D., and H. Szymczak. "Two states of magnetic frustration in $La_{0.7}Sr_{0.3}MnO_3$ amorphous films with fractal structures: New knowledge about clusters and cluster ensembles." *Journal of Applied Physics* 133, no. 8 (2023).
[62] Schmidt, M., Zimmer, F. M., & Magalhaes, S. G. (2015). Weak randomness in geometrically frustrated systems: spin-glasses. *Physica Scripta*, *90*(2), 025809.
[63] Nayak, S., Manna, P. K., Singh, B. B., & Bedanta, S. (2021). Effect of spin glass frustration on exchange bias in NiMn/CoFeB bilayers. *Physical Chemistry Chemical Physics*, *23*(11), 6481-6489.
[64] Ali, M., Adie, P., Marrows, C. H., Greig, D., Hickey, B. J., & Stamps, R. L. (2007). Exchange bias using a spin glass. *Nature Materials*, *6*(1), 70-75.
[65] Yuan, F. T., Lin, J. K., Yao, Y. D., & Lee, S. F. (2010). Exchange bias in spin glass (FeAu)/NiFe thin films. *Applied Physics Letters*, *96*(16).
[66] Liu, Pei, Bing Lv, Yongzuo Wang, Peng Chen, Lei Wang, and Cunxu Gao. "Respective dependencies of conventional and spontaneous exchange biases on spin glass." *Physical Review B* 109, no. 18 (2024): 184412.
[67] Meirzadeh, Elena, Sae Young Han, and Xavier Roy. "Exchange Bias from Frustrated Spins." (2021): 1295-1297.
[68] Mitsumoto, Kota, Chisa Hotta, and Hajime Yoshino. "Spin-orbital glass transition in a model of a frustrated pyrochlore magnet without quenched disorder." *Physical Review Letters* 124, no. 8 (2020): 087201.
[69] Rao, C. N. R. "Charge, Spin, and Orbital Ordering in the Perovskite Manganates, $Ln_{1-x}A_xMnO_3$ (Ln= Rare Earth, A= Ca or Sr)." *The Journal of Physical Chemistry B* 104, no. 25 (2000): 5877-5889.
[70] Goodenough, J. B. (1955). Theory of the role of covalence in the perovskite-type manganites [La, M (II)] Mn O 3. *Physical Review*, *100*(2), 564.
[71] Montenegro, F. C., U. A. Leitao, M. D. Coutinho-Filho, and S. M. Rezende. "Crossover from random field to spin-glass behavior in $Fe_xZn_{1-x}F_2$." *Journal of applied physics* 67, no. 9 (1990): 5243-5245.
[72] de Araújo, J. H., J. B. M. da Cunha, A. Vasquez, L. Amaral, J. T. Moro, F. C. Montenegro, S. M. Rezende, and Mauricio D. Coutinho Filho. "Mössbauer study of spin-glass $Fe_xZn_{1-x}F_2$ system." *Hyperfine Interactions* 67 (1991): 507-511.
[73] Fisher, Daniel S., Geoffrey M. Grinstein, and Anil Khurana. "Theory of random magnets." *Physics Today* 41, no. 12 (1988): 56-67.
[74] Matsuno, K., T. Katsufuji, S. Mori, Minoru Nohara, A. Machida, Y. Moritomo, K. Kato et al. "Charge Ordering and Spin Frustration in $AlV_{2-x}Cr_xO_4$." *Physical review letters* 90, no. 9 (2003): 096404.
[75] Sun, Fei, Rui Wang, Cynthia Aku-Leh, H. X. Yang, Rui He, and Jimin Zhao. "Double charge ordering states and spin ordering state observed in a $RFe_2O_4$ system." *Scientific Reports* 4, no. 1 (2014): 6429.
[76] Paul, Arpita, and Turan Birol. "Cation order control of correlations in double perovskite $Sr_2VNbO_6$." *Physical Review Research* 2, no. 3 (2020): 033156.
[77] Lindquist, Kurt P., Teresa Lee, Xianghan Xu, and Robert J. Cava. "Expanding the Family of Magnetic Vacancy-Ordered Halide Double Perovskites." *Chemistry of Materials* 36, no. 15 (2024): 7610-7618.
[78] Levy, Amir, Michael McEldrew, and Martin Z. Bazant. "Spin-glass charge ordering in ionic liquids." *Physical Review Materials* 3, no. 5 (2019): 055606.
[79] Binder, K., & Young, A. P. (1986). Spin glasses: Experimental facts, theoretical concepts, and open questions. *Reviews of Modern physics*, *58*(4), 801.

[80] Reiner, M. (1964). The deborah number. *Physics today*, *17*(1), 62-62.
[81] Dormann, J. L., Fiorani, D., & Tronc, E. (1997). Magnetic relaxation in fine-particle systems. *Advances in chemical physics*, *98*, 283-494.
[82] Fisher, D. S., & Huse, D. A. (1988). Nonequilibrium dynamics of spin glasses. *Physical Review B*, *38*(1), 373.
[83] Vincent, E., Hammann, J., Ocio, M., Bouchaud, J. P., & Cugliandolo, L. F. (2007, April). Slow dynamics and aging in spin glasses. In Complex Behaviour of Glassy Systems: Pro-ceedings of the XIV Sitges Conference Sitges, Barcelona, Spain, 10–14 June 1996 (pp. 184-219). Berlin, Heidelberg: Springer Berlin Heidel-berg.
[84] Taniguchi, T., Yamazaki, T., Yamanaka, K., Tabata, Y., & Kawarazaki, S. (2007). Critical phenomena in canonical spin glass AuMn from Hall effect measurements. Journal of magnetism and magnetic materials, 310(2), 1526-1528.
[85] Semeno, A. V., M. A. Anisimov, A. V. Bogach, S. V. Demishev, M. I. Gilmanov, V. B. Filipov, N. Yu Shitsevalova, and V. V. Glushkov. "Role of spin-glass behavior in the formation of exotic magnetic states in $GdB_6$." *Scientific Reports* 10, no. 1 (2020): 18214.
[86] Brodale, G. E., R. A. Fisher, W. E. Fogle, N. E. Phillips, and J. Van Curen. "The effect of spin-glass ordering on the specific heat of CuMn." *Journal of magnetism and magnetic materials* 31 (1983): 1331-1333.
[87] Thota, S., V. Narang, S. Nayak, S. Sambasivam, B. C. Choi, Tapati Sarkar, Mikael Svante Andersson, Roland Mathieu, and M. S. Seehra. "On the nature of magnetic state in the spinel $Co_2SnO_4$." *Journal of Physics: Condensed Matter* 27, no. 16 (2015): 166001.
[88] Fogle, W. E., Boyer, J. D., Phillips, N. E., & Van Curen, J. (1981). Calorimetric investigation of spin-glass ordering in CuMn. *Physical Review Letters*, *47*(5), 352.
[89] Heringer, M. A. V., Mariano, D. L., Freitas, D. C., Baggio-Saitovitch, E., Continentino, M. A., & Sanchez, D. R. (2020). Spin-glass behavior in $Co_3Mn_3(O_2BO_3)_2$ ludwigite with weak disorder. *Physical Review Materials*, *4*(6), 064412.
[90] Andriushchenko, P., Kapitan, D., & Kapitan, V. (2022). A new look at the spin glass problem from a deep learning perspective. *Entropy*, *24*(5), 697.
[91] Fenner, L. A., Wills, A. S., Bramwell, S. T., Dahlberg, M., & Schiffer, P. (2009). Zero-point entropy of the spinel spin glasses $CuGa_2O_4$ and $CuAl_2O_4$. In *Journal of Physics: Conference Series* (Vol. 145, No. 1, p. 012029). IOP Publishing.
[92] Jazandari, M., Abouie, J., & Vashaee, D. (2025). What Really Drives Thermopower: Specific Heat or Entropy as the Unifying Principle Across Magnetic, Superconducting, and Nanoscale Systems. *arXiv preprint arXiv:2506.06745*.
[93] Langer, S. A., Sethna, J. P., & Grannan, E. R. (1990). Nonequilibrium entropy and entropy distributions. *Physical Review B*, *41*(4), 2261.
[94] Piyakulworawat, C., Thennakoon, A., Yang, J., Yoshizawa, H., Ueta, D., Sato, T. J., ... & Lee, S. H. (2024). Zero-point entropies of spin-jam and spin-glass states in a frustrated magnet. *Physical Review B*, *109*(10), 104420.
[95] Gopal, E. (2012). *Specific heats at low temperatures*. Springer Science & Business Media.
[96] Jäckle, J. (1984). Residual entropy in glasses and spin glasses. *Physica B+ C*, *127*(1-3), 79-86.
[97] Mugiraneza, Sam, and Alannah M. Hallas. "Tutorial: a beginner's guide to interpreting magnetic susceptibility data with the Curie-Weiss law." *Communications Physics* 5, no. 1 (2022): 95.
[98] Joy, P. A., & Date, S. K. (2000). Comparison of the zero-field-cooled magnetization behavior of some ferromagnetic and ferrimagnetic systems. *Journal of magnetism and magnetic materials*, *218*(2-3), 229-237.
[99] Pramanik, P., Ghosh, S., Yanda, P., Joshi, D. C., Pittala, S., Sundaresan, A., ... & Seehra, M. S. (2019). Magnetic ground state, field-induced transitions, electronic structure, and optical band gap of the frustrated antiferromagnet GeCo 2 O 4. *Physical Review B*, *99*(13), 134422.
[100] Vincent, E. (2022). Spin glass experiments. *arXiv preprint arXiv:2208.00981*.
[101] Mathieu, R., Jönsson, P., Nam, D. N. H., & Nordblad, P. (2001). Memory and superposition in a spin glass. *Physical Review B*, *63*(9), 092401.
[102] Scalliet, C., & Berthier, L. (2019). Rejuvenation and memory effects in a structural glass. *Physical review letters*, *122*(25), 255502.
[103] Nordblad, P., Svedlindh, P., Lundgren, L., & Sandlund, L. (1986). Time decay of the remanent magnetization in a CuMn spin glass. *Physical Review B*, *33*(1), 645.
[104] Pal, S., Kumar, K., Banerjee, A., Roy, S. B., & Nigam, A. K. (2020). Field-cooled state of the canonical spin glass revisited. *Physical Review B*, *101*(18), 180402.
[105] Pal, S., Roy, S. B., & Nigam, A. K. (2022). Nonequilibrium magnetic response in concentrated spin-glass AuFe (11%) alloy. *arXiv preprint arXiv:2206.00286*.
[106] Baity-Jesi, M., Calore, E., Cruz, A., Fernandez, L. A., Gil-Narvion, J. M., Gonzalez-Adalid Pemartin, I., ... & Yllanes, D. (2023). Memory and rejuvenation effects in spin glasses are governed by more than one length scale. *Nature Physics*, 1-8.

[107] Bouchaud, J. P., Cugliandolo, L. F., Kurchan, J., & Mézard, M. (1998). Out of equilibrium dynamics in spin-glasses and other glassy systems. *Spin glasses and random fields*, *12*, 161.

[108] Vincent, E. (2007). Ageing, rejuvenation and memory: The example of spin-glasses. In *Ageing and the glass transition* (pp. 7-60). Berlin, Heidelberg: Springer Berlin Heidelberg.

[109] Jonason, K., Vincent, E., Hammann, J., Bouchaud, J. P., & Nordblad, P. (1998). Memory and chaos effects in spin glasses. *Physical Review Letters*, *81*(15), 3243.

[110] Hissariya, R., Sathe, V. G., & Mishra, S. K. (2023). Antisite disorder mediated exchange bias effect and spin-glass state in $La_{2-x}Sm_xNiMnO_6$. *Journal of Magnetism and Magnetic Materials*, *578*, 170769.

[111] Mulder, C. A. M., A. J. Van Duyneveldt, and J. A. Mydosh. "Susceptibility of the Cu Mn spin-glass: Frequency and field dependences." *Physical Review B* 23, no. 3 (1981): 1384.

[112] Martien, Dinesh. *Introduction to ac susceptibility. Quantum Design*. 2000. https://qdusa.com/siteDocs/appNotes/1078-201.pdf (retriever February 23, 2025)

[113] Chakrabarty, Tanmoy, Avinash V. Mahajan, and Susanta Kundu. "Cluster spin glass behavior in geometrically frustrated $Zn_3V_3O_8$." *Journal of Physics: Condensed Matter* 26, no. 40 (2014): 405601.

[114] Palmer, R. G. (1982). Broken ergodicity. *Advances in Physics*, *31*(6), 669-735.

[115] Vincent, E., Hammann, J., Ocio, M., Bouchaud, J. P., & Cugliandolo, L. F. (2007, April). Slow dynamics and aging in spin glasses. In *Complex Behaviour of Glassy Systems: Proceedings of the XIV Sitges Conference Sitges, Barcelona, Spain, 10–14 June 1996* (pp. 184-219). Berlin, Heidelberg: Springer Berlin Heidelberg.

[116] Vincent, E. (2007). Ageing, rejuvenation and memory: The example of spin-glasses. In *Ageing and the glass transition* (pp. 7-60). Berlin, Heidelberg: Springer Berlin Heidelberg.

[117] Janke, W. (2008). Rugged free-energy landscapes–an introduction. In *Rugged Free Energy Landscapes: Common Computational Approaches to Spin Glasses, Structural Glasses and Biological Macromolecules* (pp. 1-7). Berlin, Heidelberg: Springer Berlin Heidelberg.

[118] Fernández, J. F., & Alonso, J. J. (2013). Numerical results for the Edwards-Anderson spin-glass model at low temperature. *Physical Review B—Condensed Matter and Materials Physics*, *87*(13), 134205.

[119] Boettcher, S. (2005). Stiffness of the Edwards-Anderson model in all dimensions. *Physical review letters*, *95*(19), 197205.

[120] Baños, R. A., Cruz, A., Fernandez, L. A., Gil-Narvion, J. M., Gordillo-Guerrero, A., Guidetti, M., ... & Yllanes, D. (2012). Thermodynamic glass transition in a spin glass without time-reversal symmetry. *Proceedings of the National Academy of Sciences*, *109*(17), 6452-6456.

[121] Picco, M., Ricci-Tersenghi, F., & Ritort, F. (2000). Chaotic, memory and cooling rate effects in spin glasses: Is the Edwards-Anderson model a good spin glass?. *arXiv preprint cond-mat/0005541*.

[122] Altieri, A., & Baity-Jesi, M. (2023). An introduction to the theory of spin glasses. *arXiv preprint arXiv:2302.04842*.

[123] Fröhlich, J., & Zegarlinski, B. (1987). Some comments on the Sherrington-Kirkpatrick model of spin glasses. *Communications in mathematical physics*, *112*(4), 553-566.

[124] Bates, E., Sloman, L., & Sohn, Y. (2019). Replica symmetry breaking in multi-species Sherrington–Kirkpatrick model. *Journal of Statistical Physics*, *174*(2), 333-350.

[125] Guerra, F. (2003). Broken replica symmetry bounds in the mean field spin glass model. *Communications in mathematical physics*, *233*(1), 1-12.

[126] Thouless, D. J., Anderson, P. W., & Palmer, R. G. (1977). Solution of'solvable model of a spin glass'. *Philosophical Magazine*, *35*(3), 593-601.

[127] Bravi, B., Sollich, P., & Opper, M. (2016). Extended Plefka expansion for stochastic dynamics. *Journal of Physics A: Mathematical and Theoretical*, *49*(19), 194003.

[128] Chen, W. K., & Panchenko, D. (2018). On the TAP free energy in the mixed p-spin models. *Communications in Mathematical Physics*, *362*(1), 219-252.

[129] Aspelmeier, T., & Moore, M. A. (2019). Realizable solutions of the Thouless-Anderson-Palmer equations. *Physical Review E*, *100*(3), 032127.

[130] Talagrand, M. (2003). *Spin glasses: a challenge for mathematicians: cavity and mean field models* (Vol. 46). Springer Science & Business Media

[131] Mézard, M., & Parisi, G. (2001). The Bethe lattice spin glass revisited. *The European Physical Journal B-Condensed Matter and Complex Systems*, *20*(2), 217-233.

[132] Mézard, M., Parisi, G., & Virasoro, M. A. (1987). *Spin glass theory and beyond: An Introduction to the Replica Method and Its Applications* (Vol. 9). World Scientific Publishing Company.

[133] Braunstein, A., Mézard, M., & Zecchina, R. (2005). Survey propagation: An algorithm for satisfiability. *Random Structures & Algorithms*, *27*(2), 201-226.

[134] Bray, A. J., & Moore, M. A. (1987). Chaotic nature of the spin-glass phase. *Physical review letters*, *58*(1), 57.

[135] Newman, C. M., & Stein, D. L. (2004). Short-range spin glasses: The metastate approach. *arXiv preprint cond-mat/0411426*.
[136] Hukushima, K., & Nemoto, K. (1996). Exchange Monte Carlo method and application to spin glass simulations. *Journal of the Physical Society of Japan*, *65*(6), 1604-1608.
[137] Rozada, I., Aramon, M., Machta, J., & Katzgraber, H. G. (2019). Effects of setting temperatures in the parallel tempering Monte Carlo algorithm. *Physical Review E*, *100*(4), 043311.
[138] Moore, M. A. (2021). Droplet-scaling versus replica symmetry breaking debate in spin glasses revisited. *Physical Review E*, *103*(6), 062111.
[139] Vedula, B., Moore, M. A., & Sharma, A. (2025). Nature of spin glass order in physical dimensions. *Physical Review E*, *111*(3), 034102.
[140] Fisher, D. S., & Huse, D. A. (1988). Equilibrium behavior of the spin-glass ordered phase. *Physical Review B*, *38*(1), 386.
[141] Palassini, M., & Young, A. P. (2000). Nature of the spin glass state. *Physical Review Letters*, *85*(14), 3017.
[142] Derrida, B., & Mottishaw, P. (2024). Generalizations of Parisi's replica symmetry breaking and overlaps in random energy models. *Comptes Rendus. Physique*, *25*(G1), 329-351.
[143] Krzakala, F., & Martin, O. C. (2000). Spin and link overlaps in three-dimensional spin glasses. *Physical Review Letters*, *85*(14), 3013.
[144] Krzakała, F., & Parisi, G. (2004). Local excitations in mean-field spin glasses. *Europhysics Letters*, *66*(5), 729.
[145] Kondor, I., & Papp, G. (2024). Energy landscapes of small SK spin glasses. *arXiv preprint arXiv:2412.13580*.
[146] Lamarcq, J., Bouchaud, J. P., Martin, O. C., & Mézard, M. (2002). Non-compact local excitations in spin-glasses. *Europhysics Letters*, *58*(3), 321.
[147] Continentino, M., & Malozemoff, A. P. (1986). Dynamical susceptibility of spin glasses in the fractal cluster model. *Physical Review B*, *34*(1), 471.
[148] Lundgren, L., Nordblad, P., & Svedlindh, P. (1986). Relaxation behavior of fractal-cluster spin glasses. *Physical Review B*, *34*(11), 8164.
[149] Hukushima, K. (1999). Domain-wall free energy of spin-glass models: Numerical method and boundary conditions. *Physical Review E*, *60*(4), 3606.
[150] Amoruso, C., Marinari, E., Martin, O. C., & Pagnani, A. (2003). Scalings of domain wall energies in two dimensional Ising spin glasses. *Physical review letters*, *91*(8), 087201.
[151] Melchert, O., & Hartmann, A. K. (2007). Fractal dimension of domain walls in two-dimensional Ising spin glasses. *Physical Review B—Condensed Matter and Materials Physics*, *76*(17), 174411.
[152] Risau-Gusman, S., & Romá, F. (2008). Fractal dimension of domain walls in the Edwards-Anderson spin glass model. *Physical Review B—Condensed Matter and Materials Physics*, *77*(13), 134435.
[153] Wang, W., Moore, M. A., & Katzgraber, H. G. (2018). Fractal dimension of interfaces in Edwards-Anderson spin glasses for up to six space dimensions. *Physical Review E*, *97*(3), 032104.
[154] Khoshbakht, H., & Weigel, M. (2018). Domain-wall excitations in the two-dimensional Ising spin glass. *Physical Review B*, *97*(6), 064410.
[155] Wang, W., Machta, J., & Katzgraber, H. G. (2016). Bond chaos in spin glasses revealed through thermal boundary conditions. *Physical Review B*, *93*(22), 224414.
[156] Nishimori, H., Ohzeki, M., & Okuyama, M. (2025). Temperature chaos as a logical consequence of the reentrant transition in spin glasses. *Physical Review E*, *112*(4), 044140.
[157] Li, H., He, J., & Orbach, R. L. (2025). Nature of temperature chaos in spin glasses. *Physical Review B*, *111*(10), 104309.
[158] Aguilar-Janita, M., Franz, S., Martin-Mayor, V., Moreno-Gordo, J., Parisi, G., Ricci-Tersenghi, F., & Ruiz-Lorenzo, J. J. (2024). Small field chaos in spin glasses: Universal predictions from the ultrametric tree and comparison with numerical simulations. *Proceedings of the National Academy of Sciences*, *121*(40), e2404973121.
[159] Zhai, Q., Orbach, R. L., & Schlagel, D. L. (2022). Evidence for temperature chaos in spin glasses. *Physical Review B*, *105*(1), 014434.
[160] Sasaki, M., Hukushima, K., Yoshino, H., & Takayama, H. (2005). Temperature Chaos and Bond Chaos in Edwards-Anderson Ising Spin Glasses:<? format?> Domain-Wall Free-Energy Measurements. *Physical review letters*, *95*(26), 267203.
[161] Sherrington, D. (1975). A transparent theory of the spin glass. *Journal of Physics C: Solid State Physics*, *8*(10), L208.
[162] Villain, J. (1979). Insulating spin glasses. *Zeitschrift für Physik B Condensed Matter*, *33*(1), 31-42.
[163] Maletta, H., & Crecelius, G. (1976). ONSET OF COLLECTIVE MAGNETIC BEHAVIOUR IN MAGNETICALLY DILUTED INSULATING Eu++ COMPOUNDS. *Le Journal de Physique Colloques*, *37*(C6), C6-645.

[164] Maletta, H., & Felsch, W. (1979). Insulating spin-glass system $Eu_xSr_{1-x}S$. *Physical review B*, *20*(3), 1245.
[165] Ferré, J., Pommier, J., Renard, J., & Knorr, K. (1980). Magnetic properties of an amorphous insulating spin glass: manganese aluminosilicate. Journal of Physics C: Solid State Physics, 13(19), 3697.
[166] Garcıa-Landa, B., De Teresa, J. M., Ibarra, M. R., Ritter, C., Drost, R., and Lees, M. R. (1998). Colossal magneto-resistance in $Gd_{1/2}Sr_{1/2}MnO_3$. Journal of applied physics, 83(12), 7664-7667.
[167] Nigam, A. K., & Majumdar, A. K. (1983). Magnetoresistance in canonical spin-glasses. *Physical Review B*, *27*(1), 495.
[168] Sagar, S., Ganesan, V., Joy, P. A., Thomas, S., Liebig, A., Albrecht, M., and Anantharaman, M. R. (2010). Colossal thermoelectric power in Gd-Sr manganites. Europhysics Letters, 91(1), 17008.
[169] Joy, L. K., Singh, D., Sudeep, P. M., Ganesan, V., Ajayan, P. M., Thomas, S., and Anantharaman, M. R. (2015). Size effect on the colossal thermoelectric power in charge ordered small band width manganites based on Gd-Sr. Materials Research Express, 2(5), 055504.
[170] Nagaraja, B. S., Rao, A., and Okram, G. S. (2016). Structural, electrical, magnetic and thermal studies on $Eu_{1-x}Sr_xMnO_3$ ($0.2\leq x\leq 0.5$) manganites. Journal of Alloys and Compounds, 683, 308-317
[171] Kadomtseva, A. M., Popov, Y. F., Vorob'ev, G. P., Ivanov, V. Y., Mukhin, A. A., and Balbashov, A. M. (2005). Observation of electric polarization in $Gd_{1-x}Sr_xMnO_3$ (x= 0.5, 0.6, 0.7) single crystals. Journal of Experimental and Theoretical Physics Letters, 82, 590-593.
[172] Terai, T., Sasaki, T., Kakeshita, T., Fukuda, T., Saburi, T., Kitagawa, H., ... and Honda, M. (2000). Electronic and magnetic properties of $R_{0.5}A_{0.5}MnO_3$ compounds (R= Gd, Dy, Ho, Er; A= Sr, Ca). Physical Review B, 61(5), 3488.
[173] Kusters, 1. M., Singleton, J., Keen, D. A., McGreevy, R., & Hayes, W. (1989). Magnetoresistance measurements on the magnetic semiconductor $Nd_{0.5}Pb_{0.5}MnO_3$. *Physica B: Condensed Matter*, *155*(1-3), 362-365.
[174] Hundley, M. F., Hawley, M., Heffner, R. H., Jia, Q. X., Neumeier, J. J., Tesmer, J., ... & Wu, X. D. (1995). Transport-magnetism correlations in the ferromagnetic oxide $La_{0.7}Ca_{0.3}MnO_3$. *Applied physics letters*, *67*(6), 860-862.
[175] Liu, Y., Qin, X. Y., Xin, H. X., Zhang, J., Li, H. J., and Wang, Y. F. (2007). Electrical transport and thermoelectric properties of $Y_{1-x}Ca_xCoO_3$ ($0\leq x\leq 0.1$) at high temperatures. Journal of applied physics, 101(8), 083709.
[176] Harikrishnan, S., Kumar, C. N., Bhat, H. L., Elizabeth, S., Rößler, U. K., Dörr, K., ... and Wirth, S. (2008). Investigationson the spin-glass state in $Dy_{0.5}Sr_{0.5}MnO_3$ single crystals through structural, magnetic and thermal properties. Journal of Physics: Condensed Matter, 20(27), 275234.
[177] Jakob, G., Westerburg, W., Martin, F., & Adrian, H. (1998). Small-polaron transport in $La_{0.67}Ca_{0.33}MnO_3$ thin films. *Physical Review B*, *58*(22), 14966.
[178] Snyder, G. J., Hiskes, R., DiCarolis, S., Beasley, M. R., & Geballe, T. H. (1996). Intrinsic electrical transport and magnetic properties of $La_{0.67}Ca_{0.33}MnO_3$ and $La_{0.67}Sr_{0.33}MnO_3$ MOCVD thin films and bulk material. *Physical Review B*, *53*(21), 14434.
[179] Mott, N. (2004). *Metal-insulator transitions*. CRC Press.
[180] Mott, N. F., & Davis, E. A. (2012). *Electronic processes in non-crystalline materials*. OUP Oxford.
[181] Hassen, A., Ali, A. I., Kim, B. J., Wu, Y. S., Park, S. H., and Kim, B. G. (2007). Structural, magnetic, and electric properties of $Dy_{1-x}Sr_xCoO_{3-\delta}$ ($0.65\leq x\leq 0.90$). Journal of Applied Physics, 102(12), 123905.
[182] Sagar, S., and Anantharaman, M. R. (2012). On conduction mechanism in paramagnetic phase of Gd based manganites. Bulletin of Materials Science, 35, 41-45.
[183] Chatterjee, S., & Nigam, A. K. (2002). Spin-glass-like behavior in $Y_{1-x}Sr_xMnO_3$ (x= 0.5 and 0.6). *Physical Review B*, *66*(10), 104403
[184] Nagaraja, B. S., Rao, A., & Okram, G. S. (2015). Structural, electrical, and colossal thermoelectric properties of $Dy_{1-x}Sr_xMnO_3$ manganites. *Journal of Superconductivity and Novel Magnetism*, *28*, 223-229.
[185] Nagaraja, B. S., Rao, A., Babu, P. D., & Okram, G. S. (2015). Structural, electrical, magnetic and thermal properties of $Gd_{1-x}Sr_xMnO_3$ ($0.2\leq x\leq 0.5$) manganites. *Physica B: Condensed Matter*, *479*, 10-20.
[186] Ahad, A., Shukla, D. K., Rahman, F., Majid, S., Okram, G. S., Sinha, A. K., & Phase, D. M. (2017). Colossal thermopower, spin states and delocalization effects in single layered $La_{2-x}Sr_xCoO_4$. *Acta Materialia*, *135*, 233-243.
[187] Joy, L. K., Shanmukharao Samatham, S., Thomas, S., Ganesan, V., Al-Harthi, S., Liebig, A., ... & Anantharaman, M. R. (2014). Colossal thermoelectric power in charge ordered lanthanum calcium manganites ($La_{0.5}Ca_{0.5}MnO_3$). *Journal of Applied Physics*, *116*(21).
[188] Mandal, P. (2000). Temperature and doping dependence of the thermopower in $LaMnO_3$. *Physical Review B*, *61*(21), 14675.
[189] Fischer, K. H. (1981). Thermopower of spin glasses. *Zeitschrift für Physik B Condensed Matter*, *42*, 245-252.
[190] Anantharaman, M. R., Sagar, S., Ganesan, V., Joy, P. A., Senoy, T., Liebig, A., & Albrecht, M. (2010). Colossal thermoelectric power in Gd-Sr manganites.

[191] Coey, J. M. D., Viret, M. V., & Von Molnar, S. (1999). Mixed-valence manganites. *Advances in physics*, *48*(2), 167-293.
[192] Kong, W. J., Lu, L., Zhu, H. W., Wei, B. Q., & Wu, D. H. (2005). Thermoelectric power of a single-walled carbon nanotubes strand. *Journal of Physics: Condensed Matter*, *17*(12), 1923.
[193] Wang, X. L., Horvat, J., Liu, H. K., Li, A. H., & Dou, S. X. (2001). Spin glass state in $Gd_2CoMnO_6$ perovskite manganite. *Solid state communications*, *118*(1), 27-30.
[194] Liu, D., Wu, S., Long, X., Wang, Y., Li, H. F., Yang, M., ... & Li, S. (2021). Hidden Frustration in Double-Perovskite $CaFeTi_2O_6$. *arXiv preprint arXiv:2105.03547*.
[195] Wu, H., Shi, X. L., Liu, W. D., Li, M., Gao, H., Zhou, W., ... & Chen, Z. G. (2021). Double perovskite $Pr_2CoFeO_6$ thermoelectric oxide: Roles of Sr-doping and Micro/nanostructuring. *Chemical Engineering Journal*, *425*, 130668.
[196] Oba, A., Kamishima, K., Kakizaki, K., & Hiratsuka, N. (2012). Substitution effect on thermoelectric properties of double Perovskite $Sr_2FeMoO_6$. *Transactions of the Materials Research Society of Japan*, *37*(2), 267-270.
[197] Hasselmann, N., Neto, A. C., & Smith, C. M. (2004). Spin-glass phase of cuprates. *Physical Review B*, *69*(1), 014424.
[198] Kanazawa, I. (2000). The spin glass mechanism in high-$T_c$ cuprates. *Physica B: Condensed Matter*, *284*, 409-410.
[199] Kanazawa, I. (2001). The chiral-like spin-glass mechanism in high-$T_c$ cuprates. *Physica C: Superconductivity*, *357*, 149-152.
[200] Panagopoulos, C., Tallon, J. L., Rainford, B. D., Xiang, T., Cooper, J. R., & Scott, C. A. (2002). Evidence for a generic quantum transition in high-$T_c$ cuprates. *Physical Review B*, *66*(6), 064501.
[201] Malinowski, A., Bezusyy, V. L., Minikayev, R., Dziawa, P., Syryanyy, Y., & Sawicki, M. (2011). Spin-glass behavior in Ni-doped $La_{1.85}Sr_{0.15}CuO_4$. *Physical Review B*, *84*(2), 024409.
[202] Stilp, E., Suter, A., Prokscha, T., Morenzoni, E., Keller, H., Wojek, B. M., ... & Božović, I. (2013). Magnetic phase diagram of low-doped $La_{2-x}Sr_xCuO_4$ thin films studied by low-energy muon-spin rotation. *Physical Review B—Condensed Matter and Materials Physics*, *88*(6), 064419.
[203] Panagopoulos, C. (2004, May). Spin glass order in high-$T_c$ cuprates. In *Fluctuations and Noise in Materials* (Vol. 5469, pp. 125-128). SPIE.
[204] Raičević, I., Jaroszyński, J., Popović, D., Panagopoulos, C., & Sasagawa, T. (2008). Evidence for Charge Glasslike Behavior in Lightly Doped $La_{2-x}Sr_xCuO_4$ at Low Temperatures. *Physical review letters*, *101*(17), 177004.
[205] Cutler, M., & Mott, N. F. (1969). Observation of Anderson localization in an electron gas. *Physical Review*, *181*(3), 1336.
[206] Matusiak, M., Gnida, D., & Kaczorowski, D. (2011). Quantum criticality in Ce 2 PdIn 8: A thermoelectric study. *Physical Review B—Condensed Matter and Materials Physics*, *84*(11), 115110.
[207] Zhang, X., Li, C., Yang, M., Zhao, Y., An, Z., Li, D., ... & Li, S. (2025). Thermoelectricity evidence for quantum criticality in clean infinite-layer nickelate films. *arXiv preprint arXiv:2508.12974*.
[208] Nakamura, T., Yashiro, K., Sato, K., & Mizusaki, J. (2009). Electrical Conductivity and Thermoelectric Power of $La_{2-x}Sr_xNiO_{4+\delta}$. *ECS Transactions*, *16*(51), 317.
[209] Gardner, J. S., Gingras, M. J., & Greedan, J. E. (2010). Magnetic pyrochlore oxides. *Reviews of Modern Physics*, *82*(1), 53-107.
[210] Lee, S. H., Ratcliff, W., Huang, Q., Kim, T. H., & Cheong, S. W. (2008). Néel to spin-glass-like phase transition versus dilution in geometrically frustrated $ZnCr_{2-2x}Ga_{2x}O_4$. *Physical Review B—Condensed Matter and Materials Physics*, *77*(1), 014405.
[211] Mittelstädt, L., Schmidt, M., Wang, Z., Mayr, F., Tsurkan, V., Lunkenheimer, P., ... & Loidl, A. (2015). Spin-orbiton and quantum criticality in $FeSc_2S_4$. *Physical Review B*, *91*(12), 125112.
[212] Theumann, A., Coqblin, B., Magalhães, S. G., & Schmidt, A. A. (2001). Spin-glass freezing in Kondo-lattice compounds. *Physical Review B*, *63*(5), 054409.
[213] Nolas, G. S., Poon, J., & Kanatzidis, M. (2006). Recent developments in bulk thermoelectric materials. *MRS bulletin*, *31*(3), 199-205.
[214] T. Holmes, A., Jaccard, D., & Miyake, K. (2007). Valence instability and superconductivity in heavy fermion systems. *Journal of the Physical Society of Japan*, *76*(5), 051002.
[215] Kalinowski, L., Kądziołka-Gaweł, M., & Ślebarski, A. (2018). Cluster spin-glass-like behavior in heavy-fermion-filled cage $Ce_3Co_4Sn_{13}$ doped with Fe: Magnetic and Mössbauer effect studies. *Physical Review B*, *98*(24), 245140.
[216] Dhar, S. K., Gschneidner Jr, K. A., Bredl, C. D., & Steglich, F. (1989). Magnetic susceptibility and low-temperature heat capacity of $CePd_3B_{0.3}$. *Physical Review B*, *39*(4), 2439.
[217] Lausberg, S., Spehling, J., Steppke, A., Jesche, A., Luetkens, H., Amato, A., ... & Steglich, F. (2012). Avoided Ferromagnetic Quantum Critical Point:<? format?> Unusual Short-Range Ordered State in CeFePO. *Physical Review Letters*, *109*(21), 216402.

[218] Yan, X., Wu, J., Lei, X., He, L., Guo, W., Kuang, X., & Yin, C. (2023). Magnetic clusters and ferromagnetic spin glass in the novel hexagonal perovskite 12R-$Ba_4SbMn_3O_{12}$. *RSC advances*, *13*(17), 11234-11240.
[219] Belik, A. A., Liu, R., & Yamaura, K. (2022). Dielectric and Spin-Glass Magnetic Properties of the A-Site Columnar-Ordered Quadruple Perovskite $Sm_2CuMn(MnTi_3)O_{12}$. *Materials*, *15*(23), 8306.
[220] Ruan, M. Y., Li, T. Y., Wang, L., & Luo, Q. C. (2023). Magnetization sign reversal, spin-glass-like state and Griffiths-like phase driven by B-site disorder in double perovskite $Gd_2CrMnO_6$. *Journal of Alloys and Compounds*, *940*, 168787.
[221] Kearins, P., Solana-Madruga, E., Ji, K., Ritter, C., & Attfield, J. P. (2021). Cluster spin glass formation in the double double perovskite $CaMnFeTaO_6$. *The Journal of Physical Chemistry C*, *125*(17), 9550-9555.
[222] . Elouafi, A., Ounza, Y., Omari, L. H., Oubla, M., Lassri, M., Sajieddine, M., & Lassri, H. (2021). Spin-glass-like behavior and magnetocaloric properties in $LaBiCaMn_2O_7$ layered perovskite. *Applied Physics A*, *127*(4), 216.
[223] Yatsuzuka, H., Haraguchi, Y., Matsuo, A., Kindo, K., & Katori, H. A. (2022). Spin-glass transition in the spin–orbit-entangled $J_{eff} = 0$ Mott insulating double-perovskite ruthenate. *Scientific Reports*, *12*(1), 2429.
[224] Patra, K. P., & Ravi, S. (2022). Re-entrant spin glass and magnetocaloric effect in frustrated double perovskite $Ho_2CoMnO_6$ flat nanorod. *Journal of Magnetism and Magnetic Materials*, *559*, 169537.
[225] Ye, X., Liu, Z., Wang, W., Hu, Z., Lin, H. J., Weng, S. C., ... & Long, Y. (2019). High-pressure synthesis and spin glass behavior of a Mn/Ir disordered quadruple perovskite $CaCu_3Mn_2Ir_2O_{12}$. *Journal of Physics: Condensed Matter*, *32*(7), 075701.
[226] Liu, R., Khalyavin, D. D., Tsunoda, N., Kumagai, Y., Oba, F., Katsuya, Y., ... & Belik, A. A. (2019). Spin-Glass Magnetic Properties of A-Site Columnar-Ordered Quadruple Perovskites $Y_2MnGa(Mn_{4-x}Ga_x)O_{12}$ with $0 \leq x \leq 3$. *Inorganic Chemistry*, *58*(21), 14830-14841.
[227] Sharma, S., Yadav, P., Sau, T., Yanda, P., Baker, P. J., da Silva, I., ... & Lalla, N. P. (2019). Evidence of a cluster spin-glass state in B-site disordered perovskite $SrTi_{0.5}Mn_{0.5}O_3$. *Journal of Magnetism and Magnetic Materials*, *492*, 165671.
[228] Guo, J., Ye, X., Liu, Z., Wang, W., Qin, S., Zhou, B., ... & Long, Y. (2019). High-pressure synthesis of A-site ordered perovskite $CaMn_3(Fe_3Mn)O_{12}$ and sequential long-range antiferromagnetic ordering and spin glass transition. *Journal of Solid State Chemistry*, *278*, 120921.
[229] Palakkal, J. P., Faske, T., Major, M., Radulov, I., Komissinskiy, P., & Alff, L. (2020). Ferrimagnetism, exchange bias and spin-glass property of disordered $La_2CrNiO_6$. *Journal of Magnetism and Magnetic Materials*, *508*, 166873.
[230] Pradheesh, R., Nair, H. S., Haripriya, G. R., Senyshyn, A., Chatterji, T., Sankaranarayanan, V., & Sethupathi, K. (2017). Magnetic glass state and magnetoresistance in $SrLaFeCoO_6$ double perovskite. *Journal of Physics: Condensed Matter*, *29*(9), 095801.
[231] Lekshmi, P. N., Pillai, S. S., Suresh, K. G., Santhosh, P. N., & Varma, M. R. (2012). Room temperature relaxor ferroelectricity and spin glass behavior in $Sr_2FeTiO_6$ double perovskite. *Journal of alloys and compounds*, *522*, 90-95.
[232] Pal, A., Singh, P., Gangwar, V. K., Ghosh, S., Prakash, P., Saha, S. K., ... & Chatterjee, S. (2019). B-site disorder driven multiple-magnetic phases: Griffiths phase, re-entrant cluster glass, and exchange bias in $Pr_2CoFeO_6$. *Applied Physics Letters*, *114*(25).
[233] Chau, N., Tho, N. D., Luong, N. H., Giang, B. H., & Cong, B. T. (2006). Spin glass-like state, charge ordering, phase diagram and positive entropy change in $Nd_{0.5-x}Pr_xSr_{0.5}MnO_3$ perovskites. *Journal of Magnetism and Magnetic Materials*, *303*(2), e402-e405.
[234] Aczel, A. A., Zhao, Z., Calder, S., Adroja, D. T., Baker, P. J., & Yan, J. Q. (2016). Structural and magnetic properties of the 5 d 2 double perovskites $Sr_2BReO_6$ (B= Y, In). *Physical Review B*, *93*(21), 214407.
[235] Gorrín, A. L., Núñez, P., de la Torre, M. L., de Paz, J. R., & Puche, R. S. (2002). Spin Glass Behavior of New Perovskites $Ba_2In_{2-x}Co_xO_5$ ($0.5 \leqslant x \leqslant 1.70$). *Journal of Solid State Chemistry*, *165*(2), 254-260.
[236] Takeuchi, J., Uemura, A., Miyoshi, K., & Fujiwara, K. (2000). Colossal magnetoresistance and spin-glass behavior of the perovskite $Nd_{0.67}Sr_{0.33}Mn_{1-x}Fe_xO_3$. *Physica B: Condensed Matter*, *281*, 489-490.
[237] Lin, H., Gawryluk, D. J., Klein, Y. M., Huangfu, S., Pomjakushina, E., von Rohr, F., & Schilling, A. (2022). Universal spin-glass behaviour in bulk $LaNiO_2$, $PrNiO_2$ and $NdNiO_2$. *New Journal of Physics*, *24*(1), 013022.
[238] Jelbert, G. R., Sasagawa, T., Fletcher, J. D., Park, T., Thompson, J. D., & Panagopoulos, C. (2008). Measurement of low energy charge correlations in underdoped spin-glass La-based cuprates using impedance spectroscopy. *Physical Review B—Condensed Matter and Materials Physics*, *78*(13), 132513.
[239] Huangfu, S., Guguchia, Z., Cheptiakov, D., Zhang, X., Luetkens, H., Gawryluk, D. J., ... & Schilling, A. (2020). Short-range magnetic interactions and spin-glass behavior in the quasi-two-dimensional nickelate $Pr_4Ni_3O_8$. *Physical Review B*, *102*(5), 054423.

[240] Cardoso, C. A., Araujo-Moreira, F. M., Awana, V. P. S., Kishan, H., Takayama-Muromachi, E., & de Lima, O. F. (2004). Magnetic properties of the $RuSr_2Ln_{1.5}Ce_{0.5}Cu_2O_{10-\delta}$ (Ln= Y, Ho and Dy) and $RuSr_2YCu_2O_{8-\delta}$ rutheno-cuprate families: a comparative study. *Physica C: Superconductivity*, *405*(3-4), 212-218.
[241] Rayaprol, S., Singh, K., Jammalamadaka, S. N., Mohapatra, N., Gaur, N. K., & Sampathkumaran, E. V. (2008). Magnetic anomalies in a new manganocuprate $Gd_3Ba_2Mn_2Cu_2O_{12}$. *Solid state communications*, *147*(9-10), 353-356.
[242] Beach, K. S. D., & Gooding, R. J. (2000). Spin twists, domain walls, and the cluster spin-glass phase of weakly doped cuprates. *The European Physical Journal B-Condensed Matter and Complex Systems*, *16*(4), 579-591.
[243] Tranquada, J. M. (1996). Stripe correlations of spins and holes in cuprate superconductors. *Journal of Superconductivity*, *9*(4), 397-399.
[244] Clark, L., Nilsen, G. J., Kermarrec, E., Ehlers, G., Knight, K. S., Harrison, A., ... & Gaulin, B. D. (2014). From spin glass to quantum spin liquid ground states in molybdate pyrochlores. *Physical review letters*, *113*(11), 117201.
[245] Taniguchi, T., Munenaka, T., & Sato, H. (2009). Spin glass behavior in metallic pyrochlore ruthenate $Ca_2Ru_2O_7$. In *Journal of Physics: Conference Series* (Vol. 145, No. 1, p. 012017). IOP Publishing.
[246] Babu, G. S., Valant, M., Page, K., Llobet, A., Kolodiazhnyi, T., & Axelsson, A. K. (2011). New $(Bi_{1.88}Fe_{0.12})(Fe_{1.42}Te_{0.58})\ O_{6.87}$ pyrochlore with spin-glass transition. *Chemistry of Materials*, *23*(10), 2619-2625.
[247] Singh, D. K., & Lee, Y. S. (2012). Nonconventional spin glass transition in a chemically ordered pyrochlore. *Physical review letters*, *109*(24), 247201.
[248] Zhou, H. D., Wiebe, C. R., Janik, J. A., Vogt, B., Harter, A., Dalal, N. S., & Gardner, J. S. (2010). Spin glass transitions in the absence of chemical disorder for the pyrochlores $A_2Sb_2O_7$ (A= Mn, Co, Ni). *Journal of Solid State Chemistry*, *183*(4), 890-894.
[249] Singh, K., & Panwar, N. (2023). Observation of Griffiths phase, spin glass behaviour and magnetocaloric effect in frustrated $Ga_2Mn_{2-x}Cr_xO_7$ (x= 0, 0.1, 0.3, 0.5) pyrochlore compounds. *Current Applied Physics*, *53*, 86-93.
[250] Dwivedi, V. K. (2023). Identification of a Griffiths-like phase and its evolution in the substituted pyrochlore iridates $Y_2Ir_{2-x}Cr_xO_7$ (x= 0.0, 0.05, 0.1, 0.2). *Physical Review B*, *107*(13), 134413.
[251] Rahman, R. A. U., Ruth, D. J., & Ramaswamy, M. (2020). Emerging scenario on displacive cubic bismuth pyrochlores (Bi, M) $MNO_{7-\delta}$ (M= transition metal, N= Nb, Ta, Sb) in context of their fascinating structural, dielectric and magnetic properties. *Ceramics International*, *46*(10), 14346-14360.
[252] Castellano, C., Berti, G., Rubio-Marcos, F., Lamura, G., Sanna, S., Salas-Colera, E., ... & Demartin, F. (2017). Anomalous local lattice disorder and distortion in $A_2Mo_2O_7$ pyrochlores. *Journal of Alloys and Compounds*, *723*, 327-332.
[253] Li, Y., Kan, X., Liu, X., Feng, S., Lv, Q., Wang, W., ... & Xu, Y. (2020). Spin-Glass Behavior in Spinel Compound $ZnCoTiO_4$. *Journal of Superconductivity and Novel Magnetism*, *33*(12), 3745-3752.
[254] Li, Y., Kan, X., Liu, X., Feng, S., Lv, Q., Rehman, K. M. U., ... & Xu, Y. (2021). Spin-glass evolution behavior in spinel compounds $Co_{2-x}Zn_xSnO_4$ (0≤ x≤ 1). *Journal of Alloys and Compounds*, *852*, 156962.
[255] Groń, T., Krok-Kowalski, J., Pacyna, A. W., & Mydlarz, T. (2009). Spin-glass-like behaviour and magnetic order in $Mn[Cr_{0.5}Ga_{1.5}]S_4$ spinels. *Journal of Physics and Chemistry of Solids*, *70*(5), 900-905.
[256] Patra, P., Naik, I., Kaushik, S. D., & Mohanta, S. (2022). Crossover from meta-magnetic state to spin-glass behaviour upon Ti-substitution for Mn in $CuMn_2O_4$. *Journal of Materials Science: Materials in Electronics*, *33*(1), 554-564.
[257] Yang, W., Kan, X., Liu, X., Wang, Z., Chen, Z., Wang, Z., ... & Shezad, M. (2019). Spin glass behavior in $Zn_{0.8-x}Ni_xCu_{0.2}Fe_2O_4$ (0≤ x≤ 0.28) ferrites. *Ceramics International*, *45*(17), 23328-23332.
[258] Chen, H., Yang, X., Zhang, P. S., Liang, L., Hong, Y. Z., Wei, Y. J., ... & Wang, C. Z. (2015). Observation of spin glass transition in spinel $LiCoMnO_4$. *Chinese Physics B*, *24*(12), 127501.
[259] Borah, R., Maji, D., & Ravi, S. (2023). Evidence of cluster glass phase, exchange bias, reversible magnetocaloric effect and study of electrical properties in $Ni_{1-x}Cd_xCr_2O_4$ (x= 0–0.30) spinel. *Applied Physics A*, *129*(4), 298.
[260] Das, A., Pal, P., Ranaut, D., Islam, K. P., Bhattacharya, G., Mehta, S., ... & Choudhury, D. (2023). Origin of the long-range ferrimagnetic ordering in cubic Mn (Co) $Cr_2O_4$ spinels. *Physical Review B*, *107*(10), L100414.
[261] Kushwaha, S., & Nagarajan, R. (2023). Spin-glass behavior and redox catalytic properties of room temperature produced $ZnCrMnO_4$. *Ceramics International*, *49*(1), 683-690.
[262] Zheng, Y., Hussain, G., Li, S., Batool, S., & Wang, X. (2022). Effects of Rhenium Substitution of Co and Fe in Spinel $CoFe_2O_4$ Ferrite Nanomaterials. *Nanomaterials*, *12*(16), 2839.
[263] Nagata, S., Koseki, N., & Ebisu, S. (2012). Spin-glass in the spinel-type $CuCrTiS_4$. *Philosophical Magazine*, *92*(23), 2957-2969.
[264] Kumar, N. R., Karthik, R., Vasylechko, L., & Selvan, R. K. (2020). Reentrant spin-glass behaviour in highly frustrated Mn-rich spinel zinc manganate. *Journal of Physics: Condensed Matter*, *32*(24), 245802.

[265] Chowdhury, M. R., Seehra, M. S., Pramanik, P., Ghosh, S., Sarkar, T., Weise, B., & Thota, S. (2022). Antiferromagnetic short-range order and cluster spin-glass state in diluted spinel $ZnTiCoO_4$. *Journal of Physics: Condensed Matter*, *34*(27), 275803.
[266] . Ishikawa, T., Ebisu, S., & Nagata, S. (2010). Spin-glass and novel magnetic behavior in the spinel-type $Cu_{1-x}Ag_xCrSnS_4$. *Physica B: Condensed Matter*, *405*(7), 1881-1889.
[267] Hanashima, K., Kodama, Y., Akahoshi, D., Kanadani, C., & Saito, T. (2013). Spin glass order by antisite disorder in the highly frustrated spinel oxide $CoAl_2O_4$. *Journal of the Physical Society of Japan*, *82*(2), 024702.
[268] Nishioka, T., Tabata, Y., Taniguchi, T., & Miyako, Y. (2000). Canonical spin glass behavior in $Ce_2AgIn_3$. *Journal of the Physical Society of Japan*, *69*(4), 1012-1015.
[269] Tien, C., Lu, J. J., & Jang, L. Y. (2002). Random magnetic moments and spin-glass-like behaviors in the heavy-fermion compound $CeNi_2Sn_2$. *Physical Review B*, *65*(21), 214416.
[270] Muro, Y., Takahashi, S., Sunahara, K., Motoya, K., Akatsu, M., & Shirakawa, N. (2007). Heavy-fermion behavior in $CeRh_2Si$. *Journal of Magnetism and Magnetic Materials*, *310*(2), e40-e41.
[271] Toliński, T., Kowalczyk, A., Falkowski, M., Synoradzki, K., Chełkowska, G., Hoser, A., & Rols, S. (2012). From Heavy Fermion and Spin-Glass Behavior to Magnetic Order in $CeT_4M$ Compounds. *Acta Physica Polonica A*, *121*(5-6), 1014-1018.
[272] Niitaka, S., Nishikawa, K., Kimura, S., Narumi, Y., Kindo, K., Hagiwara, M., & Takagi, H. (2007). High-field magnetization study of the heavy fermion oxide $LiV_2O_4$. *Journal of Magnetism and Magnetic Materials*, *310*(2), e258-e260.
[273] Anand, V. K., Hossain, Z., Adroja, D. T., & Geibel, C. (2011). Signatures of spin-glass behaviour in $PrIr_2B_2$ and heavy fermion behaviour in $PrIr_2B_2C$. *Journal of Physics: Condensed Matter*, *23*(37), 376001.
[274] Xie, D. H., Zhang, W., Liu, Y., Feng, W., Zhang, Y., Tan, S. Y., ... & Lai, X. C. (2016). Spin-cluster glass state in $U(Ga_{0.95}Mn_{0.05})_3$. *Chinese Physics B*, *25*(4), 047502.
[275] Sievers, K., Scheidt, E. W., & Stewart, G. R. (1995). Heavy-fermion-spin-glass behavior in $Sn_{(1-x)}U_xPt_3$. *Physica B: Condensed Matter*, *206*, 433-436.
[276] Mydosh, J. A. (1999). Disorder and frustration in heavy-fermion compounds. *Physica B: Condensed Matter*, *259*, 882-886.
[277] Wehrle, V., Müller, T., Schröder, A., Sürgers, C., & Löhneysen, H. V. (1992). Heavy-fermion behavior and spin-glass freezing in Si-stabilized amorphous alloys based on $UPt_3$. *Zeitschrift für Physik B Condensed Matter*, *89*, 161-167.
[278] Maksimov, I., Litterst, F. J., Menzel, D., Schoenes, J., Menovsky, A. A., Mydosh, J. A., & Süllow, S. (2002). Irreversibility lines of the heavy fermion spin glass $URh_2Ge_2$. *Physica B: Condensed Matter*, *312*, 289-291.
[279] Hemberger, J., Krimmel, A., Nicklas, M., Knebel, G., Paraskevopoulos, M., Trinkl, W., ... & Loidl, A. (1999). Magnetic properties of the spin glass $PrAu_2Si_2$. *Physica B: Condensed Matter*, *259*, 907-908.
[280] Okajima, H., Sumiyama, K., Homma, Y., Hihara, T., & Suzuki, K. (1995). Heavy fermion characteristics in amorphous $SmCu_6$ alloy. *Physica B: Condensed Matter*, *206*, 392-394.
[281] Li, D., Homma, Y., Honda, F., Yamamura, T., & Aoki, D. (2015). Low temperature spin-glass behavior in non-magnetic atom disorder compound $Pr_2CuIn_3$. *Physics Procedia*, *75*, 703-710.
[282] Nimori, S., & Li, D. (2006). Studies of the reentrant spin-glass behavior in $Dy_2PdSi_3$. *journal of the physical society of japan*, *75*(Suppl), 195-197.
[283] Sachdev, S., & Read, N. (1996). Metallic spin glasses. *Journal of Physics: Condensed Matter*, *8*(48), 9723.
[284] Hewson, A. C. (1997). *The Kondo problem to heavy fermions* (No. 2). Cambridge university press.
[285] Gracià, R. S., Nieuwenhuizen, T. M., & Lerner, I. V. (2004). Concentration dependence of the transition temperature in metallic spin glasses. *Europhysics Letters*, *66*(3), 419.
[286] Cuervo-Reyes, E. (2013). Metallic spin-glasses beyond mean-field: an approach to the impurity-concentration dependence of the freezing temperature. *Journal of Physics: Condensed Matter*, *25*(13), 136006.
[287] Markovich, V., Fita, I., Wisniewski, A., Jung, G., Mogilyansky, D., Puzniak, R., ... & Gorodetsky, G. (2010). Spin-glass-like properties of $La_{0.8}Ca_{0.2}MnO_3$ nanoparticles ensembles. *Physical Review B—Condensed Matter and Materials Physics*, *81*(13), 134440.
[288] Bag, P., Baral, P. R., & Nath, R. (2018). Cluster spin-glass behavior and memory effect in $Cr_{0.5}Fe_{0.5}Ga$. *Physical Review B*, *98*(14), 144436.
[289] Krause, J. K., Long, T. C., Egami, T., & Onn, D. G. (1980). Low-temperature specific heat of the metallic glasses $Fe_x\ Ni_{80-x}P_{14}B_6$ and $(Fe_yNi_{100-y})_{79}\ P_{13}B_8$ for the spin-glass and spin-cluster-glass regimes. *Physical Review B*, *21*(7), 2886.
[290] Adler, J., van Enter, A. C., & Harris, A. B. (1988). Transmission of order in a correlated spin glass. *Physical Review B*, *38*(16), 11405.

[291] Benka, G., Bauer, A., Schmakat, P., Säubert, S., Seifert, M., Jorba, P., & Pfleiderer, C. (2022). Interplay of itinerant magnetism and spin-glass behavior in $Fe_xCr_{1-x}$. *Physical review materials*, *6*(4), 044407.
[292] Pal, S., Thiebes, Y., Niewa, R., Nigam, A. K., & Roy, S. B. (2023). Nonequilibrium magnetic response in concentrated spin-glass $Au_{0.89}Fe_{0.11}$ alloy. *Journal of Magnetism and Magnetic Materials*, *570*, 170504.
[293] FORD, P., & MYDOSH, J. (1974). TRANSPORT PROPERTIES OF SOME SPIN GLASS SYSTEMS. *Le Journal de Physique Colloques*, *35*(C4), C4-241.
[294] Mehmood, S., & Ali, Z. (2023). DFT study of the spin glass and ferrimagnetism in quadruple perovskites $CaCu_3B_2Ir_2O_{12}$ (B= Mn, Fe, Co, and Ni) for spintronic applications. *Applied Physics A*, *129*(1), 76.
[295] Trexler, M. M., & Thadhani, N. N. (2010). Mechanical properties of bulk metallic glasses. *Progress in Materials Science*, *55*(8), 759-839.
[296] Shao, L., Xue, L., Luo, Q., Wang, Q., & Shen, B. (2019). The role of Co/Al ratio in glass-forming GdCoAl magnetocaloric metallic glasses. *Materialia*, *7*, 100419.
[297] Du, J., Zheng, Q., Brück, E., Buschow, K. H. J., Cui, W. B., Feng, W. J., & Zhang, Z. D. (2009). Spin-glass behavior and magnetocaloric effect in Tb-based bulk metallic glass. *Journal of magnetism and magnetic materials*, *321*(5), 413-417.
[298] Jin, F., Pang, C. M., Wang, X. M., & Yuan, C. C. (2023). The role of rare earth elements in tailorable thermal and magnetocaloric properties of RE-Co-Al (RE= Gd, Tb, and Dy) metallic glasses. *Journal of Non-Crystalline Solids*, *600*, 121992.
[299] Yi, K., Tang, Q., Wu, Z., Gu, J., & Zhu, X. (2023). Structural, magnetic, and electrical transport properties of half-metallic double perovskite $La_2CrNiO_6$ oxides. *Journal of Alloys and Compounds*, *933*, 167742.
[300] Duong, T. V., Dung, D. D., Nguyen, Q. V., Duong, A. T., Nguyen, C. X., Nguyen, H. T., & Sung, C. L. (2023). Thermoelectric, Magnetic Properties and Re-entrant Spin-glass State in MBE Grown FeAs Film on $LaAlO_3$ (100) Substrate. *ECS Journal of Solid State Science and Technology*.
[301] Geohegan, J. A., & Bhagat, S. M. (1981). Magnetic phases of amorphous transition metal-metalloid alloys. *Journal of Magnetism and Magnetic Materials*, *25*(1), 17-32.
[302] Uemura, Y. J., Yamazaki, T., Harshman, D. R., Senba, M., & Ansaldo, E. J. (1985). Muon-spin relaxation in AuFe and CuMn spin glasses. *Physical Review B*, *31*(1), 546.
[303] Mizoguchi, T., McGuire, T. R., Kirkpatrick, S., & Gambino, R. J. (1977). Measurement of the Spin-Glass Order Parameter in Amorphous $Gd_{0.37}Al_{0.63}$. *Physical Review Letters*, *38*(2), 89.
[304] Marchenkov, V. V., Irkhin, V. Y., & Semiannikova, A. A. (2022). Unusual kinetic properties of usual Heusler alloys. *Journal of Superconductivity and Novel Magnetism*, *35*(8), 2153-2168.
[305] Elphick, K., Frost, W., Samiepour, M., Kubota, T., Takanashi, K., Sukegawa, H., ... & Hirohata, A. (2021). Heusler alloys for spintronic devices: review on recent development and future perspectives. *Science and technology of advanced materials*, *22*(1), 235-271.
[306] Hames, F. A., & Crangle, J. (1971). Ferromagnetism in Heusler-type alloys based on platinum-group or palladium-group metals. *Journal of Applied Physics*, *42*(4), 1336-1338.
[307] Kroder, J., Manna, K., Kriegner, D., Sukhanov, A. S., Liu, E., Borrmann, H., ... & Felser, C. (2019). Spin glass behavior in the disordered half-Heusler compound IrMnGa. *Physical Review B*, *99*(17), 174410.v
[308] Sharma, A. K., Jena, J., Rana, K. G., Markou, A., Meyerheim, H. L., Mohseni, K., ... & Parkin, S. S. (2021). Nanoscale noncollinear spin textures in thin films of a $D_{2d}$ Heusler compound. *Advanced Materials*, *33*(32), 2101323.
[309] Ibarra, R., Lesne, E., Ouladdiaf, B., Beauvois, K., Sukhanov, A. S., Wawrzyńczak, R., ... & Markou, A. (2022). Noncollinear magnetic order in epitaxial thin films of the centrosymmetric MnPtGa hard magnet. *Applied Physics Letters*, *120*(17).
[310] Gupta, S., Chakraborty, S., Bhasin, V., Pakhira, S., Biswas, A., Mudryk, Y., ... & Mazumdar, C. (2023). 4 d element induced improvement of structural disorder and development of weakly reentrant spin-glass behavior in NiRuMnSn. *Physical Review B*, *108*(5), 054405.
[311] Xu, Q., Zhou, S., Wu, D., Uhlarz, M., Tang, Y. K., Potzger, K., ... & Schmidt, H. (2010). Cluster spin glass behavior in Bi ($Fe_{0.95}Co_{0.05}$) $O_3$. *Journal of Applied Physics*, *107*(9).
[312] Zhang, W., Sun, Y., Wang, H., Li, Y., Zhang, X., Sui, Y., ... & Wu, G. (2014). The spin glass behavior in the Heusler alloy $Cu_2VAl$. *Journal of alloys and compounds*, *589*, 230-233.
[313] Zhang, Y., Li, J., Tian, F., Cao, K., Wang, D., Ren, S., ... & Song, X. (2019). Spin cluster size dependence of exchange bias effect in $Mn_{50}Ni_{40}Ga_{10}$ Heusler alloys. *Intermetallics*, *107*, 10-14.
[314] Liu, C., Jing, C., Zhang, Y., Liu, Y., Sun, J., Huang, Y., ... & Li, Z. (2017). Exchange bias and spin glass transition in quaternary MnCuNiSn Heusler alloy. *Journal of Magnetism and Magnetic Materials*, *444*, 61-67.
[315] Cao, K., Tian, F., Huang, S., Zhang, Y., Zhao, Q., Yao, K., ... & Yang, S. (2021). Tuning the exchange bias effect via thermal treatment temperature in bulk $Ni_{50}Mn_{42}In_3Sb_5$ Heusler alloys. *Applied Physics Express*, *14*(10), 105502.

[316] Hiroi, M., Rokkaku, T., Matsuda, K., Hisamatsu, T., Shigeta, I., Ito, M., ... & Terada, N. (2009). Ferromagnetism and spin-glass transitions in the Heusler compounds $Ru_{2-x}Fe_xCrSi$. *Physical Review B—Condensed Matter and Materials Physics*, *79*(22), 224423.
[317] Khorwal, A. K., Dash, S., Kumar, A., Lukoyanov, A. V., Shreder, E. I., Bitla, Y., ... & Patra, A. K. (2022). Evidence for canonical spin glass behaviour in polycrystalline $Mn_{1.5}Fe_{1.5}Al$ Heusler alloy. *Journal of Magnetism and Magnetic Materials*, *546*, 168752.
[318] Sokolovskiy, V., Buchelnikov, V., & Entel, P. (2014). Theoretical prediction of the spin glass behavior in the low-temperature phase of $Ni_2Mn_{1.36}In_{0.64}$ Heusler alloy. *physica status solidi (c)*, *11*(5-6), 1110-1115.
[319] Kaštil, J., Kamarád, J., Friák, M., Míšek, M., Dutta, U., Kral, P., ... & Arnold, Z. (2023). Cluster spin glass in off-stoichiometric $Ni_{2.01}Mn_{1.58}Sn_{0.41}$ alloy. *Intermetallics*, *161*, 107995.
[320] Li, Y., Lu, N. L., Shi, S., Jin, Y. F., Han, Z. D., Lin, Y. C., ... & Jiang, X. F. (2018). Suppression of reentrant spin glass and induced zero-field-cooled exchange bias by lattice contraction in NiMnSbAl alloys. *Journal of Alloys and Compounds*, *766*, 791-795.
[321] Khan, M., & Albagami, A. (2017). Cooling field dependent exchange bias in $Mn_2Ni_{1.4}Ga_{0.6}$: A reentrant spin glass system with short range ferromagnetic ordering. *Journal of Alloys and Compounds*, *727*, 100-106.
[322] Hiroi, M., Sano, H., Tazoko, T., Shigeta, I., Ito, M., Koyama, K., ... & Kindo, K. (2017). Magnetic and electrical properties of Heusler compounds $Ru_2Cr_{1-x}X_xSi$ (X= V, Ti). *Journal of Alloys and Compounds*, *694*, 1376-1382.
[323] Zhang, Y. J., Zeng, Q. Q., Wei, Z. Y., Hou, Z. P., Liu, Z. H., Liu, E. K., ... & Wu, G. H. (2018). Cluster spin glass state caused by antiphase boundaries in NiFeGa alloys. *Journal of Alloys and Compounds*, *749*, 134-139.
[324] Yu, G., Chen, H., Ma, S., Luo, X., Liu, C., Chen, C., ... & Zhong, Z. (2022). Robust topological Hall effect in a reentrant spin glass system $Mn_{1.89}Pt_{0.98}Ga_{1.12}$. *Physica B: Condensed Matter*, *640*, 414043.
[325] Ma, L., Wang, W. H., Lu, J. B., Li, J. Q., Zhen, C. M., Hou, D. L., & Wu, G. H. (2011). Coexistence of reentrant-spin-glass and ferromagnetic martensitic phases in the $Mn_2Ni_{1.6}Sn_{0.4}$ Heusler alloy. *Applied Physics Letters*, *99*(18).
[326] Pan, H., Ma, L., Li, G. K., Jia, L. Y., Zhen, C. M., Hou, D. L., ... & Wu, G. H. (2017). Large exchange bias effect in the super spin glass state of $Mn_{50}Ni_{38}Al_{12}$ alloy. *Intermetallics*, *86*, 116-120.
[327] Borgohain, B., Siwach, P. K., Singh, N., Awana, V. P. S., & Singh, H. K. (2019). Martensitic ferromagnetism and spin glass behaviour in $Ni_{47}Mn_{36}Cr_4Sn_{13}$ ribbons. *Intermetallics*, *111*, 106492.
[328] Raji, G. R., Uthaman, B., Rajan, R. K., MP, S., Thomas, S., Suresh, K. G., & Varma, M. R. (2016). Martensitic transition, spin glass behavior and enhanced exchange bias in Si substituted $Ni_{50}Mn_{36}Sn_{14}$ Heusler alloys. *RSC Advances*, *6*(38), 32037-32045.
[329] Zhang, L., Zhang, J., Li, K., Zhou, C., Yao, Y., Tan, T. T., ... & Carpenter, M. A. (2020). Glassy Magnetic Transitions and Accurate Estimation of Magnetocaloric Effect in Ni–Mn Heusler Alloys. *ACS Applied Materials & Interfaces*, *12*(39), 43646-43652.
[330] Nakai, Y., Sakuma, M., & Kunitomi, N. (1987). Magnetic Phase Diagram of Competing Antiferromagnet Au–Cr. *Journal of the Physical Society of Japan*, *56*(1), 301-310.
[331] Coles, B. R., Sarkissian, B. V. B., & Taylor, R. H. (1978). The role of finite magnetic clusters in Au-Fe alloys near the percolation concentration. *Philosophical Magazine B*, *37*(4), 489-498.
[332] Sharma, S., Amaladass, E. P., & Mani, A. (2017). Defect induced magnetism and super spin glass state in reactive ion beam deposited nano-structured AlN thin films. *Materials & Design*, *131*, 204-209.
[333] Sahu, B., Fobasso, R. D., & Strydom, A. M. (2022). Spin–glass behavior in Shastry–Sutherland lattice of $Tm_2Cu_2In$. *Journal of Magnetism and Magnetic Materials*, *543*, 168599.
[334] Maletta, H. (1982). Magnetic ordering in $Eu_xSr_{1-x}S$, a diluted Heisenberg system with competing interactions. *Journal of Applied physics*, *53*(3), 2185-2190.
[335] Westerholt, K., & Bach, H. (1981). $Eu_xSr_{1-x}S_ySe_{1-y}$: A Model System for Studying Competing Magnetic Interactions. *Physical Review Letters*, *47*(26), 1925.
[336] Zirngiebl, E., Güntherodt, G., & Maletta, H. (1984). Raman scattering in the spin glass system $Eu_xSr_{1-x}S$ due to spin correlation effects. *Solid state communications*, *52*(2), 131-134.
[337] Mydosh, J. A. (1978). Spin glasses—recent experiments and systems. *Journal of Magnetism and Magnetic Materials*, *7*(1-4), 237-248.
[338] Maletta, H. (2005, May). The ferromagnetic to spin glass crossover in $Eu_xSr_{1-x}$. In *Heidelberg Colloquium on Spin Glasses: Proceedings of a Colloquium held at the University of Heidelberg 30 May–3 June, 1983* (pp. 90-102). Berlin, Heidelberg: Springer Berlin Heidelberg.
[339] Lei, H., Bozin, E. S., Wang, K., & Petrovic, C. (2011). Antiferromagnetism in semiconducting $KFe_{0.85}Ag_{1.15}Te_2$ single crystals. *Physical Review B—Condensed Matter and Materials Physics*, *84*(6), 060506.
[340] Sun, F., Guo, Z., Liu, N., Lin, K., Wang, D., & Yuan, W. (2017). $KFeCuTe_2$: a new compound to study the removal of interstitial Fe in layered tellurides. *Dalton Transactions*, *46*(11), 3649-3654.

[341] Wang, B., Guo, Z., Sun, F., Deng, J., Lin, J., Wu, D., & Yuan, W. (2019). The transition between antiferromagnetic order and spin-glass state in layered chalcogenides $KFeAgCh_2$ (Ch= Se, S). *Journal of Solid State Chemistry*, *272*, 126-130.

[342] Yuan, D., Liu, N., Li, K., Jin, S., Guo, J., & Chen, X. (2017). Structure evolution and spin-glass transition of layered compounds $ALiFeSe_2$ (A= Na, K, Rb). *Inorganic Chemistry*, *56*(21), 13187-13193.

[343] Oledzka, M., Ramanujachary, K. V., & Greenblatt, M. (1996). Physical properties of quaternary mixed transition metal sulfides: $ACuFeS_2$ (A= K, Rb, Cs). *Materials research bulletin*, *31*(12), 1491-1499.

[344] Eggenkamp, P. J. T., Vennix, C. W. H. M., Story, T., Swagten, H. J. M., Swüste, C. H. W., & de Jonge, W. J. M. (1994). Magnetic study of the diluted magnetic semiconductor $Sn_{1-x}Mn_xTe$. *Journal of Applied Physics*, *75*(10), 5728-5730.

[345] Pekarek, T. M., Watson, E. M., Garner, J., Shand, P. M., Miotkowski, I., & Ramdas, A. K. (2007). Nonlinear magnetization behavior near the spin-glass transition in the layered III-VI diluted magnetic semiconductor $Ga_{1-x}Mn_xS$. *Journal of applied physics*, *101*(9), 09D511.

[346] Ochoa, J. M., Bindilatti, V., Ter Haar, E., Coaquira, J. A. H., de Souza Brito, G. E., Gratens, X., & Sagredo, V. (2004). Spin glass behavior in $MnIn_2Se_4$ and $Zn_{1-x}Mn_xIn_2Se_4$ magnetic semiconductors. *Journal of Magnetism and Magnetic Materials*, *272*, 1308-1309.

[347] Tirgar, M., Barati Abgarmi, H., & Abouie, J. (2025). Magnetic interactions and cluster formation: Boosting surface thermopower in topological insulators. *Physical Review B*, *112*(11), 115158.

[348] Choi, J., Choi, S., Choi, J., Park, Y., Park, H. M., Lee, H. W., ... & Cho, S. (2004). Magnetic properties of Mn-doped Bi2Te3 and Sb2Te3. *physica status solidi (b)*, *241*(7), 1541-1544.

[349] Choi, J., Lee, H. W., Kim, B. S., Park, H., Choi, S., Hong, S. C., & Cho, S. (2006). Magnetic and transport properties of Mn-doped $Bi_2Se_3$ and $Sb_2Se_3$. *Journal of Magnetism and Magnetic Materials*, *304*(1), e164-e166.

[350] Yu, S., Zhao, G., Peng, Y., Wang, X., Liu, Q., Yu, R., ... & Jin, C. (2020). (Ba, K)(Zn, Mn)$_2$Sb$_2$: A New Type of Diluted Magnetic Semiconductor. *Crystals*, *10*(8), 690.

[351] Gu, G., Zhao, G., Lin, C., Li, Y., Jin, C., & Xiang, G. (2018). Asperomagnetic order in diluted magnetic semiconductor (Ba, Na)(Zn, Mn)$_2$As$_2$. *Applied Physics Letters*, *112*(3).

[352] Maksymowicz, L. J., Lubecka, M., & Jabłoński, R. (2000). Magnetic semiconductor thin films of $CdCr_{2-2x}In_{2x}Se_4$ in spin glass state. *Journal of magnetism and magnetic materials*, *215*, 579-581.

[353] Bandaranayake, R. J., Lin, J. Y., Jiang, H. X., & Sorensen, C. M. (1997). Synthesis and properties of $Cd_{1-x}Mn_xS$ diluted magnetic semiconductor ultrafine particles. *Journal of magnetism and magnetic materials*, *169*(3), 289-302.

[354] Paul, K. B. (2018). The magnetic state in the binary $Dy_xGe_{1-x}$ ($x \leq 0.02$) alloy semiconductor. *Journal of Magnetism and Magnetic Materials*, *460*, 471-479.

[355] Mihalache, V., Cernea, M., & Pasuk, I. (2017). Relationship between ferromagnetism and, structure and morphology in un-doped ZnO and Fe-doped ZnO powders prepared by hydrothermal route. *Current Applied Physics*, *17*(8), 1127-1135.

[356] Massey, M. C., Manuel, I., Edwards, P. S., Parker, D., Pekarek, T. M., & Haraldsen, J. T. (2018). Magnetic impurity bands in $Ga_{1-x}Mn_xS$: Towards understanding the anomalous spin-glass transition. *Physical Review B*, *98*(15), 155206.

[357] Khaliq, A., Minikayev, R., Arciszewska, M., Avdonin, A., Brodowska, B., Khan, A., ... & Kilanski, L. (2022). Spin-glass like magnetic ordering in $Ge_{1-xy}(Sn_xMn_y)Te$ multiferroics. *Journal of Magnetism and Magnetic Materials*, *544*, 168695.

[358] Górska, M., Kilanski, L., Podgórni, A., Dobrowolski, W., Szymczak, R., Anderson, J. R., ... & Slynko, E. I. (2016, December). Magnetic properties of clusters in IV-VI and II-IV-$V_2$ diluted magnetic semiconductors. In *International Symposium on Clusters and Nanomaterials* (Vol. 10174, pp. 113-123). SPIE.

[359] Peng, J., Yang, X., Lu, Z., Huang, L., Chen, X., He, M., ... & Liu, J. M. (2023). Ferromagnetism induced by magnetic dilution in van der Waals material metal thiophosphates. *Advanced Quantum Technologies*, *6*(3), 2200105.

[360] Janik, E., Dynowska, E., Bak-Misiuk, J., Wojtowicz, T., Karczewski, G., Kossut, J., ... & Ando, K. (1998). Zinc-blende $Mg_{1-x}Mn_xTe$—a new diluted magnetic semiconductor system. *Journal of crystal growth*, *184*, 976-979.

[361] Yu, S., Peng, Y., Zhao, G., Zhao, J., Wang, X., Zhang, J., ... & Jin, C. (2023). Colossal negative magnetoresistance in spin glass Na (Zn, Mn) Sb. *Journal of Semiconductors*, *44*(3), 032501.

[362] Gnatenko, Y. P., & Bukivskij, P. M. (2015). Optical observation of spin-glass formation in $Cd_{1-X}Mn_XTe$ compound. *Materials Research Express*, *2*(9), 095903.

[363] Skipetrov, E. P., Bogdanov, E. V., Kovalev, B. B., Skipetrova, L. A., Knotko, A. V., Emelyanov, A. V., ... & Slynko, V. E. (2022). Electronic structure and unusual magnetic properties of diluted magnetic semiconductors $Pb_{1-x-y}Sn_xSc_yTe$. *Journal of Alloys and Compounds*, *893*, 162330.

[364] Mounkachi, O., El Moussaoui, H., Masrour, R., Ilali, J., Mediouri, K. E., Hamedoun, M., ... & Benyoussef, A. (2014). High freezing temperature in $SnO_2$ based diluted magnetic semiconductor. *Materials Letters*, *126*, 193-196.

[365] Pekarek, T. M., Watson, E. M., Shand, P. M., Miotkowski, I., & Ramdas, A. K. (2010). Spin-glass ordering in the layered III-VI diluted magnetic semiconductor $Ga_{1-x}Mn_xS$. *Journal of Applied Physics*, *107*(9).
[366] Shand, P. M., Christianson, A. D., Pekarek, T. M., Martinson, L. S., Schweitzer, J. W., Miotkowski, I., & Crooker, B. C. (1998). Spin-glass ordering in the diluted magnetic semiconductor $Zn_{1-x}Mn_xTe$. *Physical Review B*, *58*(19), 12876.
[367] Salmani, E. M., Laghrissi, A., Lamouri, R., Ez-Zahraouy, H., & Benyoussef, A. (2017). The electronic, magnetic and optical properties of ZnO doped with doubles impurities (Cr, Fe): An LDA-SIC and Monte Carlo study. *Journal of Magnetism and Magnetic Materials*, *422*, 348-355.
[368] Lumetzberger, J., Ney, V., Zhakarova, A., Daffe, N., Primetzhofer, D., & Ney, A. (2023). Room-temperature spin glass behavior in zinc ferrite epitaxial thin films. *arXiv preprint arXiv:2301.11277*.
[369] Yamahara, H., Seki, M., Adachi, M., Takahashi, M., Nasu, H., Horiba, K., ... & Tabata, H. (2015). Spin-glass behaviors in carrier polarity controlled $Fe_{3-x}Ti_xO_4$ semiconductor thin films. *Journal of Applied Physics*, *118*(6), 063905.
[370] Yu, J., Wu, W., Zhang, H., Shao, R., Zhang, F., Wang, H., ... & Wang, W. (2022). Robust spin glass state with exceptional thermal stability in a chemically complex alloy. *Physical Review Materials*, *6*(9), L091401.
[371] Wang, X. L., James, M., Horvat, J., & Dou, S. X. (2002). Spin glass behaviour in ferromagnetic $La_2CoMnO_6$ perovskite manganite. *Superconductor Science and Technology*, *15*(3), 427.